\documentclass{article}
\pdfoutput=1
\usepackage{amsfonts,amssymb,amsmath}
\usepackage[dvips]{epsfig}
\usepackage[T1]{fontenc}
\usepackage[latin1]{inputenc}
\usepackage{graphicx}
\usepackage[english]{babel}
\usepackage{amsmath}
\usepackage{amssymb}
\usepackage{amsfonts}
\usepackage{color}

\begin{document}

\title{\bf   Surrounded  Vaidya Solution by Cosmological Fields}
\author{Y. Heydarzade\thanks{%
email: heydarzade@azaruniv.ac.ir}~~and~~F. Darabi\thanks{%
email: f.darabi@azaruniv.ac.ir} ,
\\{\small Department of Physics, Azarbaijan Shahid Madani University, 53714-161,
Tabriz, Iran}}
\maketitle
\begin{abstract}
In the present work,
we study the general
surrounded Vaidya  solution by the various cosmological fields and its nature describing  the possibility of the
formation of naked singularities or black holes. 
   Motivated by the fact that {\it real} astrophysical black
holes as \textit{non-stationary} and  \textit{non-isolated} objects
 are  living in non-empty backgrounds, we focus on the black hole subclasses
of this general solution describing
  a dynamical evaporating-accreting  black holes in the  dynamical cosmological backgrounds
of dust, radiation, quintessence, cosmological constant-like and phantom fields, the so called   ``\textit{surrounded Vaidya black hole}''. Then, we
analyze the timelike geodesics associated with the obtained surrounded
black holes and we find that some new correction terms arise relative to the case of Schwarzschild black hole. Also, we address some of the subclasses of the obtained surrounded black hole solution  for both dynamical
and stationary limits. Moreover,  we classify the obtained solutions according to their behaviors under imposing the positive
energy condition and discuss how this condition imposes some severe and important
restrictions on the black hole and its background field dynamics.
\end{abstract}
%%%%%%%%%%%%%%%%%%%%%%%%%%%%%%%%%%%%%%%%%%%%%%%%%%%%%%%%%%%%%%%%
%%%%%%%%%%%%%%%%%%%%%%%%%%%%%%%%%%%%%%%%%%%%%%%%%%%%%%%%%%%%%%%%%%%%%%%%%%%%%%%%%
%
\section{Introduction}
Nowadays, we know that black holes are
not just a mathematically possible solution to Einstein's
field equations, rather they seem to be  some  realistic astrophysical objects. It is more than a
decade that we have obtained good evidences indicating that most of the galaxies,
as our Milky Way, host many stellar active black holes, as well
as a super-massive active black hole, in their centers.
On the other hand, due to black hole evaporation \cite{haw} and accretion-absorbtion
processes \cite{acc, shap}, it is accepted that the mass and other parameters of black holes
are not fixed, and should change with time. Therefore, generally speaking, real black holes are non-stationary, and the stationary black holes such as Schwarzschild
and Reissner-Nordström are only ideal models. Thus, the study of non-stationary black
holes is  meaningful and so motivating in the exploration of real black holes. There
are a lot of research works on general dynamical black holes and their properties, see the works of  Ashtekar $\&$ Krishnan \cite{dy1} and Hayward \cite{dy2} as instances. Actually, studying black holes from astrophysical point of view and by astrophysicists has been originated in recent decades
due to the dramatic increase in the number of black hole candidates from the sole candidate Cygnus X-1. This  study needs a deeper understanding
of the black hole physics and especially the black hole radiation
by astrophysicists and relativists. Hawking 
used a quantum field theoretical approach to explore the black hole radiation,
for the first time \cite{haw}. Afterwards, some  models to describe the
classical essence of this radiation in a language which is free from the usual quantum field theoretic tools and is
more familiar to the astrophysicists and relativists, have been introduced.
 For instance, the Vaidya solution \cite{vaidyajoon, vaidyajoon1} has provided a simple classical model  for the black hole
radiation and has been vastly investigated in this regard \cite{evap1, evap3, evap4, evap6, parikh}, see also \cite{v1,
v2, v3, v4, v5, v6, v7, v8, v9,v10, v11, v12} for more studies. 
In fact, the Vaidya solution  is one of the non-static solutions of the Einstein
field equations and can be regarded as a generalization of
the static Schwarzschild black hole solution. This solution is characterized by a dynamical mass function depending on the retarded time coordinate $u$, i.e $m=m(u)$ and an
ingoing/outgoing  flow $\sigma(u,r)$. Thus, it can be implemented as a classical model for a dynamical
black hole which is effectively evaporating  or accreting, regarding its
effective flow direction.   On the other hand, the Vaidya solution has been used for studying the process of spherical symmetric gravitational collapse and as a testing
ground for the cosmic censorship conjecture \cite{ccc,
naked1, naked2, naked3, col, kur}, see also \cite{hark}
where  a
possible astrophysical application of the model for describing the energy source of gamma-ray bursts is discussed. These studies also are motivated by the time
dependant mass parameter of this solution along with its outgoing radiation flow during
the  collapse ending by a naked singularity or a black hole. It is shown that
if the outgoing flux diverges,  the back-reaction will 
prevent the  formation of  naked singularity \cite{hic}. The observable sign of the formation of a naked singularity, by the collapse process,  appears to be the burst of a radiation possessing a non-thermal spectrum, as the Cauchy horizon is approached  \cite{vaz}. Indeed, this is in contrast to the slow 
evaporation of a  black hole via black-body spectrum of the Hawking radiation
\cite{haw}. Then, it would be important 
to carefully investigate Vaidya solution  to better understanding of  real
dynamical black holes or the typical signs of  naked singularities and to explore if there are any astrophysical objects whose properties resemble those of a naked singularity \cite{vaz}. The Vaidya  solution was  generalized to the charged case known as the Bonnor-Vaidya solution \cite{bonnor}, see also its application for example in \cite{sullivan, kaminga, pati, wanglin, xiang}.
Also,  a generalisation of the Vaidya solution is introduced
in \cite{wanguli}. This generalisation is based on the fact that the total supporting energy-momentum tensor   of spacetime, constructed from type I and type II
energy-momentum tensors \cite{type},  is linear in term of the mass function.
Consequently, any linear superposition of particular solutions to the Einstein
field equations will also a solution. Then, using this approach,  we can  construct more general solutions such
as the Bonnor-Vaidya \cite{bonnor}, Vaidya-de Sitter \cite{mallet},
radiating dyon solution \cite{dyon}, Bonnor-Vaidya-de Sitter \cite{BVdS,
Patino, Mallet*, koberlin, saida} and
the Husain solution \cite{husain}.

On the other hand, a new exact static solution to Einstein field equations has been recently introduced
 by Kiselev \cite{Kiselev}. Actually, the
Kiselev solution is nothing but the static generalization of the Schwarzschild solution to include a non-empty cosmological background, especially well
known for the quintessence background. This generalization is well motivated by the fact that  black holes in real world are not isolated and are not embedded in empty backgrounds. The black hole solutions coupled to  matter  fields, such as  Kiselev solution,  are of interest
in studying astrophysical distorted black holes \cite{dis1, dis2, dis3, dis4}, as well as in exploring the `no hair' theorems \cite{no, hus, str, wig}. Indeed, a crucial assumption for the no-hair theorem is that
the black hole is isolated, i.e., the spacetime is asymptotically
flat and contains no other sources. However, in real world astrophysical situations this requirement is not fulfilled, for examples,
 for black holes in binary systems, for black
holes  surrounded by plasma, or black holes having an accretion disk or jets
 in their vicinity. All these situations indicate that a black
hole may put on different
types of wigs. For these cases, the standard no-hair theorem for the isolated
black holes can be questioned, see for examples \cite{hus, bek}. In a recent
 research, the authors of \cite{jamil} discussed on distinguishing rotating Kiselev black hole from naked singularity using spin precession of test gyroscope.
In general, since black holes possess strong gravitational attraction such that their nearby matter, even light,
cannot escape from their gravitational field, they  cannot be observed directly
and  there are some different
ways to detect them in binary systems as well as at the centers of
their host galaxies. The  most promising way is the accretion process. In the language
of astrophysics, the accretion is defined as
the inward flow of matter fields surrounding a compact object, such as black
holes and neutron stars,  due to the gravitational attraction.
Then, the process of accretion into black holes  is one of the
most interesting research fields in relativistic astrophysics \cite{accretion1*, accretion2*, accretion3*, accretion4*, accretion5*}. This process may be described by a
perfect fluid coupled to general relativity representing  a plasma which
obeys
the equations of ideal or resistive magnetohydrodynamics  or a fluid coupled to
radiation.  Such  accretion processes along with their detailed physical
descriptions, can be found in \cite{accretion1}
and references therein, see also \cite{accretion6*, accretion6**, accretion7*, accretion8*, accretion10*, accretion11*, accretion12*, accretion13*}. On the other hand, there are also other
kind of accretion processes related to the  black holes surrounded by  exotic matter fields as potential
models of  dark energy,
whose existence and features  are motivated by the problems in the standard
model of cosmology.
A number of theoretical and observational studies confirmed that our universe in its
early stages experienced an inflation process while it is undergoing an accelerated expansion in the late time. In
order to explain these events, an energy component, known  as the
dark energy, is required to be  introduced
to the framework of general theory of relativity.
The cosmological constant is a leading candidate for
dark energy while there are other proposals including the dynamical scalar fields such as
quintessence and phantom fields.
In the Bousso's work \cite{bousso}, one finds that ``\textit{Q-space exhibits
thermodynamic properties similar to those of the de Sitter
horizon. The horizon radius in Q-space grows linearly
with time, and consequently the temperature slowly decreases.
We find that this behavior is consistent with the
first law of thermodynamics: the temperature and entropy
respond appropriately to the flux of quintessence
stress-energy across the horizon}", 
 for a cosmological setup, where ``Q-space" stands for quintessence-space. This important result
along with the observational data confirming a dark energy fluid responsible
for the accelerating expansion of Universe with the
equation of state parameter $\omega<-1/3$, has motivated the community to study
in detail the black hole solutions in the quintessence background.  For some
recent  studies of Kiselev black holes,  see \cite{gosh} for its generalization to rotating case,  \cite{kis1, kis2, kis3}
for quasinormal modes and Hawking radiation, \cite{gader, prad, az, kof} for thermodynamical
studies, \cite{geo1, geo2, geo3, geo4, geo5, geo6} for trajectories and particle dynamics around this black hole, \cite{ac} for accretion process and \cite{lens}
for gravitational lensing among the
others.
 One should also note that the Kiselev solution can be implemented for more
generic backgrounds of dust, radiation, quintessence, cosmological constant
and phantom fields as well as for any realistic combination of these cosmological fields.
Then,  by the presence of such  fields
around the black holes, one may have interest to  explore some  interesting
facts such as whether black holes have hair or scalar wigs
\cite{accretion2},  how black holes affect
these  cosmological surrounding fields and what are the  consequences
or what are the  influences of these surrounding fields  on
the features, behaviors and abundance of black holes.
In this regard,
one may find the reference  \cite{accretion3} as a good review including
 various scenarios of accretion process into
black holes, see also \cite{accretion6**, accretion3**} for charged black
hole accretion.
Among the all of the accretion processes, the most interesting one are related
to those that the accretion of the surrounding fields enforcing a
black hole to shrink. These surrounding field include the scalar fields or fluid
violating the weak energy condition, i.e $\rho>0~\&~  \rho+p>0$ \cite{accretion3}.
Specific scenarios involving the accretion of phantom
energy have shown that the black-hole area decreases with
the accretion \cite{accretion4, accretion5, accretion6, gao}. For example, in
 \cite{accretion4},  it is shown that black holes will gradually vanish as the universe approaches a cosmological big rip state. The big rip scenario for
a cosmos occurs when its filling dark energy is  the phantom energy with  $p< -\rho$. In this scenario,
the cosmological phantom field disrupts finally all bounded objects of the
universe up to sub-nuclear scales. For the test-field approximation, one may find  the accretion process of a scalar field violating the energy conditions leading  the decrease in the black holes area
  in \cite{accretion6, accretion7}. Moreover, the shrink of the black hole area through the
accretion of a phantom scalar field has been confirmed in
full nonlinear general relativity \cite{accretion8}. In this
regard, the shrink of the black hole area
 by the accretion of a potentially surrounding field is
an interesting phenomena in the sense that it can be an alternative
process for black hole evaporation through the Hawking radiation or even
be an auxiliary for speeding up it. One  physical explanation for a black hole
mass diminishing may be is that accreting particles of a phantom
scalar field have a total negative energy \cite{phantom}. Similar
 particles possessing negative energies are created through the Hawking
radiation process and also  in the   energy extraction process from a black
hole by the Penrose mechanism.
The effect of phantom-like dark energy onto a charged
Reissner-Nordström black hole is studied in \cite{accretion9**} and it is found that accretion is possible only through the outer horizon. On the other hand, for scalar fields regarding the energy conditions, there is a possibility  indicating that the accretion of a scalar field can be  partial such that the amount of accreted scalar field depends on features of the incident wave packet, i.e. the wave number and the width of the packet. This has
been studied both in the test-field approximation \cite{accretion10} and in full general relativity \cite{accretion8}. In this line, some
  studies in the test-field limit indicate that a scalar field can also be sustained by a black hole without being accreted \cite{accretion11}.

In the present work,
following the approach of \cite{Kiselev, Heydarzade} introduced for the static black holes, and motivated by the facts that real astrophysical black
holes are neither \textit{stationary} nor \textit{isolated} and are not embedded in empty backgrounds, we wish to find a  more realistic dynamical solution for the classical description of the evaporating-accreting black holes in  generic dynamical backgrounds.
The organization of the paper is as follows. In section 2, we introduce the general
surrounded Vaidya  solution, its nature describing  the possibility of the
formation of naked singularities or black holes, interaction of its possible black holes
with their backgrounds as well as its timelike geodesic analysis in the general
form. Then, in subsections 3 to 7, we
investigate in detail the special classes of this solution as the surrounded Vaidya black
hole by the dust, radiation, quintessence, cosmological constant-like and phantom
fields, respectively. The paper ends with a conclusion, in section 8.
%%%%%%%%%%%%%%%%%%%%%%%%%%%%%%%%%%%%%%%%%%%%%%%%%%%%%%%%%%%%%
\section{The General Surrounded Vaidya 
Solutions}
In this section, we are looking for the general surrounded  Vaidya  solutions by the approach of \cite{Kiselev, Heydarzade}.
Then, we consider the general spherical symmetric spacetime metric in the form of
\begin{equation}\label{metricvaidya}
ds^{2}=-f(u,r)du^2+2\epsilon dudr+r^2d\Omega^2,~~~\epsilon=\pm1,
\end{equation}
where
$d\Omega^2=d\theta^2+sin^2\theta d\phi^2$ is the two dimensional unit sphere
and $f(u,r)$ is a generic metric function depending on both of the advanced/retarded time
coordinate $u$ and the radial coordinate $r$. The cases, $\epsilon=-1$ and
$\epsilon=+1$
represent the outgoing and ingoing flows corresponding to
the effectively evaporating and accreting Vaidya black hole solutions, respectively.
Using the metric (\ref{metricvaidya}), we obtain nonvanishing components of the
Einstein tensor as
\begin{eqnarray}\label{gmunu}
&&{G^{0}}_{0}={G^{1}}_{1}=\epsilon G_{01}=\epsilon G_{10}=\frac{1}{r^2}\left(f^{\prime}r-1+f  \right),\nonumber\\
&&{G^{1}}_{0}=\epsilon G_{00}+fG_{01}=-\frac{\dot f}{r},\nonumber\\
&&{G^{2}}_{2}=\frac{1}{r^2}G_{22}=\frac{1}{r^2}\left(rf^{\prime}+\frac{1}{2}r^2 f^{\prime\prime}\right),\nonumber\\
&&{G^{3}}_{3}=\frac{1}{r^2 sin^2 \theta}G_{33}=\frac{1}{r^2}\left(rf^{\prime}+\frac{1}{2}r^2 f^{\prime\prime}\right),
\end{eqnarray}
where dot and prime signs represent the derivatives with respect to the
time coordinate $u$ and the radial coordinate $r$, respectively.
Then, the total energy-momentum supporting this spacetime should have the following non-diagonal form
\begin{equation}\label{*energyT*}
{T^{\mu}}_{\nu}=
\begin{pmatrix}{T^{0}}_{0} & 0 & 0 & 0 \\
{T^{1}}_{0} & {T^{1}}_{1} & 0 & 0 \\
0 & 0 & {T^{2}}_{2} & 0 \\
0 & 0 & 0 &  {T^{3}}_{3}\\
\end{pmatrix},
\end{equation}
where also must obey the symmetries in Einstein tensor ${G^{\mu}}_{\nu}$.
With respect to the field equations in (\ref{gmunu}), the equalities  ${G^{0}}_{0}={G^{1}}_{1}$ and ${G^{2}}_{2}={G^{3}}_{3}$ require ${T^{0}}_{0}={T^{1}}_{1}$ and
 ${T^{2}}_{2}={T^{3}}_{3}$, respectively. Then, for the nature of the Vaidya solution in the presence of a dynamical background, one can consider a total energy-momentum tensor supporting the Einstein field equations in the following form
\begin{equation}\label{EMtotal}
{T^{\mu}}_{\nu}={\tau^{\mu}}_{\nu}+{\mathcal{T}^{\mu}}_{\nu},
\end{equation}
where ${\tau^{\mu}}_\nu$  is the energy-momentum tensor
associated to the Vaidya null radiation-accretion as%
\begin{equation}\label{null}
{\tau^{\mu}}_{\nu}=\sigma k^{\mu}k_{\nu},
\end{equation}
such that  $\sigma=\sigma(u,r)$ is the measure of the energy flux or the energy density of the outgoing radiation-ingoing accretion flow \cite{schwartz} and  $k_\mu={\delta^{0}}_{\mu} $ is a null vector field while ${\mathcal{T}^{\mu}}_{\nu}$ is the energy-momentum tensor of the surrounding
fluid defined as  in \cite{Kiselev}
 \begin{eqnarray}\label{energysur}
&&{\mathcal{T}^{0}}_{0}=-\rho_s(u,r),\nonumber\\
&&{\mathcal{T}^{i}}_{j}=-\rho_{s}(u,r)\alpha\left[-(1+3\beta)\frac{r_i r^j}{r_n r^n}+\beta{\delta^{i}}_{j}\right],
\end{eqnarray}
where subscript $``s"$ stands for the surrounding field which can be a dust,
radiation, quintessence, cosmological constant, phantom field or even  any
complex
field constructed by the combination
of these fields\footnote{In the sections 3-7, we will use the subscripts ``d,
r, q, c'' and  ``p'', instead of the general subscript ``s'', for  denoting the surrounding dust, radiation, quintessence,
cosmological constant-like and phantom fields, respectively.}. This form of energy-momentum for the surrounding fluid is
implying that the spatial profile of the Vaidya solution surrounding energy-momentum tensor is proportional to the time component, describing the dynamical energy density $\rho_s(u,r)$,
 with
the arbitrary constant parameters $\alpha$ and $\beta$ depending the internal structure of the surrounding
fields.  The isotropic averaging over the angles results in \cite{Kiselev}
\begin{equation}\label{av}
<{\mathcal{T}^{i}}_{j}>=\frac{\alpha}{3}\rho_{s}(u,r){\delta^{i}}_{j}=p_{s}(u,r){\delta^{i}}_{j},
\end{equation}
since we considered $<r^{i}r_{j}>=\frac{1}{3}{\delta^{i}}_{j}r_n r^n$. Then,
we have the barotropic equation of state
for the surrounding field as\begin{equation}\label{jimbl}
p_s(u,r)=\omega_s \rho_s(u,r), ~~~\omega_s=\frac{1}{3}\alpha,
\end{equation}
where $p_s(u,r)$ and $\omega_s$ are the dynamical pressure and the constant
equation of state parameter
of the surrounding field, respectively\footnote{One
should note that the fluid in Eq.(6)  is not a perfect fluid.    Actually, the effective
``averaged''
energy-momentum as  ${T^{\mu}}_{\nu}=\left(-\rho, <{\mathcal{T}^{i}}_{j}>
\right)$ where $<{\mathcal{T}^{i}}_{j}>=\frac{\alpha}{3}\rho_{s}(u,r){\delta^{i}}_{j}=p_{s}(u,r){\delta^{i}}_{j}$,
can be treated as an \textit{effective perfect fluid}.}. Thus, regarding the Einstein tensor
components in (\ref{gmunu}) and the total energy-momentum tensor given by the equations (\ref{*energyT*})-(\ref{energysur}),
we have ${\mathcal{T}^{0}}_{0}={\mathcal{T}^{1}}_{1}$ and ${\mathcal{T}^{2}}_{2}={\mathcal{T}^{3}}_{3}$.
These exactly provide the so called principle of additivity and linearity considered in \cite{Kiselev}
in order to determine the  free parameter $\beta$ of the energy momentum-tensor ${\mathcal{T}^{\mu}}_{\nu}$ of the surrounding field as
\begin{equation}\label{mingl}
\beta=-\frac{1+3\omega_s}{6\omega_s}.
\end{equation}
Then, by substituting $\alpha$ and $\beta$ parameters in (\ref{jimbl}) and (\ref{mingl}) into (\ref{energysur}), the non-vanishing components of the surrounding energy-momentum tensor ${\mathcal{T}^{\mu}}_{\nu}$ will be
\begin{eqnarray}\label{lala}
&&{\mathcal{T}^{0}}_{0}={\mathcal{T}^{1}}_{1}=-\rho_s(u,r),\nonumber\\
&&{\mathcal{T}^{2}}_{2}={\mathcal{T}^{3}}_{3}=\frac{1}{2}\left(1+3\omega_s\right)\rho_s(u,r).
\end{eqnarray}
 Now, by having the Einstein tensor components and the corresponding total
 energy-momentum
tensor ${T^{\mu}}_{\nu}$, one can  obtain the associated field equations. Then, the  ${G^{0}}_{0}={T^{0}}_{0}$ and ${G^{1}}_{1}={T^{1}}_{1}$ components of the Einstein field equations give
the following differential equation
\begin{equation}\label{G00}
\frac{1}{r^2}\left(f^{\prime}r-1+f  \right)=-\rho_s.
\end{equation}
Similarly, the ${G^{1}}_{0}={T^{1}}_{0}$ component leads to
\begin{equation}\label{G10}
-\frac{\dot f}{r}=\epsilon \sigma,
\end{equation}
and ${G^{2}}_{2}={T^{2}}_{2}$ and ${G^{3}}_{3}={T^{3}}_{3}$ components read as
\begin{equation}\label{G22}
\frac{1}{r^2}\left(rf^{\prime}+\frac{1}{2}r^2 f^{\prime\prime}\right)=\frac{1}{2}(1+3\omega_{s} )\rho_{s}.
\end{equation}
Thus, we see that there are three unknown dynamical functions  $f(u,r)$, $\sigma(u,r)$ and $\rho_s(u,r)$ which can
be determined analytically by  the above three differential equations.
Simultaneous solving the differential equations (\ref{G00}) and (\ref{G22}), one obtains the following  solution for the metric function
\begin{equation}\label{fmetric}
f(u,r)=1-\frac{2M(u)}{r}-\frac{N_s(u)}{{r}^{{3\omega_s +1}}
},
\end{equation}
with the energy density $\rho_s (u,r)$
of the surrounding field in the form of
\begin{equation}\label{rho}
\rho_s (u,r)=- \frac{3\omega_s N_{s}(u)}{{r}^{{3(\omega_s +1)}}},
\end{equation}
where $M(u)$ and $N_s(u)$ are integration  coefficients representing the  Vaidya dynamical mass and the surrounding dynamical field structure parameter, respectively.

On the other hand, respecting to the  weak energy condition imposing the positivity of  any kind of energy density of the surrounding
field, i.e $\rho_s\geq0$,  demands
\begin{equation}\label{WEC}
\omega_s N_{s}(u)\leq0.
\end{equation}
This implies that for the surrounding field with a positive equation of state
parameter $\omega_s$, it is needed
to have $N_{s}(u)\leq0$ and  conversely for a negative $\omega_s$, it is
required to have $N_{s}(u)\geq0$. Then, this condition determines the gravitational
nature of the
term associated to surrounding field  in the metric function $f(u,r)$.

Regarding the metric function (\ref{fmetric}),  the spacetime metric (\ref{metricvaidya}) reads as
\begin{equation}\label{mjoon}
ds^{2}=-\left(1-\frac{2M(u)}{r}-\frac{N_{s}(u)}{{r}^{{3\omega_s +1}}
}\right)du^2+2\epsilon dudr+r^2d\Omega^2,
\end{equation}
representing an effectively evaporating-accreting Vaidya spacetime in a dynamical background.
  One may realize the following distinct subclasses of this general solution as
\begin{itemize}
\item  {\bf The solution by setting $f=f(u,r)$ and $\rho_s=\rho_s(r)$ in the field equations (\ref{G00})
to (\ref{G22}). }

These considerations lead to $M=M(u)$ and $N_s=constant$ in the metric function $f(u,r)$  and $\sigma\neq 0$
for the black hole's radiation density. In this case, there is no dynamics in the surrounding field
and consequently there is no accretion to the black hole. Indeed, this case represents an evaporating black hole solution with $\epsilon=-1$ in
a static background. Then, the  evaporating black hole in an empty background,
 i.e $\rho_s=0$  \cite{vaidyajoon},
and (anti)-de Sitter space, i.e $\rho_s = \rho_{\Lambda}=constant$  \cite{mallet, Mallet2, vick},  are special subclasses of our general solution.
Some interesting physical features of these solutions can be found in the
references \cite{evap1, xiang, kim, york, zheng, naresh, cao,  ren, saa, hamid, liu, Mallet3}.

\item {\bf The solution by setting   $f=f(r)$ and $\rho_s=\rho_s(r)$ in the field equations (\ref{G00}) to (\ref{G22}).}

These considerations lead to $M=constant$, $N_s=constant$ in the metric function and $\sigma=0$
for the radiation-accretion density. This case represents a non-dynamical back hole in a static background
and consequently, there are no accretion and evaporation. The Schwarzschild black hole as well as its generalization to
(anti)-de Sitter background are two special subclasses of our general  solution. For a general
background, not just the (anti)-de Sitter background,  it is  interesting that using the following coordinate transformation
\begin{equation}\label{tr}
du=dt+\frac{\epsilon dr}{1-\frac{2M}{r}-\frac{N_{s}}{{r}^{{3\omega_s +1}} }},
\end{equation}
one can obtain the general static solution of the Schwarzschild black hole surrounded by a surrounding field
as
\begin{equation}\label{METRIC}
ds^2=-\left(1-\frac{2M}{r}-\frac{N_s}{r^{3\omega_s +1}}
\right)dt^2
+\frac{dr^2}{
 1-\frac{2M}{r}-\frac{N_s}{r^{3\omega_s +1}}
}
+r^2 d\Omega^2,
\end{equation}
which was  found by Kiselev \cite{Kiselev}. Then, the Kiselev solution
 also can be obtained as a subclass of our general  dynamical solution (\ref{mjoon}) in the stationary limit.
\item {\bf The solution for $\epsilon=+1$ with changing the background field
 parameters
as $\omega_s\rightarrow\frac{1}{3}(2k-1)$ and $N_s(u)\rightarrow-\frac{2g(u)}{2k-1}$.}

By this considerations, we recover the Husain solution describing a null
fluid collapse \cite{husain} as
\begin{equation}\label{husain}
ds^{2}=-\left(1-\frac{2M(u)}{r}+\frac{2g(u)}{(2k-1){r}^{{2k}}
}\right)du^2+2\epsilon dudr+r^2d\Omega^2,
\end{equation}
with the energy density
\begin{equation}\label{rhohusain}
\rho_s (u,r)=\frac{2g(u)}{{r}^{{2k+2}}}.
\end{equation}
\end{itemize}
This solution and it various applications are widely studied in the literature, see for instances \cite{hus1,
hus2} and \cite{hus3, prabir} where a barotropic equation of state is considered
for the collapse study. There is a difference in the method obtaining the solutions
in the present work and in \cite{husain} as well as in the other mentioned
works. The solution (\ref{husain}), as in \cite{husain}, is obtained by the  ``{\it
pre-imposed}''  equation of state $p=k\rho^a$, whereas in our approach, the effective equation of state is resulting from the isotropic averaging over the angles for the surrounding field distribution. Our approach is motivated by the present anisotropy in the Einstein tensor components (\ref{gmunu}) and the corresponding total energy-momentum tensor (\ref{*energyT*}),
such that the surrounding fluid behaves effectively as a perfect fluid  with
the effective (averaged) equation of state $p_s(u,r)=\omega_s \rho(u,r)$, see (\ref{av}). As the advantage of this averaging method, one can substitute for $\omega_s$ the same known cosmological field equation of state parameters $\frac{1}{3},0,-1,-\frac{2}{3}$ and $-\frac{4}{3}$ for the radiation, dust, cosmological, quintessence and phantom fields, respectively, when  the black hole is embedded in these cosmological backgrounds. Substituting the same values of cosmological parameters for $k$, through $p=k\rho^a$ even for $a=1$,
in (\ref{husain})  gives different solutions with respect to (\ref{mjoon}) for the general dynamical case as well
as for the known static solution in \cite{Kiselev}  in the stationary limit, by doing a similar transformation to (\ref{tr}). For example, throwing a bunch of dust with the mass of $M_{dust}(=g)$ to the black hole with  mass $M$, one expects a resulting metric
for the final black hole as $f(r)=1-\frac{2M_{eff}}{r}$ where $M_{eff}=M+M_{dust}$, whereas substituting $k=0$  in the metric (\ref{husain}) gives $f(r)=1-\frac{2M}{r}-2M_{dust}$
which  seems to be incorrect due to  the   gravitational potential form of the final black hole and  also the dimensional consideration. One also
realizes that, as we will see in the next sections of the paper, there is a possibility
of the formation of both the naked singularities and black holes in different
backgrounds for the solution with $\epsilon=-1$.
\subsection{The Analysis of Naked Singularity or Black Hole Formations }
In order to investigate the formation of naked singularity or black hole
associated to the obtained solution  (\ref{mjoon}),
we follow the approach of \cite{prabir}.
The  equation for the  radial null geodesics using
 the metric (\ref{metricvaidya}), or (\ref{mjoon}), can be obtained by setting $ds^{2}=0$ and
$d\Omega_{2}^{2}=0$ as\begin{equation}
\frac{du}{dr}=\frac{2\epsilon}{f(u,r)}.
\end{equation}
This system has a singularity at  $r=0,~u=0$. Defining the function $X$ 
as  $X=\frac{u}{r}$ gives us the possibility of studying the limiting behavior of $X$ as we
approach the singularity located at $r=0,~u=0$,  along the radial null
geodesics. Denoting this limiting value of $X$   by $X_{0}$, we have
\begin{eqnarray}\label{26}
\begin{array}{c}
X_{0}\\\\
{}
\end{array}
\begin{array}{c}
=lim~~ X \\
\begin{tiny}u\rightarrow 0\end{tiny}\\
\begin{tiny}r\rightarrow 0\end{tiny}
\end{array}
\begin{array}{c}
=lim~~ \frac{u}{r} \\
\begin{tiny}u\rightarrow 0\end{tiny}\\
\begin{tiny}r\rightarrow 0\end{tiny}
\end{array}
\begin{array}{c}
=lim~~ \frac{du}{dr} \\
\begin{tiny}u\rightarrow 0\end{tiny}\\
\begin{tiny}r\rightarrow 0\end{tiny}
\end{array}
\begin{array}{c}
=lim~~ \frac{2\epsilon}{f(u,r)}. \\
\begin{tiny}u\rightarrow 0\end{tiny}~~~~\\
\begin{tiny}r\rightarrow 0\end{tiny}~~~~
\end{array}
\end{eqnarray}
Using the metric function (\ref{fmetric}) in (\ref{26}), we obtain
\begin{eqnarray}\label{27}
\begin{array}{c}
\frac{2\epsilon}{X_{0}}=lim~\left(1-\frac{2M(u)}{r}-\frac{N_s(u)}{{r}^{{3\omega_s +1}}}\right).\\ 
\begin{tiny}u\rightarrow 0\end{tiny}\\
\begin{tiny}r\rightarrow 0\end{tiny}
\end{array}
\end{eqnarray}
Now, following the method of \cite{prabir} for our case, we consider $M(u)=m u$~and~$N_s(u)=n u^{3\omega_s+1}$, where 
$m$ and $n$ are constants. Thus, using (\ref{27}), we
obtain the following algebraic equation in terms of $X_{0}$  
\begin{equation}\label{28}
n  X_{0}^{3\omega_s+2}+2m X_{0}^{2}-X_{0}+2\epsilon=0.
\end{equation}
 A black hole will be formed if one obtains only non-positive
solutions of this equation. However, if we find a positive real
root for (\ref{28}), then this system describes a
naked singularity and consequently provides counterexamples for the cosmic censorship conjecture by Penrose \cite{ccc}. It is difficult to find exact solutions for $X_{0}$ in (\ref{28}) for the generic values of $n, m, \epsilon$ and $\omega_s$ parameters.  However, as a result, one can find that there are possibilities
of the formation of both the naked singularities and black holes for the various backgrounds of dust, radiation,
quintessence, cosmological constant-like and phantom backgrounds for some
particular ranges of $m$ and $n$ parameters. We postpone the  detailed study of this equation   for the mentioned backgrounds, to the sections
3 to 7.
%%%%%%%%%%%%%%%%%%%%%%%%%%%%%%%%%%%%%%%%%%%%%%%%%%%%%%%%%%%%%%%%%%%%%%%%%
\subsection{The Analysis of the Black Hole-Background Field Interactions }
Because in this work,  we are mainly interested in the possible
interactions between a dynamical black hole and its surrounding background,
hence  regarding the possibility of formation of black holes as mentioned in the previous subsection
  and as we will see in the sections 3-7, here we consider only the case that black holes are formed and we analyze in detail
the general radiation-accretion profile for the corresponding systems and
classify the possible situations under the positive energy condition.

Substituting the metric function (\ref{fmetric}) in the equation (\ref{G10}) gives the radiation-accretion density
of the effectively evaporating-accreting black hole as
\begin{equation}\label{sigma}
\sigma(u,r)=\epsilon\left(\frac{2\dot M(u)}{r^{2}}
+\frac{\dot N_{s}(u)}{r^{3\omega_s +2}}
\right),
\end{equation}
where the first and second terms in RHS are the radiation-accretion density corresponding to the mass change
of the black hole and the  dynamics of the surrounding
field, respectively.   This shows that for construction of a realistic
effectively evaporating-accreting black hole model, one needs to implement such a solution including a dynamical black hole in a
dynamical background described by the energy-momentum (\ref{lala}).
Considering (\ref{sigma}), the following points can be realized.
\begin{itemize}
\item By turning off the background field dynamics, i.e. $\dot N_s(u)=0$,  we recover the energy flux associated
to the mass change of the central black hole corresponding to the original
Vaidya solution \cite{vaidyajoon}. See  \cite{schwartz} for more discussion
on the properties of the original
Vaidya solution. 
\item For the background field possessing  $\omega_s>0$,  if $\dot M(u)$ and $\dot N_s(u)$ have a same order of magnitude,   the surrounding background field contribution to the total density $\sigma(u,r)
$ is dominant near the black
hole while at far distances from the black hole it decreases faster than the
contribution  of the black hole mass changing term.
In contrast,  for the background field possessing  $\omega_s<0$,    the surrounding background field contribution is dominant at large distances while the black hole contribution is dominant near the black hole itself.
Then, from the astrophysical point of view, the detected amount of the  radiation-accretion density by the observer not only depends on the distance from the black hole
but
also depends on the nature of background field.
\end{itemize}
Considering  the  positive  energy density condition (by the weak energy condition) on the total radiation-accretion density $\sigma(u,r)$
in (\ref{sigma}) requires
\begin{equation}\label{mmi}
\epsilon\left(\frac{2\dot M(u)}{r^{2}}+\frac{\dot N_{s}(u)}{r^{3\omega_s +2}}\right)
\geq0.
\end{equation}
This inequality confines the dynamical behaviours of the black hole  and its background field at any time and distance $(u,r)$. In the case of a static background,
as in the Vaidya's original solution \cite{vaidyajoon}, it is required that $\epsilon$
and $\dot M(u)$  have the same signs to have positive energy density. This shows that for a
radiating black hole with $\dot M(u)<0$ we have $\epsilon=-1$ which represents
the outgoing null flow, while for an accreting black hole it is required to have $\epsilon=+1$, representing the ingoing null flow. In the presence of
the background dynamics, it is not mandatory that $\epsilon$ and $\dot M(u)$
take the same signs and the satisfaction of the positive energy density condition
can be achieved  even by their opposite signs depending on the background field parameters $\dot{N}_s(u)$ and
$\omega_s$.
Based on the relation (\ref{mmi}), the dynamical behaviour 
of the background field is governed by
\begin{equation}\label{nnk}
\begin{cases}\dot N_{s}(u)\leq -2\,{r}^{3\omega_s }\,\dot M(u),~~~ & \epsilon=-1,
\\
\\
\dot N_{s}(u)\geq -2\,{r}^{3\omega_s }\,\dot M(u), &\epsilon=+1. 
\end{cases} 
\end{equation}
Then, at any distance $r$ from the black hole, the
background field must obey the above conditions. One astrophysical importance
of such a physical constraint is that the observer knows the dynamical range
of the background field at any distance and that, prior to any observation,
he knows how to include or remove the background
field contribution if he is only interested  in black hole's contribution,
or vice versa.  Interestingly, for the special case of $\dot N_{s}(u)= -2\,{r}^{3\omega_s }\,\dot M(u)$,
there is no pure radiation-accretion density, i.e $\sigma(u,r)=0$.
This case corresponds to two possible physical situations. The first one is related to the situation where for any particular distance $r_0$, the background $\dot N(u)$ and black hole $\dot M(u)$ behave such that their contributions cancel
out each others  leading to $\sigma(u,r_0)=0$. The second situation is related to
the case that for the given dynamical behaviors of the black hole and its background, one can always find the particular time dependent distance 
\begin{equation}\label{jimjim}
r_{*}(u)=\left(-\frac{\dot N_{s}(u)}{2\dot M(u)}\right)^{\frac{1}{3\omega_{s}}},
\end{equation}
 possessing zero energy density $\sigma(u,r_{*}(u))$. For the case of constant rates of $\dot{N}_{s}(u)$ and  $\dot{M}(u)$, the
distance $r_{*}$ is fixed to a particular value. To have a particular distance at which the density $\sigma(u,r_{*})$ is zero, the positivity of $r_*$ also requires that
$\dot M(u)$ and $\dot N_{s}(u)$ have opposite signs.
 For the cases in which $r_*$ is not positive, the lack of a positive real
value radial coordinate is interpreted as follows: the radiation-accretion density $\sigma(u,r)$ never and nowhere vanishes.

In the case of being the positive radial coordinate $r_*$, for the given radiation-accretion behaviors of the black hole and its surrounding field, i.e  $\dot M(u)$ and $\dot N_{s}(u)$, it is possible
to find a distance at which we have no any radiation-accretion energy density contribution.
In other words, it turns out that the rate of outgoing radiation energy density of
the black hole is exactly balanced by the rate of ingoing absorption rate of surrounding field at the
distance $r_{*}$ and vice versa. Beyond or within this particular distance, the various
general situations can be realized in the Tables 1 and 2  for the black hole (BH) and its surrounding field (SF). One practical
 importance
of (\ref{jimjim}) for an astrophysicist is that a particle detector at this distance will detect
vanishing radiation-accretion density.
\vspace{0.6cm}
\begin{table}[ht]
\begin{center}
\tabcolsep=0.08cm
\begin{tabular}{|c|c|c|c|c|c|c|c|c|c|}\hline
&$\epsilon$ & $\omega_s$ & $\dot M$ & $\dot N$ &$r_*$& $\sigma({r<r_*})$ &$\sigma({r=r_*})$
&$\sigma({r>r_*})$  & Physical effect \\\hline
I&-1 & + & + & + &- &-& -& - & Not Physical  \\\hline
II&-1 & + & - & + & +&-  &0 & + & Absorbtion of BH's radiation by SF\\\hline
III&-1 & + & + & - & +&+ & 0& - & Accretion of SF by BH\\\hline
IV&-1 & + & - & - & - &+  & +& + & Accretion/Decay of SF by Evaporating/Vanishing BH   \\\hline
V&-1 & - & + & + &- &-& -& - & Not Physical  \\\hline
VI&-1 & - & - & + & +&+  & 0& - & Absorbtion of BH's radiation by SF \\\hline
VII&-1 & - & + & - &+ &- &0 & + & Accretion of SF by BH  \\\hline
VIII&-1 & - & - & - & -&+ & +&+& Accretion/Decay of SF by Evaporating/Vanishing BH  \\\hline
\end{tabular}
\end{center}
\caption{General BH and SF parameters for $\epsilon=-1$.}
\label{1}
\end{table}
\begin{table}[ht]
\begin{center}
\tabcolsep=0.08cm
\begin{tabular}{|c|c|c|c|c|c|c|c|c|c|}\hline
&$\epsilon$ & $\omega_s$ & $\dot M$ & $\dot N$ & $r_*$ &$\sigma({r<r_*})$ &$\sigma({r=r_*})$
&$\sigma({r>r_*})$  & Physical Process \\\hline
I&+1 & + & + & + & - & +& +& + & Accretion of BH and SF \\\hline
II&+1 & + & - & + & + & +  &0 & - & Absorbtion of BH's radiation by SF \\\hline
III&+1 & + & + & - & +&- & 0& + & Accretion of SF by BH \\\hline
IV&+1 & + & - & - & - &- & -& - & Not Physical  \\\hline
V&+1 & - & + & + & - &+& +& + & Accretion of BH and SF  \\\hline
VI&+1 & - & - & + & + &-  & 0& + & Absorbtion of BH's radiation by SF\\\hline
VII&+1 & - & + & - & + &+ &0 & - & Accretion of SF by BH \\\hline
VIII&+1 & - & - & - & - &- & -&-& Not Physical  \\\hline
\end{tabular}
\end{center}
\caption{General BH and SF parameters for $\epsilon=+1$.}
\label{2}
\end{table}

\vspace{1cm}
Then, regarding these tables and Eq.(\ref{mmi}), we find the following results.
\begin{itemize}
\item The cases possessing negative values of $r_*$ (the cases I, IV, V
and VIII) mean that the radiation-accretion density does not vanish somewhere and forever.
Among these cases, the ones which have positive $\sigma(u,r)$ are only physical, i.e the cases IV and VIII for $\epsilon=-1$, and I and V for $\epsilon=+1$.
Then, one realizes that how the weak energy condition causes in practice
the nonphysical
events to be hidden to an astrophysicist aiming to investigate a black hole and
his surrounding field.  \item The remaining positive values of $r_*$, corresponding to a zero radiation-accretion
density, are physically viable and their corresponding physical processes are listed in the last column. These properties are determined according to the behaviours of the parameters $\epsilon$,  $\omega_s$, and quantities  $\dot M(u)$, $\dot N_{s}(u)$, and $\sigma(u,r)$. Those values of $r_*$ corresponding to the negative energy density $\sigma(u,r)$ represent no physical situation about the evaporation-absorption or accretion. The real features of those regions are hidden by the weak energy condition. Then, it is physically reasonable
to do any astrophysical experiment in the regions respecting the energy condition. \item For $\omega_s>-\frac{2}{3}$, the particular distance $r_*$,  where
$\sigma(u,r)$ vanishes,  corresponds to two possible cases as $r_{*}(u)=\left(-\frac{\dot N_{s}(u)}{2\dot M(u)}\right)^{\frac{1}{3\omega_{s}}}$ and $r_*=\infty$. In
the first case, for $-\frac{2}{3}<\omega_s<0$ with $|\dot N_s(u)|\ll |\dot M(u)|$ and for $\omega_s\geq0$ with $|\dot M(u)|\ll |\dot N_{s}(u)|$, we
have $r_*\rightarrow \infty$.  This means that the first situation  indicates that black hole evolves very faster than its background while the second indicates
that black hole evolves very slow relative to its background. By satisfaction
of these dynamical conditions to hold $r_*\rightarrow\infty$, the positive energy density is respected everywhere in the spacetime. Then, in practice,
an astrophysical observer can detect a radiation-accretion density resulting from the interaction
of the black hole with its surrounding field even at far distances, in which
for $-\frac{2}{3}<\omega_s<0$ and $\omega_s\geq0$ the main contribution in
the detected radiation-accretion density belongs to the black hole and surrounding
field, respectively. In other cases, the positive energy
density will be respected in some regions while violated beyond those regions.  \item For $\omega_s\leq-\frac{2}{3}$, the particular distance $r_*$  is given as
  $r_{*}(u)=\left(-\frac{\dot N_{s}(u)}{2\dot M(u)}\right)^{\frac{1}{3\omega_{s}}}$. Then, for a rapidly evolving black hole relative to its background, i.e $|\dot N_s(u)|\ll |\dot M(u)|$, we have $r_*\rightarrow \infty$.
This case  implies an evolving black hole in an almost static background
in which the positive energy condition is respected everywhere in this spacetime.
Then, for $\omega_s\leq-\frac{2}{3}$ representing a dark energy fluid, an astrophysicist finds that it is the black hole which has the main contribution in the radiation-accretion density.
\end{itemize}

\subsection{Timelike Geodesics for the Surrounded Black Holes}
The geodesics for our metric (\ref{metricvaidya}), or (\ref{mjoon}),
will all lie on  a plane due to the spherical symmetry in which for the sake
of simplicity, one can choose $\theta=\pi/2$. The geodesic equations for
the above spacetime metric can be derived by varying the following action
\begin{equation}
I=\int \mathcal{L}d\tau=\frac{1}{2}\int\left(-f(u,r)\overset{*}u^2+2\epsilon\overset{*}
u \overset{*} r+r^2 \overset{*}\varphi^2 \right)d\tau,
\end{equation}
where the star sign denotes the derivative with respect to the proper time $\tau$.
Then, we have the following three equations 
\begin{equation}\label{phi}
\overset{*} \varphi=\frac{L}{r^2},
\end{equation}
and 
\begin{equation}\label{r}
-\frac{1}{2}f^{\prime}\overset{*} u^2 +r\overset{*} \varphi^2
-\epsilon \overset{**}u=0,
\end{equation}
and
\begin{equation}\label{u}
\epsilon \overset{**} r=\frac{1}{2}\dot f \overset{*} u^2+f\overset{**}
u+f^{\prime}\overset{*} u\overset{*} r,
\end{equation}
for $\varphi, r$ and $u$ variables respectively,  where $L$
is the conserved angular momentum per unit mass and dot and prime signs denote the derivative with respect to $u$ and $r$, respectively\footnote{One can also reach at these equations using the geodesic
equation $\frac{d^2 x^\mu}{d\tau^2}-\Gamma^{\mu}_{\alpha\beta}\frac{x^\mu}{d\tau}
\frac{x^\nu}{d\tau}=0$ where $x^\mu$ and
$\Gamma^{\mu}_{\alpha\beta}$ represent the adopted 
coordinates in the metric (\ref{metricvaidya}) and the corresponding Christoffel
symbols, respectively.}.
Using (\ref{phi}) in (\ref{r}), one finds
\begin{equation}\label{m}
f\overset{**} u=\epsilon f\frac{L^2}{r^3}-\frac{1}{2}\epsilon f f^\prime \overset{*} u^2.
\end{equation}
On the other,  using the timelike geodesics condition as $g_{\mu\nu}\dot
x^{\mu}\dot x^{\nu}=-1$, one finds
\begin{equation}\label{g}
f^\prime \overset{*} r \overset{*} u=-\frac{1}{2}\epsilon f^\prime +\frac{1}{2}\epsilon
ff^\prime-\frac{1}{2}\epsilon f^\prime \frac{L^2}{r^2}
\overset{*} u^2,
\end{equation}
where the equation (\ref{phi})  has been used. Then, by substituting (\ref{m}) and (\ref{g}) in (\ref{u}), we arrive at
the following general equation of  motion in term of the metric function for the radial coordinate
\begin{equation}
\overset{**} r=\frac{1}{2}\epsilon\dot f\overset{*} u^2-\frac{1}{2}f^\prime
-\frac{1}{2}f^\prime\frac{L^2}{r^2}+f\frac{L^2}{r^3}.
\end{equation}
Then, using the metric function $f(u,r)=1-\frac{2M(u)}{r}-\frac{N_s(u)}{{r}^{{3\omega_s +1}}}$, this equation takes the following form
\begin{eqnarray}\label{rdot}
\overset{**} r&=&-\frac{M(u)}{r^2}+\frac{L^2}{r^3}-\frac{3M(u)L^{2}}{r^4}\nonumber\\
&&-\frac{(3\omega_s+1)N(u)}{2r^{3\omega_s+2}}-\frac{3(\omega_s+1)N(u)L^{2}}{2r^{3\omega_s+4}}\nonumber\\
&&+\frac{1}{2}\epsilon\dot f\overset{*} u^2.
\end{eqnarray}
Then, one realizes the following three interesting points.
\begin{enumerate}
\item The terms in the first line are exactly the same as that of the standard
Schwarzschild black hole in which the first term represents the Newtonian gravitational force,
the second term represents a repulsive centrifugal force and the third term
is the  relativistic correction of the Einstein GR which accounts for the
perihelion precession. 
\item The terms in the second line  are new correction terms due to the presence of the background
field which surrounds the Vaidya black hole, in which its first term is similar
to the term of gravitational potential in the first line, while its second
term is similar to the relativistic correction of GR. Then, regarding  (\ref{rdot}) one realizes
that for the more realistic non-empty backgrounds, the geodesic equation
of any object depends strictly not only on the mass of the central object of the system
and the conserved angular momentum of the orbiting body,  but also on the background field nature. The new correction
terms may be small in general in comparison to their Schwarzschild counterparts
(the first and third terms in the first line). However,
one can show that there are possibilities that these  terms are comparable
to them.
Then, in order to find a situation where these  forces are comparable to the Newtonian gravitational force and the GR correction term in (\ref{rdot}), we define the distances $D_{s_{1}}$
and $D_{s_{2}}$   which  satisfy $|\frac{a_{s_1}}{a_{N}}|\simeq1$ and $|\frac{a_{s_2}}{a_{L}}|\simeq1$,
  respectively, where $a_N$, $a_L$ are the Newtonian and the relativistic
correction accelerations, respectively, and $a_{s_1}$ and $a_{s_2}$ are defined as
\begin{equation}\label{mhs}
a_{s_1}=\frac{(3\omega_s+1)N(u)}{2r^{3\omega_s+2}},~~~~~a_{s_2}=\frac{3(\omega_s+1)N(u)L^{2}}{2r^{3\omega_s+4}}. \end{equation}
Then, the distances $D_{s_{1}}$
and $D_{s_{2}}$ will be given by 
\begin{equation}\label{dis}
D_{s_{1}}^{3\omega_s}=\left(\frac{|(3\omega_s +1)N_s(u)|}{2M(u)}\right),~~~~~
D_{s_{2}}^{3\omega_s}=\left(\frac{|(\omega_s +1)N_s(u)|}{2M(u)}\right).
\end{equation}
We give the detailed study of these particular distances for the various
cosmological backgrounds, in the sections 3 to 7.
\item The new correction term in the third
line is also a non-Newtonian gravitational force originated from the dynamics of black hole and its surrounding field.
It is  associated with the  radiation-accretion power of the black hole and its surrounding
field\footnote{In the stationary limit, where there is no dynamics for the
black hole and its surrounding field, this term vanishes while the terms
in the first and second lines in (\ref{rdot}) still exist.}.  Calling this acceleration as the \textit{induced acceleration} $a_{i}$, where the subscript $i$ stands for ``induced'', we have 
\begin{equation}\label{abs}
a_{i}=\frac{1}{2}\epsilon\dot f\overset{*} u^{2}
=-\epsilon\left(  \frac{\dot M(u)}{r} +\frac{\dot N(u)}{2r^{3\omega_s+1}}\right)\overset{*} u^{2},
\end{equation}
 in which, following Lindquist, Schwartz and Misner
\cite{lind}, one can define the generalized ``\textit{total apparent flux}'' as $\mathcal{A}_{F}=\epsilon\left( \dot M(u) +\frac{\dot  N(u)}{2r^{3\omega_s}}\right){\overset{*}u}^{2}=\mathfrak{L}+\frac{\mathfrak{N}}{2r^{3\omega_{s}}}$ where $\mathfrak{L}$ and $\mathfrak{N}$ are the apparent fluxes associated
to the black hole and its surrounding field radiation-accretion rates, respectively. Using these definitions, (\ref{abs}) takes the following form 
\begin{equation}
a_{i}=-\frac{\mathfrak{L}}{r}-\frac{\mathfrak{N}}{2r^{3\omega_s+1}}. 
\end{equation}
As mentioned in \cite{lind}, this new correction
term may be small in general in comparison to the Newtonian term. However,
one can show that there are possibilities that these two terms are comparable.
Then, in order to find a situation where this induced force is comparable to the Newtonian gravitational force  in (\ref{rdot}), we define the distance $R$ which  satisfies $a_i\simeq a_N$, where $a_N$ is the Newtonian gravitational acceleration.
Then, this distance will be given by the solutions of the following equation for different values of $M$, $\omega_s$ and apparent fluxes $\mathfrak{L}$ and $\mathfrak{N}$ as
\begin{equation}\label{eq}
\mathfrak{L}R^{3\omega_s}+\frac{1}{2}\mathfrak{N}\simeq MR^{3\omega_s-1}.
\end{equation}
Finding the general solutions to this equation in terms of the generic $\mathfrak{L}, \mathfrak{N}, M$ and $\omega_s$ parameters is not simple.
However,  one can find that there
are possible solutions for the various backgrounds
of dust, radiation, quintessence, cosmological constant-like and phantom fields for some particular
ranges of the parameters. We give the detailed study of this equation for the mentioned backgrounds, in the sections 3 to 7.
\end{enumerate}
%%%%%%%%%%%%%%%%%%%%%%%%%%%%%%%%%%%%%%%%%%%%%%%%%%%%%%%%%%%%%%%%%
\section{Evaporating-Accreting Vaidya Black Hole Surrounded by the Dust
Field}
\subsection{Naked Singularity or Black Hole Formation Analysis}
\begin{figure}
\begin{center} 
\includegraphics[scale=0.4]{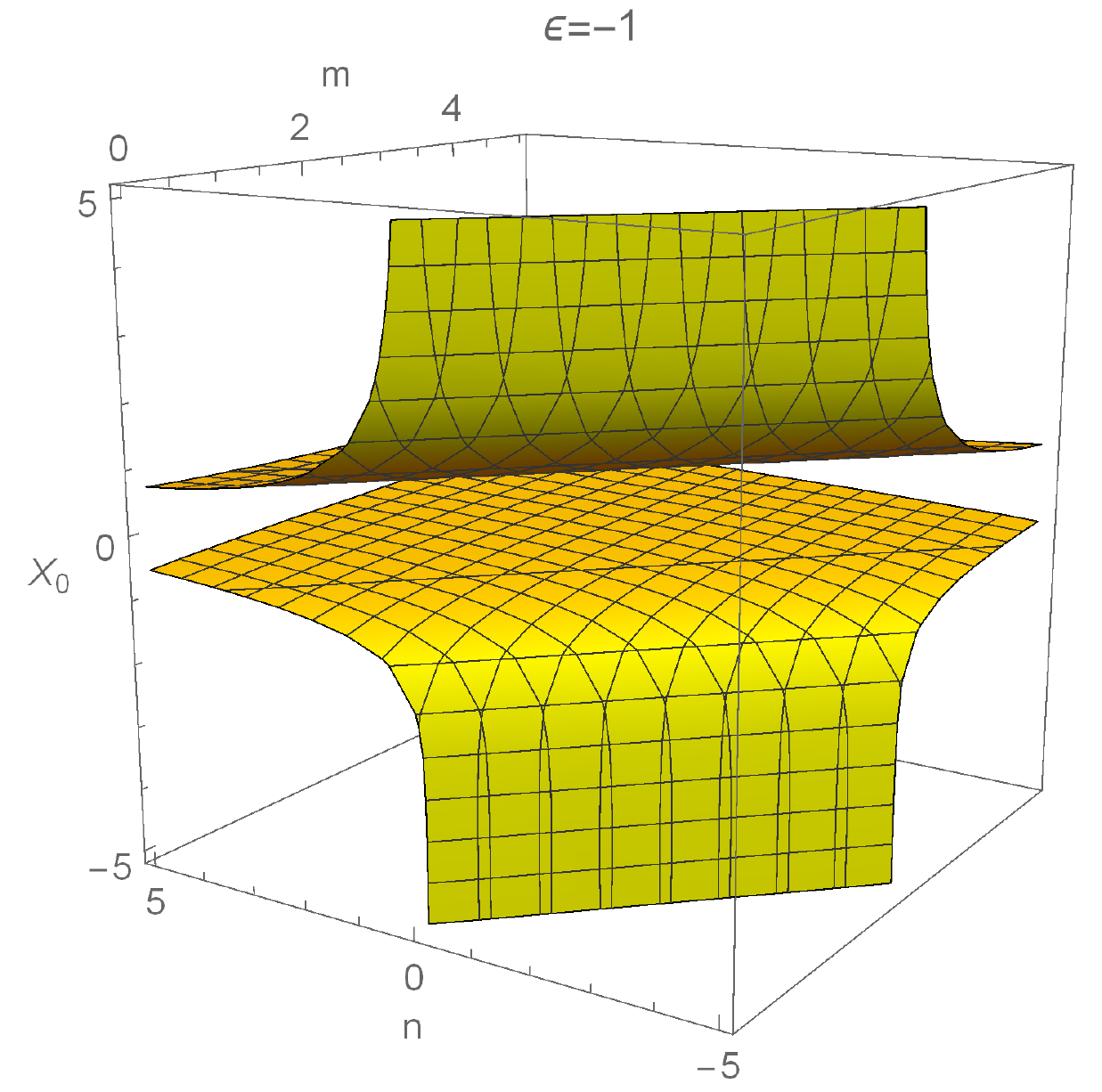}
\includegraphics[scale=0.4]{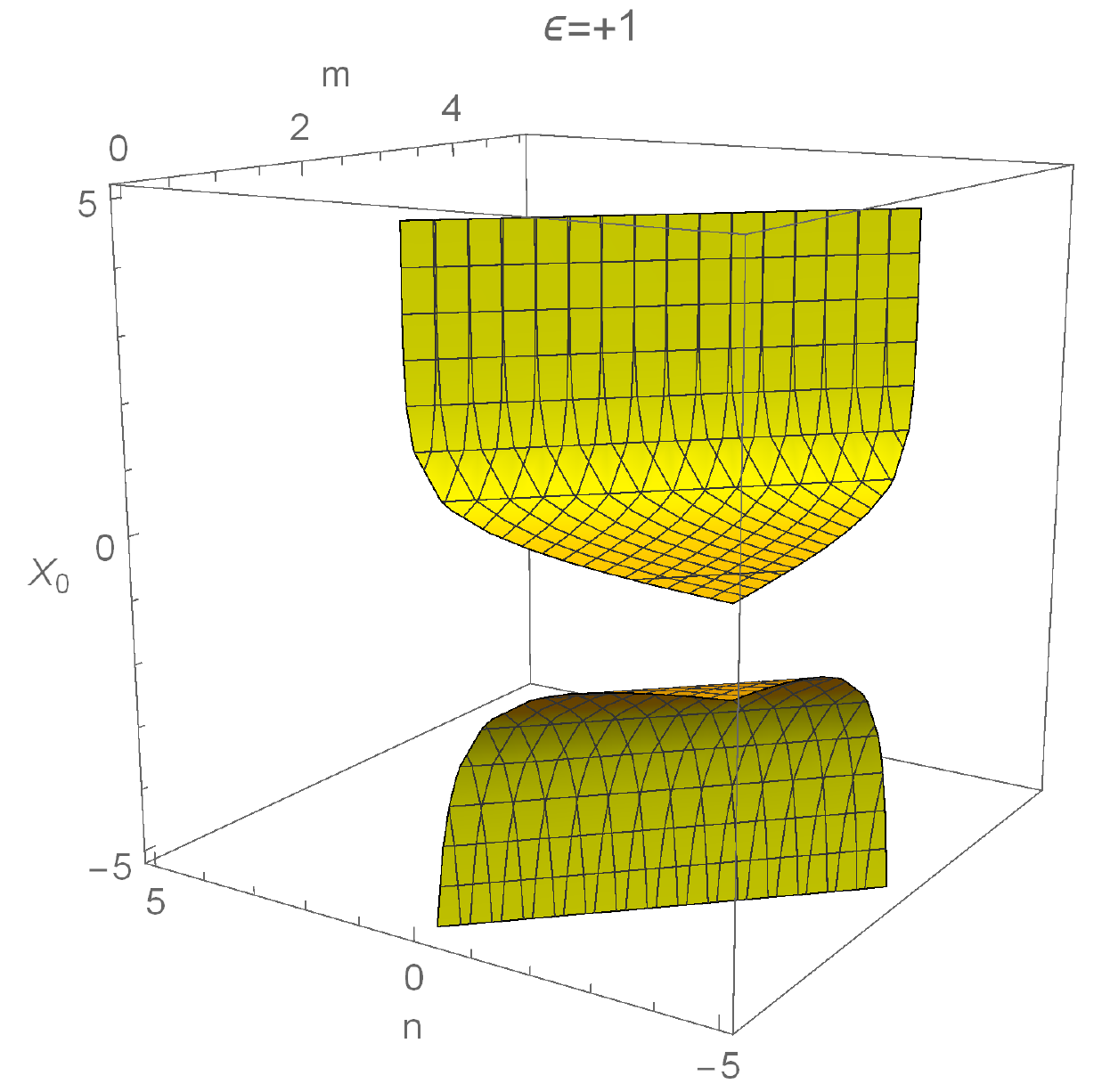}
\caption{\label{dn} The variation $X_0$ versus typical
values of the $m$ and $n$ parameters in (\ref{ddd}) for the dust background.}
\end{center}
\end{figure}

For this case, the equation (\ref{28}) takes the following form
\begin{equation}\label{ddd}
(n+2m) X_{0}^{2}-X_{0}+2\epsilon=0.
\end{equation}
Then, one can obtain the following set of solutions to (\ref{ddd})
\begin{eqnarray}
&&X_{01}=\frac{1-\sqrt{1+16m+8n}}{2(2m+n)}, 
~~~X_{02}=\frac{1+\sqrt{1+16m+8n}}{2(2m+n)},~~\epsilon=-1,\nonumber\\
&&X_{01}=\frac{1-\sqrt{1-16m-8n}}{2(2m+n)}, 
~~~X_{02}=\frac{1+\sqrt{1-16m-8n}}{2(2m+n)},~~\epsilon=+1.
\end{eqnarray}
Thus, one finds that some particular conditions on the parameters $m$ and
$n$ are required for having positive
or negative solutions.
In Figure \ref{dn}, we have plotted the solutions of  (\ref{ddd}) for some typical ranges of $m$ and $n$ parameters.  Then, regarding this figure, one realizes the possibility of the formation of  both  naked singularities and black holes in the dust background depending on the value of parameters.
%%%%%%%%%%%%%%%%%%%%%%%%%%%%%%%%%%%
\subsection{Black Hole-Dust Background Field Interactions}
For the dust surrounding field, we set the equation of state parameter of the dust field as $\omega_d=0$ \cite{Kiselev, AVikman}. Then, the metric (\ref{mjoon}) takes the following form
\begin{equation}
ds^{2}=-\left(1-\frac{2M(u)+N_d(u)}{r}\right)du^2+2\epsilon dudr+r^2d\Omega^2,
\end{equation}
where $N_{d}(u)$ denotes the normalization parameter for the dust field surrounding
the back hole, with
the dimension of $\left[N_d \right]=l$ where $l$ denotes the length. It is seen that the effectively radiating-accreting  black hole in the dust background appears as an effectively radiating-accreting  black hole with an effective
mass $M_{eff}(u)=2M(u)+N_d(u)$. In this case, the presence of new mass   term changes the thermodynamics,
causal structure
and Penrose diagrams just up to a re-scaling in the original Vaidya solution.

The radiation-accretion density in the dust background is given by
\begin{equation}\label{sigma*}
\sigma(u,r)=\epsilon\left(\frac{2\dot M(u)+\dot N_{d}(u)}{r^{2}}\right).
\end{equation}
For the Vaidya's original solution in an empty background, i.e $N_d(u)=0$,
 or even in
a static background, i.e $\dot N_d(u)=0$,  the positive  energy density condition, i.e $\sigma(u,r)\geq0$, requires that  $\epsilon$ and $\dot M(u)$  always have the same signs. This means that for $\epsilon =+ 1$,  $M(u)$ is
a monotone increasing mass function while for the case of
$\epsilon=-1$, $M(u)$ is  a monotone decreasing mass function. In our general solution for the Vaidya black hole in
the dust background,  the condition $\sigma(u,r)\geq0$ imposed on (\ref{sigma*}) is satisfied for more general situations indicated  in the Table \ref{table3}.
\vspace{0.5cm}
\begin{table}[ht]
\begin{center}
\tabcolsep=0.08cm
\begin{tabular}{|c|c|c|c|c|}\hline
$\epsilon$  & $\dot M$  & $\dot N_d$  & Condition& Physical Process \\\hline
 -1&- &- & \ No Condition&  Accretion/Decay of SF by Evaporating/Vanishing BH  \\\hline
-1 & + & - & $|\dot N_{d}(u)|\geq |2\,\dot M(u)| $& Accretion of SF by BH\\\hline
-1 & - & + &$|\dot N_{d}(u)|\leq |2\,\dot M(u)| $&Absorbtion of BH's radiation by SF\\\hline
+1 & + & - & $|\dot N_{d}(u)|\leq |2\,\dot M(u)|$ &Accretion of SF by BH\\\hline
+1 & - & + &$ |\dot N_{d}(u)|\geq |2\,\dot M(u)|$ &Absorbtion of BH's radiation by SF\\\hline
 +1&+ &+ & No Condition&Accretion of BH and SF \\\hline
\end{tabular}
\caption{\label{table3} BH and its surrounding dust field parameters for $\epsilon=\pm 1$. For these cases, the positive energy condition is satisfied everywhere in spacetime.}
\end{center}
\end{table}
\vspace{0.5cm}

\begin{figure}
\centering
\includegraphics[scale=0.62]{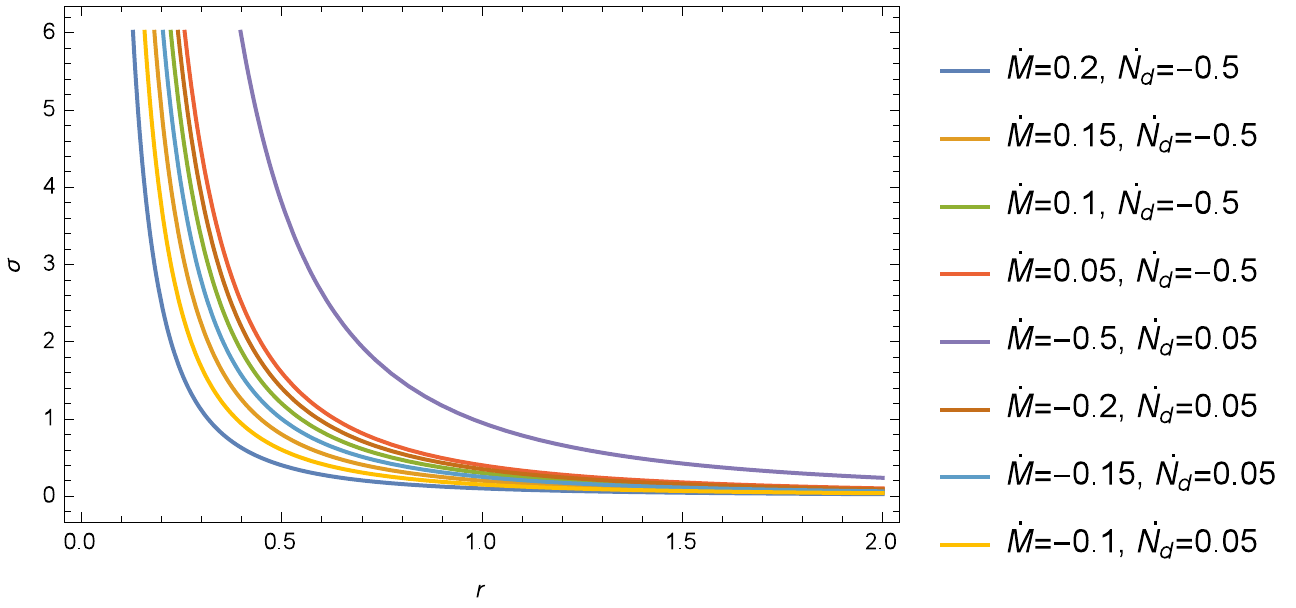}
\includegraphics[scale=0.62]{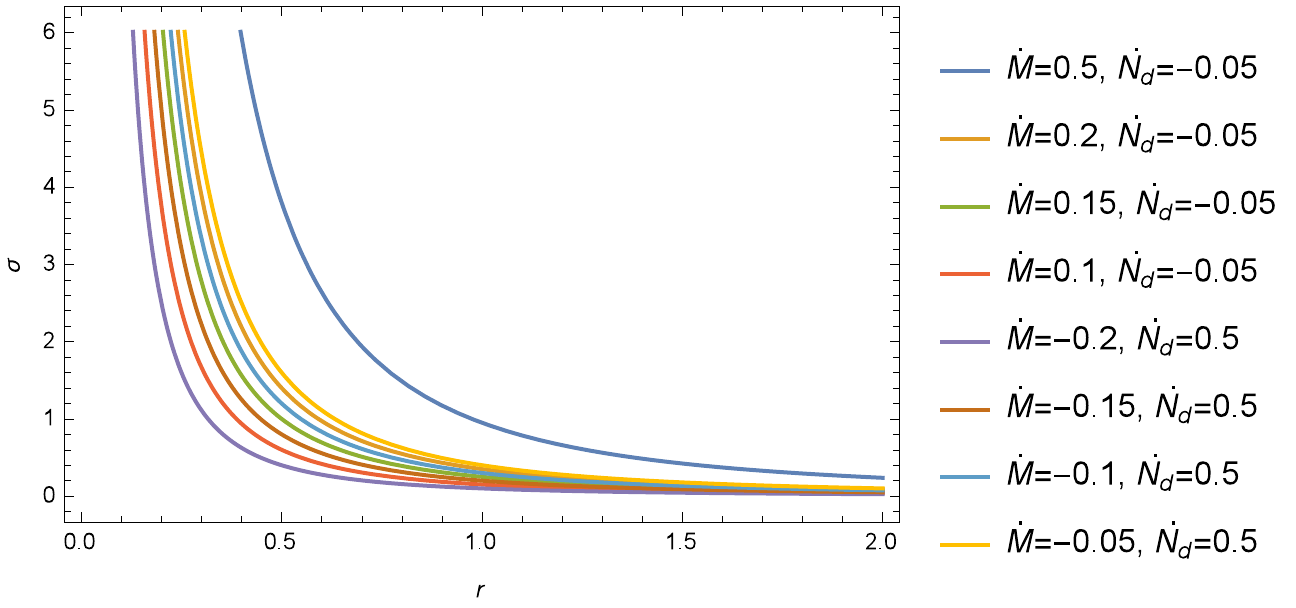}
\caption{\label{ds}
\newline
\textbf{\textit{Left Fig}}: The radiation-accretion  density $\sigma$ versus the distance $r$ for some typical constant values of $\dot M$ and $\dot N_{d}$  for $\epsilon=-1$ in the dust background. The four upper   cases and the four lower cases correspond to the conditions $|\dot N_{d}(u)|\geq |2\,\dot M(u)|$  and  $|\dot N_{d}(u)|\leq |2\,\dot M(u)|$, respectively. By these conditions, it is clear that $\sigma(r)$ is a decreasing function but is positive, and consequently the positive energy condition is satisfied everywhere in spacetime.
\newline
\textit{\textbf{Right Fig.}} The radiation-accretion  density $\sigma$ versus the distance $r$ for some typical constant values of
$\dot M$ and $\dot N_{d}$  for $\epsilon=+1$ in the dust background. The
four upper   cases and the four lower cases correspond to the conditions  $|\dot N_{d}(u)|\leq |2\,\dot M(u)|$ and $|\dot N_{d}(u)|\geq |2\,\dot M(u)|$,
 respectively. By these conditions, it is clear that $\sigma(r)$ is a decreasing function but is positive, and consequently the positive energy condition is satisfied everywhere in spacetime.}
\end{figure}

Interestingly, for the special case of $\dot N_{d}(u)= -2\dot M(u)$,
there is no pure radiation-accretion density, i.e $\sigma(u,r)=0$, and the
energy-momentum tensor (\ref{EMtotal}) will be diagonalized. This means
that the black hole and its surrounding background completely cancel out the effects
of each others. For $\dot N_d(u)\neq -2\dot M(u)$,  regarding (\ref{sigma*}),
we find that for $r_*\rightarrow\infty $, the radiation-accretion density vanishes, i.e   $\sigma(u,r)\rightarrow 0$. This means that for the  effective emission case, the out going  radiation can penetrate through the dust background  so far from the black hole and for the effective accretion case by the black hole, the black hole affect its so far surrounding objects.
Regrading the conditions in the Table \ref{table3} for $\epsilon=-1 $ and $\epsilon=+1$, the behaviour of radiation-accretion
density $\sigma$ in (\ref{sigma*}) is plotted for some typical values of $\dot M$ and $ \dot
N_{d}$ in the Figure \ref{ds}.
Using these plots, one can compare the  radiation-accretion density values for the various situations.
%%%%%%%%%%%%%%%%%%%%%%%%%%%%%%%%%%%%%%%%%%%%%%%%%%
\subsection{Timelike Geodesics for the Black Hole in the Dust Field Background}
\begin{figure}
\begin{center}
\includegraphics[scale=0.4]{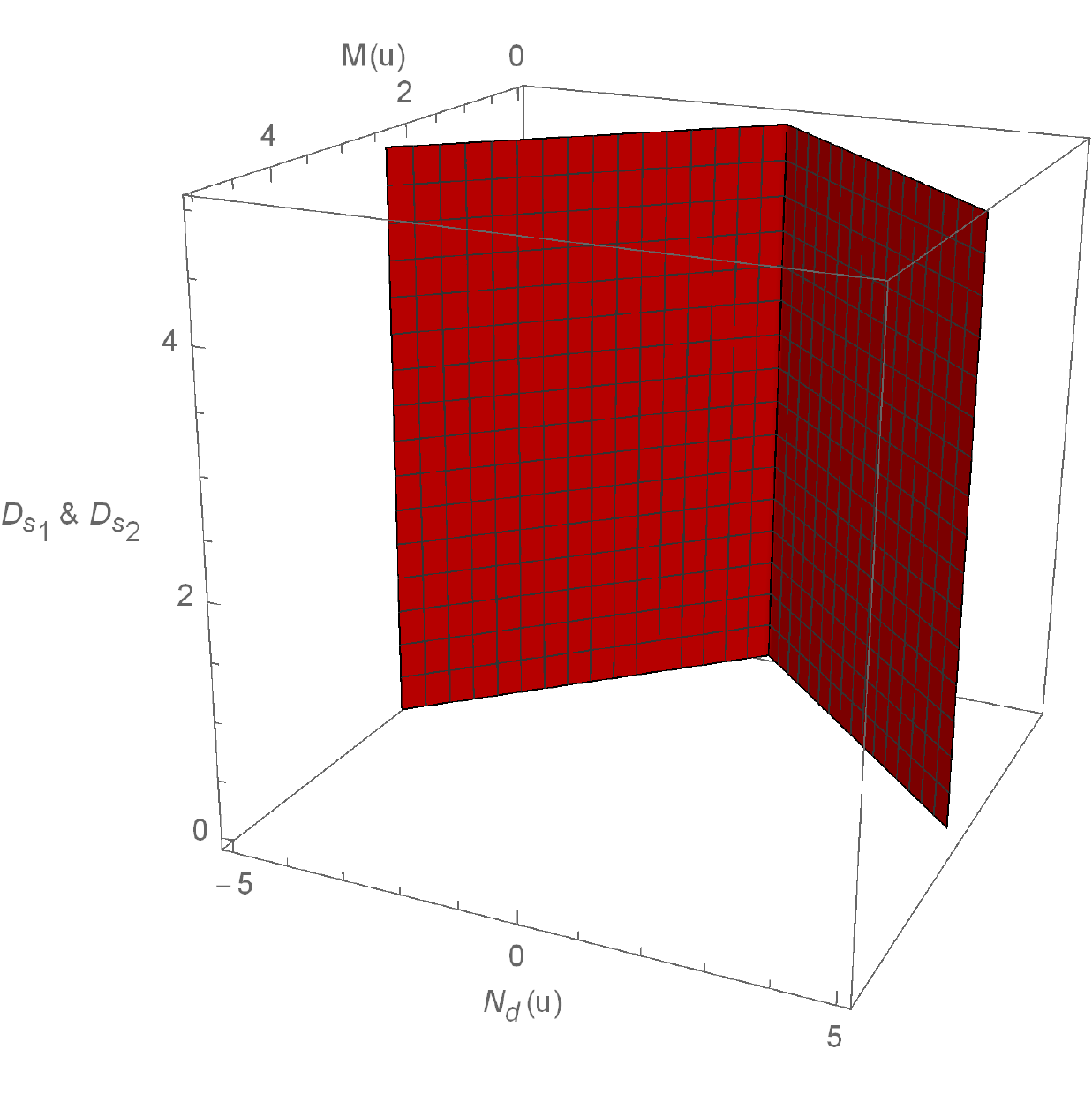}
\end{center}
\caption{\label{gad} The variation of $D_{s_1}$
and $D_{s_2}$ versus typical
values of the $M(u)$ and $N_d(u)$ parameters for the dust background.}
\end{figure}

\begin{figure}
\begin{center}
\includegraphics[scale=0.35]{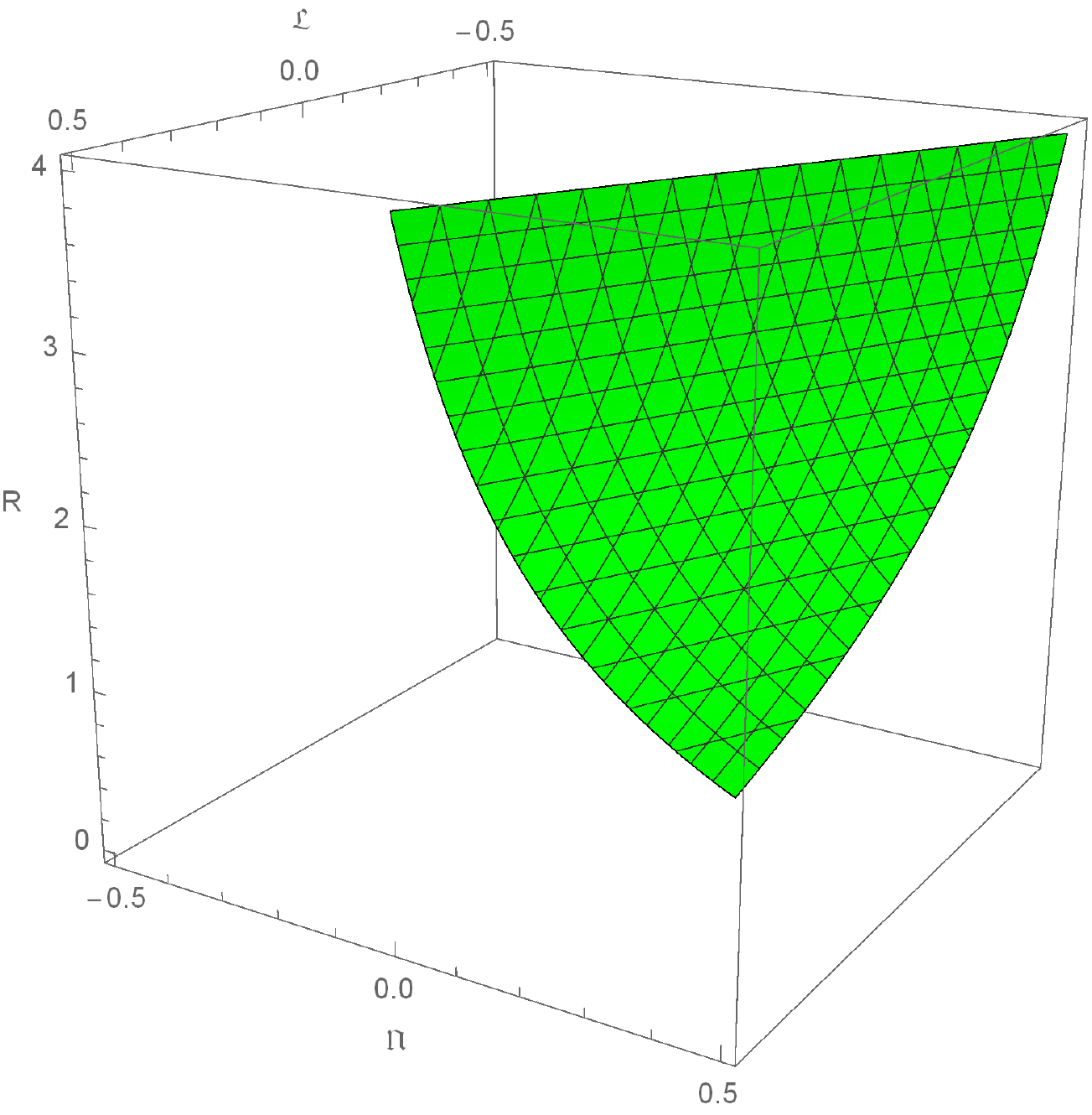}
\caption{\label{gdd} The variation of $R$ versus typical values of the $\mathfrak{L}$ and $\mathfrak{N}$ parameters in (\ref{eqd})
for the dust background. We have set $M=1$ without loss of generality.}
\end{center}
\end{figure}

For this case, we have $D_{s_1}=D_{s_2}$ and both the particular situations associated with  $|\frac{a_{s_1}}{a_{N}}|\simeq1$ and $|\frac{a_{s_2}}{a_{L}}|\simeq1$ 
are met for $M(u)=\frac{|N_d(u)|}{2}$
in the whole spacetime. In   Figure \ref{gad}, we have plotted  the possibility
of being these particular
situations for some typical ranges of $M(u)$ and $N_d(u)$ parameters. Then,
one realizes the possibility of equality of the Newtonian force as well as GR correction terms to the corresponding dust background field contributions.
\\ Also, for this case, the equation (\ref{eq}) associated with $a_i\simeq a_N$ takes the following form
\begin{equation}\label{eqd}
\mathfrak{L} +\frac{1}{2}\mathfrak{N}\simeq MR^{-1}.
\end{equation}
One can find the following solutions to (\ref{eqd})
\begin{equation}
R\simeq \frac{2M}{2\mathfrak{L}+\mathfrak{N}}.
\end{equation}
Then, one realizes that how this particular distance depends on the parameters $\mathfrak{L}, \mathfrak{N}$ and $M$.
In the Figure \ref{gdd}, we have plotted the solutions of  (\ref{eqd}) for some typical ranges of $\mathfrak{L}$ and $\mathfrak{N}$ parameters.
This figure indicates that depending the parameter values, there are locations
where the induced force, resulting from the radiation-accretion phenomena
in the dust background, is equal to the Newtonian
gravitational force.

\section{Evaporating-Accreting Vaidya Black Hole Surrounded by the Radiation Field}
\subsection{Naked Singularity or Black Hole Formation Analysis}
\begin{figure}
\begin{center}
\includegraphics[scale=0.35]{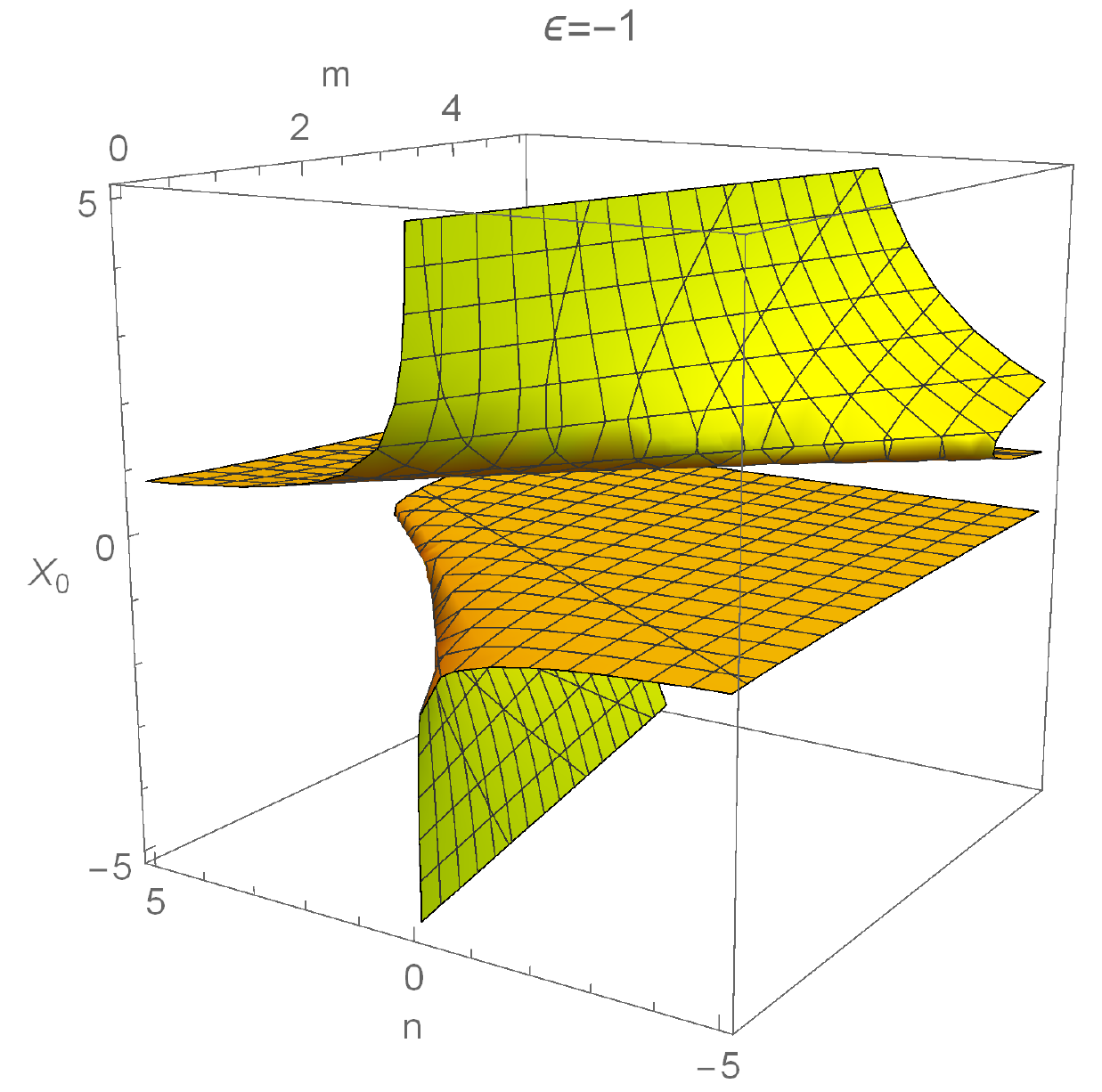}
\includegraphics[scale=0.35]{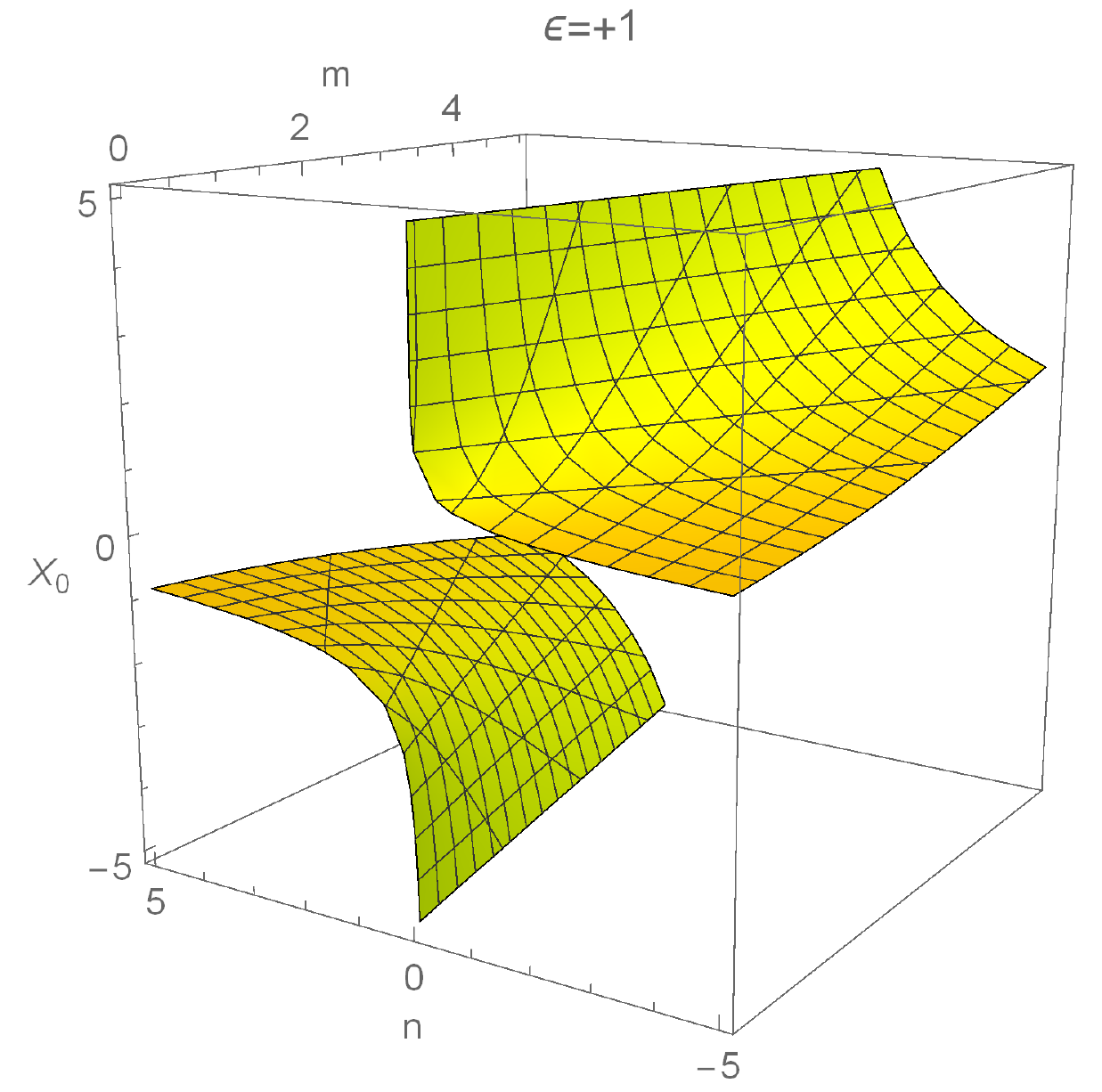}
\caption{ \label{nr} The variation of $X_0$ versus typical
values of the $m$ and $n$  parameters in (\ref{rrr}) for the radiation background.}
\end{center}
\end{figure}

For this case, the equation (\ref{28}) takes the following form
\begin{equation}\label{rrr}
nX_{0}^3 +2m X_{0}^{2}-X_{0}+2\epsilon=0.
\end{equation}
Then, we obtain the following solutions to (\ref{rrr})
\begin{eqnarray}
&&X_{01}=-\frac{2m}{3n}-\frac{-4m^2 -3n}{3n\Delta}+\frac{\Delta}{3n}, \nonumber\\
&&X_{02}=-\frac{2m}{3n}+\frac{(1+i\sqrt3)(-4m^2 -3n)}{6n\Delta}-\frac{(1-i\sqrt3)\Delta}{6n}, \nonumber\\
&&X_{03}=-\frac{2m}{3n}+\frac{(1-i\sqrt3)(-4m^2 -3n)}{6n\Delta}-\frac{(1+i\sqrt3)\Delta}{6n}, \end{eqnarray}
where $\Delta$ is given by 
\begin{eqnarray}
&&\Delta=\Delta_{-}=\left(-8m^3-9mn+27n^2+3\sqrt3\sqrt{-m^2n^2-16m^3n^2-n^3-18mn^3+27n^4}\right)^{\frac{1}{3}},
~~\epsilon=-1,\nonumber\\
&&\Delta=\Delta_{+}=\left(-8m^3-9mn-27n^2+3\sqrt3\sqrt{-m^2n^2+16m^3n^2-n^3+18mn^3+27n^4}\right)^{\frac{1}{3}},
~~\epsilon=+1.
\end{eqnarray}
 Then, one finds that some particular conditions are needed on the parameters $m$ and $n$  for having positive or negative solutions.
In Figure \ref{nr}, we have plotted the solutions of  (\ref{rrr}) for some typical ranges of $m$ and $n$ parameters.
This figure indicates the possibility of the formation of both the naked singularities and black
holes in the radiation background depending on the value of parameters.

\subsection{Black Hole-Radiation Background Field Interactions}
For the radiation surrounding field, we set the equation of state parameter of the
radiation field as $\omega_r=\frac{1}{3}$
 \cite{Kiselev, AVikman}. Then, the metric (\ref{mjoon})
takes the following form
\begin{equation}
ds^{2}=-\left(1-\frac{2M(u)}{r}-\frac{N_{r}(u)}{r^2}\right)du^2+2\epsilon
dudr+r^2d\Omega^2,
\end{equation}
where $N_{r}(u)$ is the normalization parameter for the radiation field
surrounding the black hole, with
the dimension of $[N_r ]=l^2$. Regarding the positive energy condition on the surrounding radiation field, represented by the relation
(\ref{WEC}), it is required that $N_r(u)\leqslant0$. Then, by defining the positive parameter $\mathcal{N}_r(u)=-N_r(u)$, we have
\begin{equation}
ds^{2}=-\left(1-\frac{2M(u)}{r}+\frac{\mathcal{N}_{r}(u)}{r^2}\right)du^2+2\epsilon
dudr+r^2d\Omega^2.
\end{equation}
This metric looks like a radiating charged
Vaidya black, namely the Bonnor-Vaidya
black hole \cite{bonnor}, with the
dynamical charge  $Q(u)=\sqrt{\mathcal{N}_r(u)}$,  see also \cite{dyon} for the radiating dyon solution. This result can be
 interpreted as the positive contribution of the characteristic feature
of the surrounding radiation field  to the effective charge term of the Vaidya
black hole with the  $\frac{1}{r^2}$ gravitational contribution. The appearance of an effective charge in the black hole solution
changes the causal structure and Penrose diagrams of this black hole solution in comparison to the neutral Vaidya black holes.
A similar effect in the causal structure of spacetime happens when one adds charge to the static Schwarzschild black hole leading to Reissner-Nordström black hole.
Then, turning off the background radiation field which surrounds the dynamical
Vaidya
black
hole  is equal to turning off the charge
in the static Reissner-Nordström case.

\begin{figure}
\centering
\includegraphics[scale=0.62]{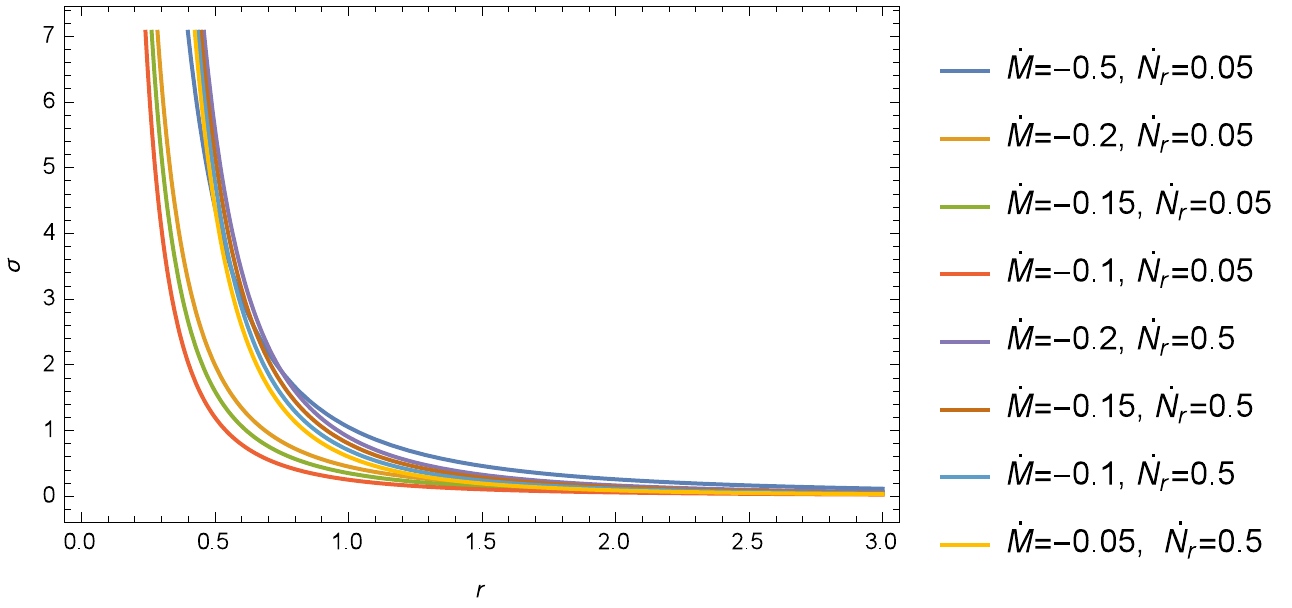}
\includegraphics[scale=0.62]{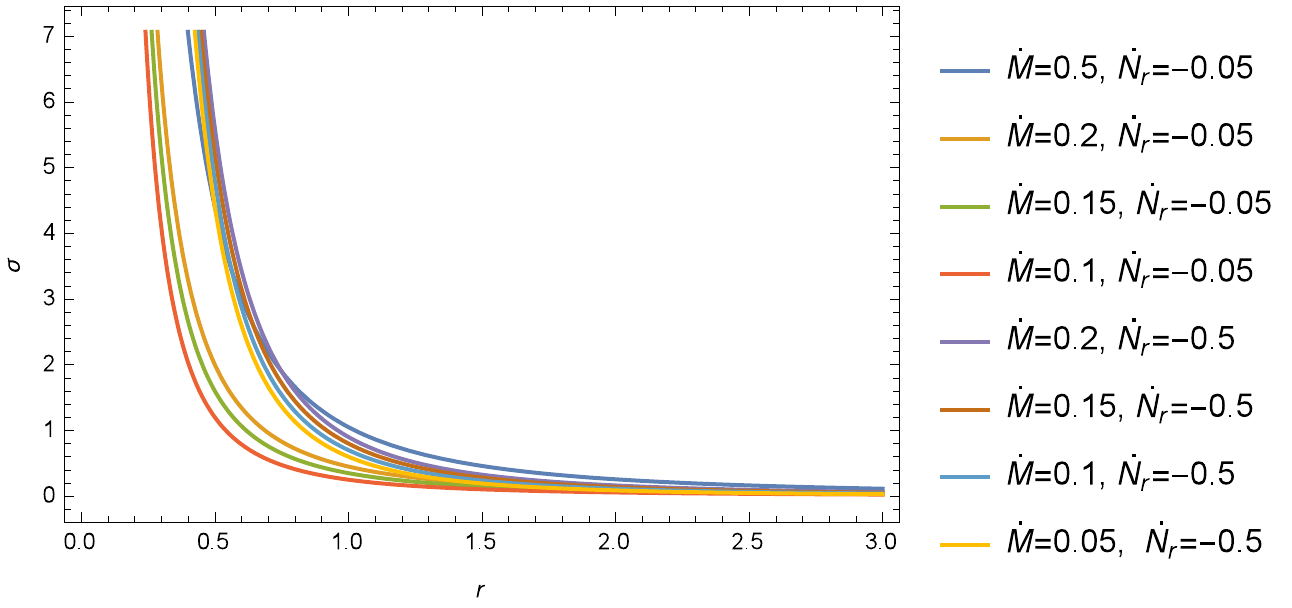}
\caption{\label{sr}
\newline
\textbf{\textit{Left Fig}.} The radiation-accretion  density $\sigma$ versus the distance $r$ for some typical constant
$\dot M$ and $\mathcal{\dot N}_{r}$ values for $\epsilon=-1$ in the radiation background. Here, $\sigma(r)$ is a decreasing function but is positive, and consequently the positive energy condition is satisfied everywhere in spacetime.
\newline
\textbf{\textit{Right Fig.}} The radiation-accretion  density $\sigma$ versus the distance $r$ for some typical constant
$\dot M$ and $\mathcal{\dot N}_{r}$ values for $\epsilon=+1$ in the radiation  background. Here, $\sigma(r)$ is a decreasing function but is positive, and consequently the positive energy condition is satisfied everywhere in
spacetime.}
\end{figure}

In this case, the total radiation-accretion density is given by
\begin{equation}\label{sigma**}
\sigma(u,r)=\epsilon\left(\frac{2\dot M(u)}{r^{2}}-\frac{\mathcal{\dot N}_{r}(u)}{r^3}
\right).
\end{equation}
 Then, we see that there is no positive $r_{*}(u)$ for $\dot M(u)$ and $\mathcal{\dot N}_r(u)$
having  opposite signs, and consequently $\sigma(u,r)$ never vanishes except
at infinity. But as $r_*\rightarrow\infty $, the radiation-accretion density again vanishes,
i.e   $\sigma(u,r_*)\rightarrow
0$. This means that for the  emission case,
the out going  radiation
can penetrate through the radiation background  so far from the black hole
and for the accretion case by the black hole, the black hole affects its so
far surrounding radiation filed.
The positivity condition of $\sigma(u,r)$ is satisfied everywhere for the situations  present in the Table \ref{table4}.
\vspace{0.5cm}
\begin{table}[ht]
\begin{center}
\tabcolsep=0.08cm
\begin{tabular}{|c|c|c|c|}\hline
$\epsilon$ &  $\dot M$ & $\mathcal{\dot N}_r$ & Physical Process \\\hline
-1   & - & + & Absorbtion of BH's radiation by SF\\\hline
+1   & + & -  & Accretion of SF by BH\\\hline
\end{tabular}
\end{center}
 \caption{BH and its surrounding radiation field parameters for $\epsilon=\pm1$.
For these cases, the positive energy condition is satisfied everywhere in
 spacetime. For any other behaviour of the $\dot M(u)$ and $\mathcal{\dot N}_r(u)$  parameters, the positive energy condition will be violated.}
 \label{table4}
\end{table}
\vspace{0.5cm}
\\
Regrading the Table \ref{table4}, the behaviour of radiation-accretion
density $\sigma$ in (\ref{sigma**}) is plotted for some typical values of $\dot M$ and $ \mathcal{\dot
N}_r$ in Figure \ref{sr}.
Using these plots, one can compare the  radiation-accretion densities for the various situations.

\subsection{Timelike Geodesics for the Black Hole in the Radiation Field Background}
\begin{figure}
\begin{center}
\includegraphics[scale=0.4]{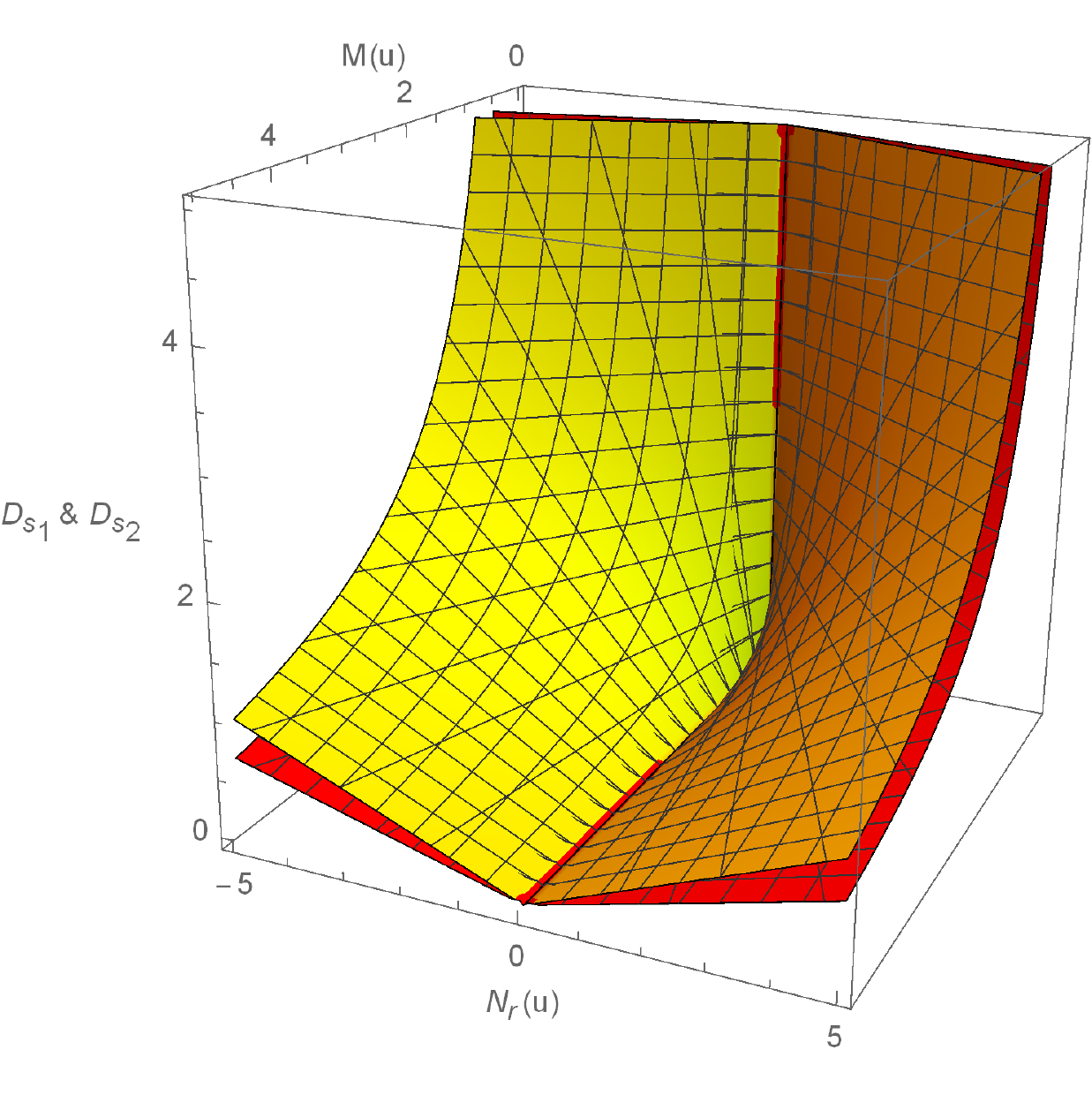}
\end{center}
\caption{\label{rddg} The variation of $D_{s_1}$
(yellow plot) and $D_{s_2}$ (red plot) versus typical
values of the $M(u)$ and $N_r(u)$ parameters for the radiation background.}
\end{figure}

For this case, the distances $D_{s_{1}}$
and $D_{s_{2}}$ associated with  $|\frac{a_{s_1}}{a_{N}}|\simeq1$ and $|\frac{a_{s_2}}{a_{L}}|\simeq1$,
respectively, will be given by 
\begin{equation}
D_{s_{1}}=\frac{|N_r(u)|}{M(u)},~~~~~
D_{s_{2}}=\frac{2|N_r(u)|}{3M(u)}.
\end{equation}
In   Figure \ref{rddg}, we have plotted the location of these particular
distances versus some typical ranges of $M(u)$ and $N_r(u)$ parameters.
Then, one realizes the possibility of
equality of the Newtonian force and GR correction terms to the corresponding radiation background
field contributions.

\begin{figure}
\begin{center}
\includegraphics[scale=0.35]{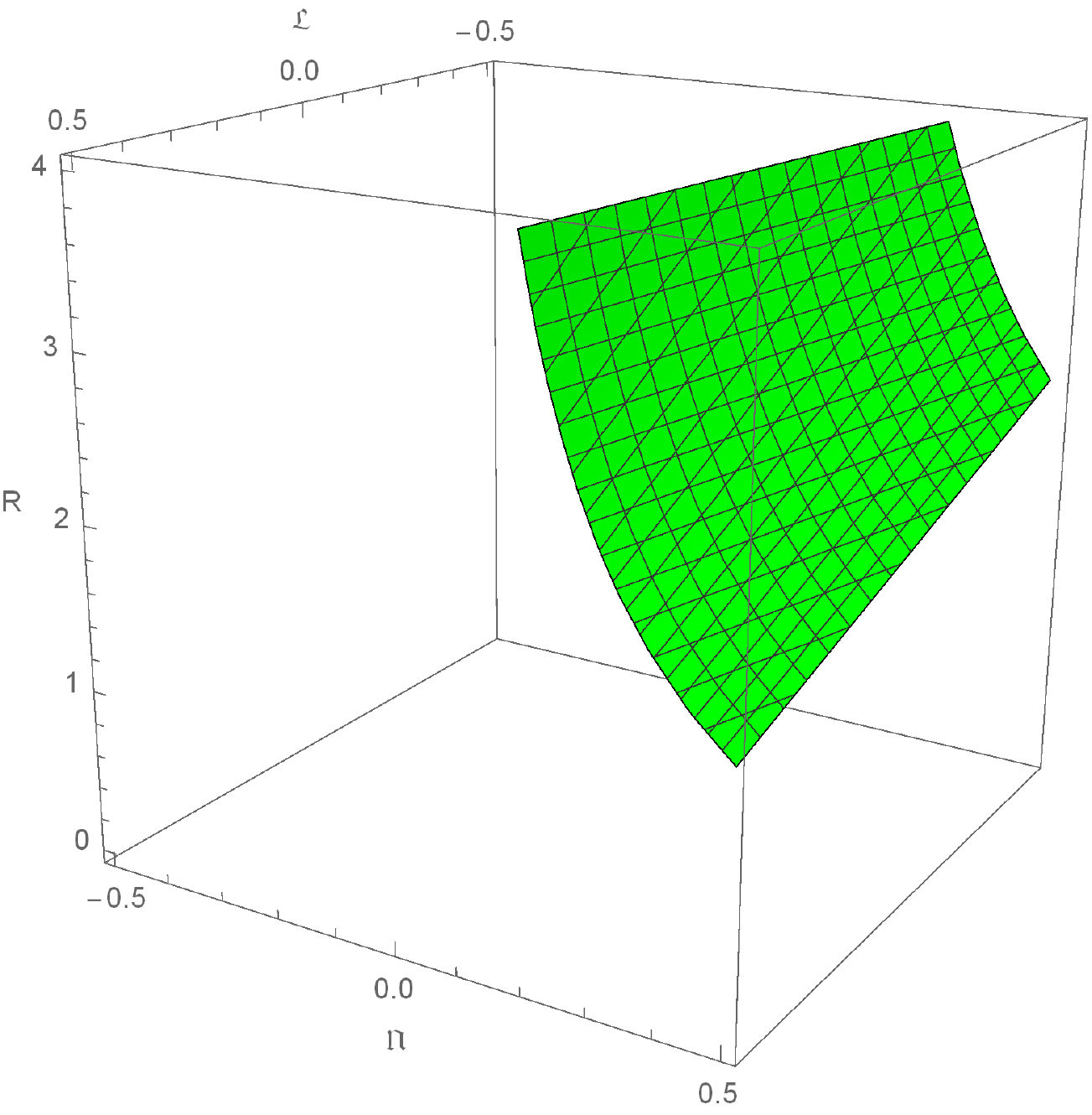}
\caption{\label{rrrr} The variation of $R$ versus typical
values of the $\mathfrak{L}$ and $\mathfrak{N}$ parameters in (\ref{eqd})
for the radiation background. We have set $M=1$ without loss of generality.}
\end{center}
\end{figure}

Moreover, for this case,  the equation (\ref{eq}) associated with $a_i\simeq a_N$ takes the following form
\begin{equation}\label{eqr}
\mathfrak{L} R+\frac{1}{2}\mathfrak{N}\simeq M.
\end{equation}
Then, one can find the following solutions to (\ref{eqr})
\begin{equation}
R\simeq \frac{2M-\mathfrak{N}}{2\mathfrak{L}}.
\end{equation}
It is seen  that how this particular distance depends on the parameters $\mathfrak{L}, \mathfrak{N}$ and $M$.
In  Figure \ref{rrrr}, we have plotted the solutions of  (\ref{eqr}) for some typical ranges of $\mathfrak{L}$ and $\mathfrak{N}$ parameters.
This figure shows that depending the parameter values, there are locations
where the induced force, resulting from the radiation-accretion phenomena
in the radiation background, is equal to the Newtonian
gravitational force.
%%%%%%%%%%%%%%%%%%%%%%%%%%%%%%%%%%%%%%%%%%%%%%%%%%%%%%%%%%%%%%%%%%%%%%%%%%%%%%%%%%%%%

\section{Evaporating-Accreting  Vaidya Black Hole Surrounded by the Quintessence Field}
\subsection{Naked Singularity or Black Hole Formation Analysis}
\begin{figure}
\begin{center}
\includegraphics[scale=0.35]{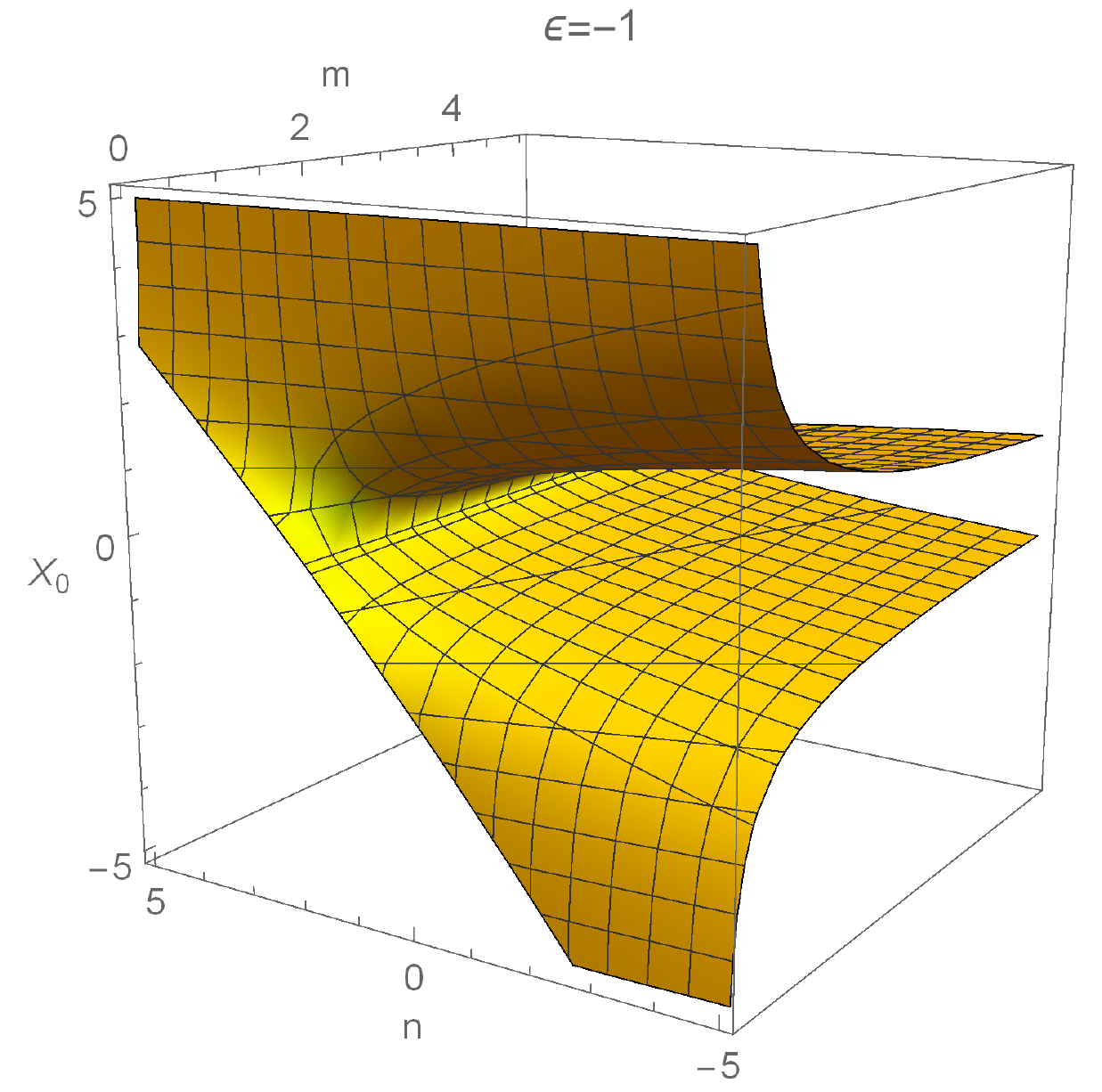}
\includegraphics[scale=0.35]{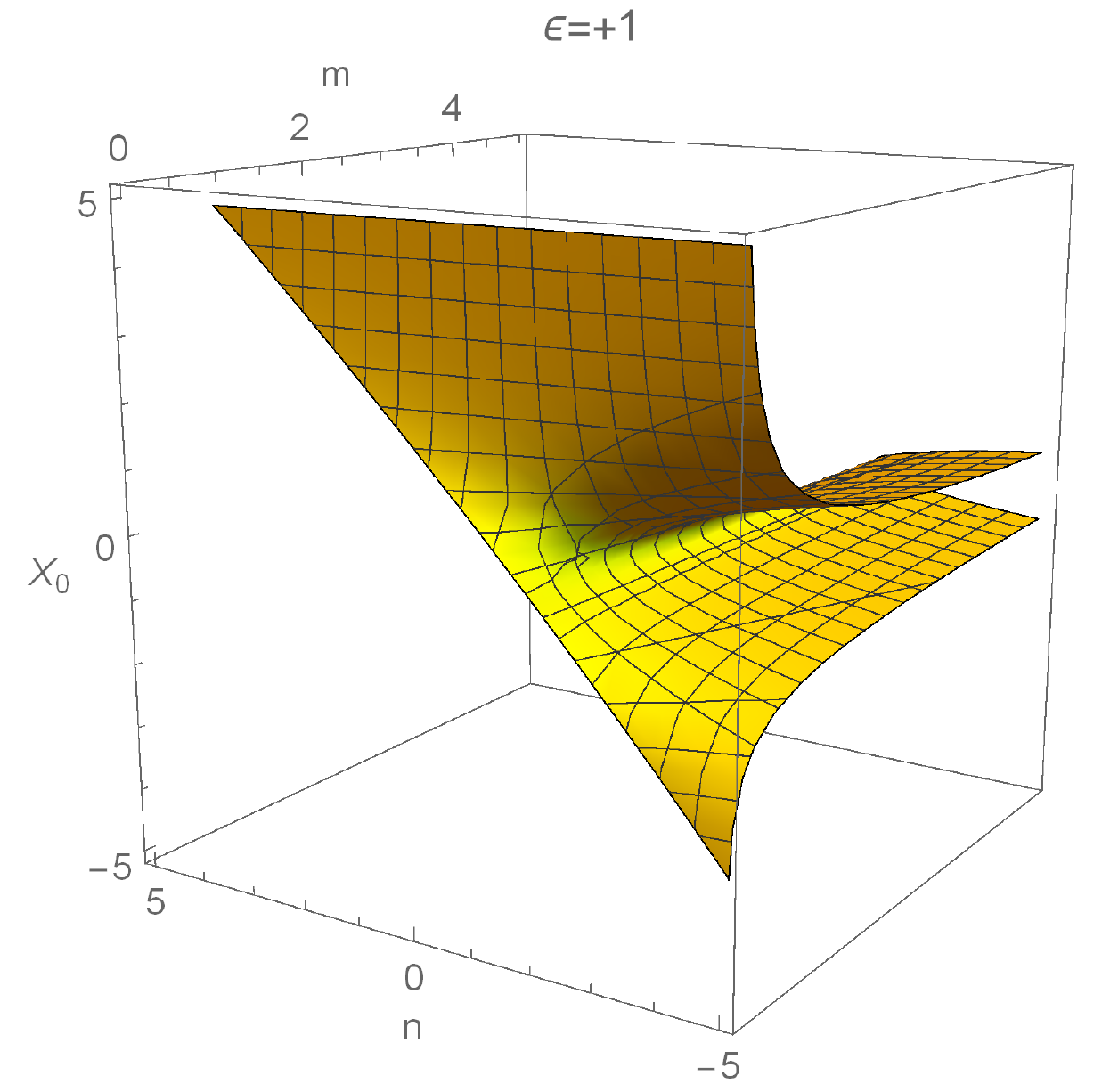}
\caption{\label{qqq} The variation $X_0$ versus typical
values of the $m$ and $n$  parameters in (\ref{qqq}) for the quintessence
background.}
\end{center}
\end{figure}

For this case, the equation (\ref{28}) takes the following form
\begin{equation}\label{qqq1}
n +2m X_{0}^{2}-X_{0}+2\epsilon=0.
\end{equation}
Then, one can find the  solutions as
\begin{eqnarray}
&&X_{01}=\frac{1-\sqrt{1+16m-8mn}}{4m}, 
~~~X_{02}=\frac{1+\sqrt{1+16m-8mn}}{4m},~~\epsilon=-1,\nonumber\\
&&X_{01}=\frac{1-\sqrt{1-16m-8mn}}{4m}, 
~~~X_{02}=\frac{1+\sqrt{1-16m-8mn}}{4m},~~\epsilon=+1.
\end{eqnarray}
Similarly, some particular conditions are required on the parameters $m$ and
$n$  for having positive
or negative solutions.
In Figure \ref{qqq}, we have plotted the solutions of  (\ref{qqq1}) for some typical ranges of $m$ and $n$ parameters.  Then, regarding this figure, one realizes the possibility of the formation of  both  naked singularities and black
holes in the quintessence background depending on the value of parameters.

\subsection{Black Hole-Quintessence Background Field Interactions}
In the cosmological context, the quintessence filed is known as the simplest
scalar field dark energy model without having theoretical problems such as Laplacian instabilities or ghosts. The energy
density and the pressure profile of the quintessence filed are generally considered to vary with time and depend on the scalar field and the potential, which are
 given by $\rho= \frac{1}{2}\dot \phi^2 + V(\phi)$ and
$p = \frac{1}{2}\dot \phi^2 -V(\phi)$, respectively. Then, the associated equation of state parameter for quintessence field lies in the range $-1<\omega_q<-\frac{1}{3}$.
 The static  Schwarzschild black hole solution surrounded by a quintessence
field was found by Kiselev \cite{Kiselev}. This solution was generalized
to the charged case and  studied in
\cite{Ghaderi1, Jamil, Jamill}.

For the quintessence surrounding field, we set the equation of state parameter
of quintessence
field as $\omega_q=-\frac{2}{3}$
 \cite{Kiselev, AVikman}. Then, the metric (\ref{mjoon})
takes the following form
\begin{equation}
ds^{2}=-\left(1-\frac{2M(u)}{r}-N_{q}(u)r\right)du^2+2\epsilon dudr+r^2d\Omega^2,
\end{equation}
where $N_q(u)$ is the normalization parameter for the quintessence field surrounding the black hole, with
the dimension of $[N_{q}] = l^{-1}$. This result shows a non-trivial contribution of the characteristic feature
of the surrounding quintessence field  to the metric of the Vaidya black hole. The presence of the background quintessence filed
changes the causal structure and Penrose diagrams of this black hole solution in comparison to the black hole in an empty background.
A rather similar effect happens when one immerses an static Schwarzschild in a (anti)-de Sitter background with the difference that here the spacetime
 tends asymptotically to quintessence rather than  (anti)-de Sitter asymptotics.

Regarding the positive energy condition for the quintessence background, represented by the relation
(\ref{WEC}), it is required to have $N_q(u)\geqslant0$. The radiation density is given by
\begin{equation}\label{sigma***}
\sigma(u,r)=\epsilon\left(\frac{2\dot M(u)}{r^{2}}+\dot N_{q}(u)\right).
\end{equation}
Then, the dynamical behaviour of the background quintessence field is
governed by
\begin{equation}
\begin{cases}\dot N_{q}(u)\leq -\,\frac{2}{r^2}\,\dot M(u),~~~ & \epsilon=-1,
\\
\\
\dot N_{q}(u)\geq -\frac{2}{r^2}\,\dot M(u), &\epsilon=+1.
\end{cases}
\end{equation}
Consequently, at any distance $r$ from the black hole, the
surrounding quintessence field must obey the above conditions. Interestingly, for the special case of $\dot N_{q}(u)= -\frac{2\dot M(u)}{r^2}$,
there is no pure radiation-accretion density, i.e $\sigma(u,r)=0$.
  This case corresponds to two possible physical situations. The first one is related to the situation where observer can be located at
any distance $r$ such that the quintessence background's and black hole's contributions cancel
out each others  leading to $\sigma(u,r)=0$ for a moment or even a period of time. Then, it is required that for an evaporating
black hole, we have an equal absorbing quintessence background or for an accreting black hole we
have an equal accreted quintessence background.  The second situation is related to
the case that for the given dynamical behaviors of the black hole and its quintessence background, one can find the particular distance
 $r_{*}=\sqrt{-\frac{2\dot M(u)}{\dot N_{q}(u)}}$ possessing zero energy density. For $|\dot N_q(u)|\ll|\dot M(u)|$, we have $r_*\rightarrow\infty$. This indicates that for an evolving black hole in an almost static quintessence
background, the positive energy condition is satisfied everywhere.
Also, the positivity of $r_*$ also requires that
$\dot M(u)$ and $\dot N_q(u)$ have opposite signs.
Then, if one realize the black and its surrounding quintessence filed behaviors, i.e  $\dot M(u)$ and $\dot N_q(u)$ values, he can find a distance at which we have no any radiation-accretion energy density contribution.
Based on these possibilities, the various situations  in the Table \ref{table5} can be realized.
\vspace{0.5cm}
\begin{table}[ht]
\begin{center}
\tabcolsep=0.08cm
\begin{tabular}{|c|c|c|c|c|c|c|c|}\hline
$\epsilon$ &  $\dot M$ & $\dot N_q$ & $r_*$ &$\sigma({r<r_*})$ &$\sigma({r=r_*})$&
$\sigma({r>r_*})$  & Physical Process \\\hline
-1  & - & - & Imaginary & +  &+ & + & Accretion/Decay of SF by Evaporating/Vanishing BH  \\\hline
-1  & + & - & + & -  &0 & + & Accretion of SF by BH  \\\hline
-1  & - & + & + & +  &0 & - & Absorbtion of BH's radiation by SF \\\hline
+1   & - & + & + &- & 0& + & Absorbtion of BH's radiation by SF \\\hline
+1   & + & - & + &+ & 0& - & Accretion of SF by BH  \\\hline
+1  & + & + & Imaginary & +  &+ & + & Accretion of BH and SF  \\\hline
\end{tabular}
\end{center}
\caption{BH and its surrounding quintessence field parameters for $\epsilon=\pm1$.
For the quintessence background, the positive energy condition may be completely
or partially respected regarding to the above situations.}
\label{table5}
\end{table}
\vspace{0.5cm}
\\
Then, regarding this table, the  positive values of $r_*$ are physically viable and their corresponding physical processes are listed in the last column. These properties are determined according to the behaviours of the parameters $\epsilon$,  $\omega_q$, and quantities  $\dot M(u)$, $\dot N_q(u)$, and $\sigma(u,r)$. Those values of $r_*$ corresponding to the negative energy density $\sigma(u,r)$ represent no physical situation about the evaporation-absorption or accretion. The real features of those regions are hidden by the weak energy condition. In the reference \cite{jiao},   the accretion into a static Kiselev black hole with a static exterior spacetime surrounded by a quintessence
field without the back-reaction effect is studied.  The obtained results in \cite{jiao} are implying that the accretion
rate and the critical points depend on the background quintessence parameter
$N_q$. Then, these features deserve to be incorporated in astrophysical studies of the accretion processes.

Regrading the Table 5, the behaviour of radiation-accretion
density $\sigma$ in (\ref{sigma***}) is plotted for some typical values of $\dot M$ and $ \dot
N_q$ in Figure 3.
Using these plots, one can compare the  radiation-accretion densities for the various situations.  \begin{figure}
\centering
\includegraphics[scale=0.62]{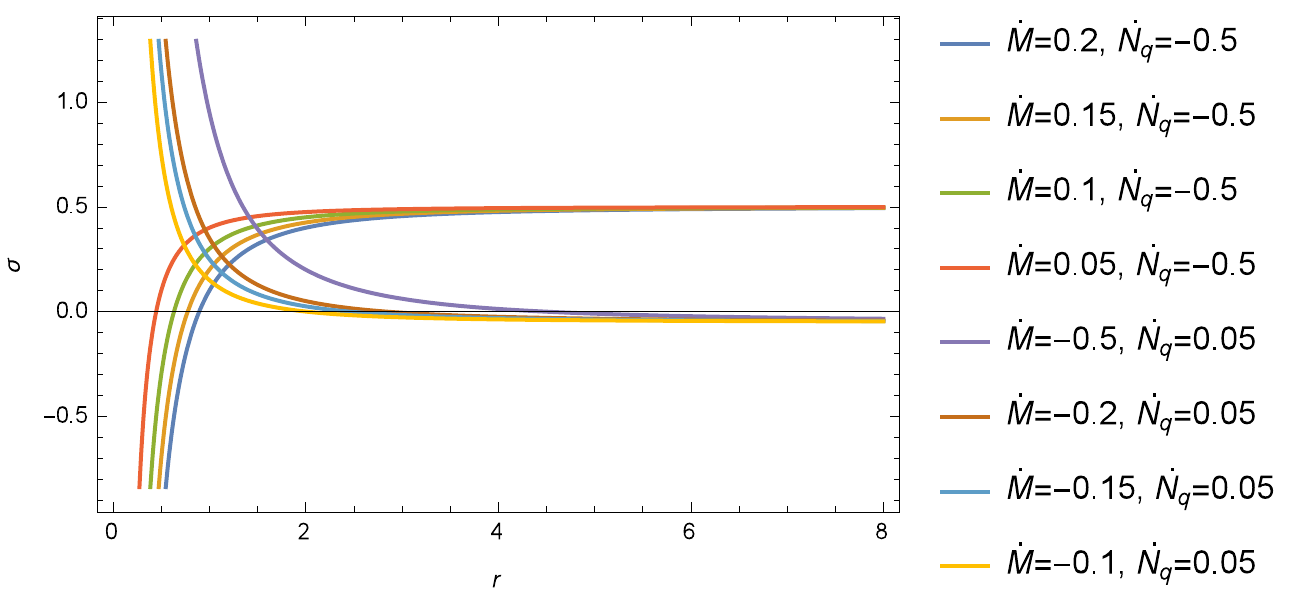}
\includegraphics[scale=0.62]{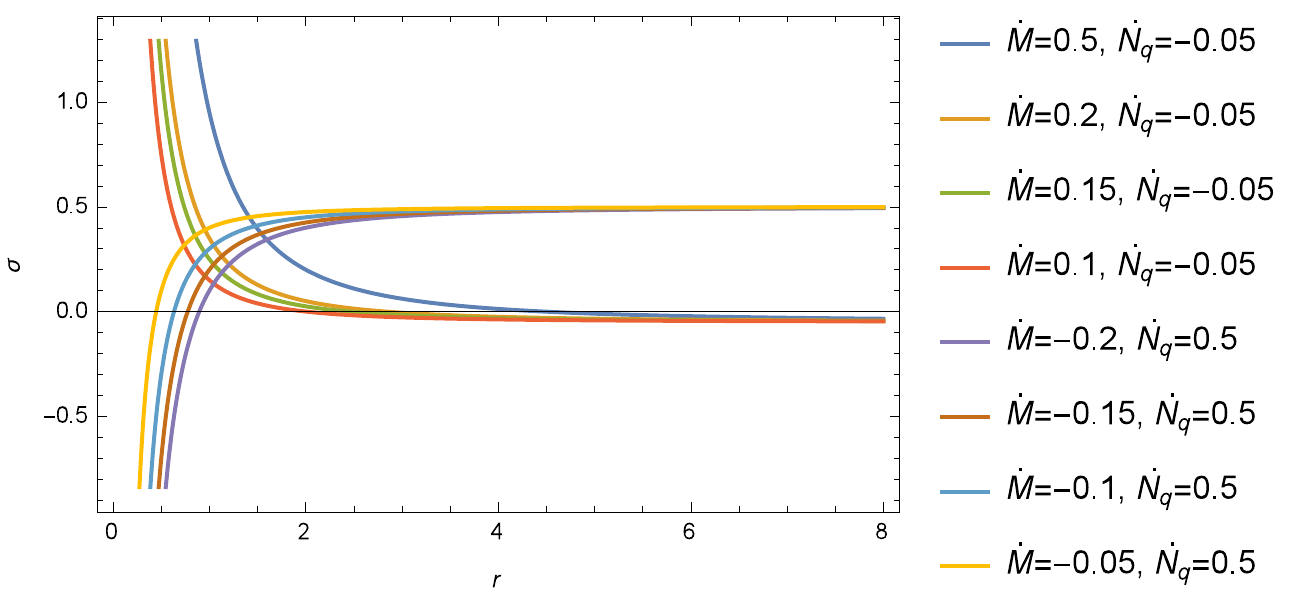}
\caption{\label{quiq}
\newline
\textbf{\textit{Left Fig.}} The radiation-accretion  density $\sigma$ versus the distance $r$ for some typical constant
$\dot M$ and $\dot N_{q}$ values for $\epsilon=-1$ in the quintessence background.
In  the four upper cases, the accretion density  is an increasing function from negative to the positive values.
In the four lower cases,   the radiation
density is a decreasing function decreases from positive values to negative values. Then, for a dynamical quintessence background, if the  condition  $|\dot N_q(u)|\ll|\dot M(u)|$ is not met, the positive energy condition is violated in some regions of spacetime.
\newline
\textbf{\textit{Right Fig.}} The radiation-accretion  density $\sigma$ versus the distance $r$ for some typical constant
$\dot M$ and $\dot N_{q}$ values for $\epsilon=+1$ in the quintessence  background.
In  the upper panel, the accretion density  is a decreasing function from positive to the negative values.
In the lower panel,   the radiation
density is an increasing function  from negative values to positive values. Then, for a dynamical quintessence background, if the  condition  $|\dot N_q(u)|\ll|\dot M(u)|$ is not met, the positive energy condition is violated in some regions of spacetime.}
\end{figure}
\subsection{Timelike Geodesics for the Black Hole in the Quintessence Field Background}
For this case, the distances $D_{s_{1}}$
and $D_{s_{2}}$ associated with  $|\frac{a_{s_1}}{a_{N}}|\simeq1$ and $|\frac{a_{s_2}}{a_{L}}|\simeq1$,
respectively, are given as 
\begin{equation}
D_{s_{1}}^2=\frac{2M(u)}{|-N_q(u)|},~~~~~
D_{s_{2}}^2=\frac{6M(u)}{|N_q(u)|}.
\end{equation}
In   Figure \ref{rqq}, we have plotted the location of these particular
distances for some typical ranges of the black hole mass $M(u)$ and background
quintessence field $N_q(u)$ parameters.
Then, one finds that there are   possibilities for the 
equality of the Newtonian force and GR correction terms to the corresponding quintessence background
field contributions.
\begin{figure}
\begin{center}
\includegraphics[scale=0.4]{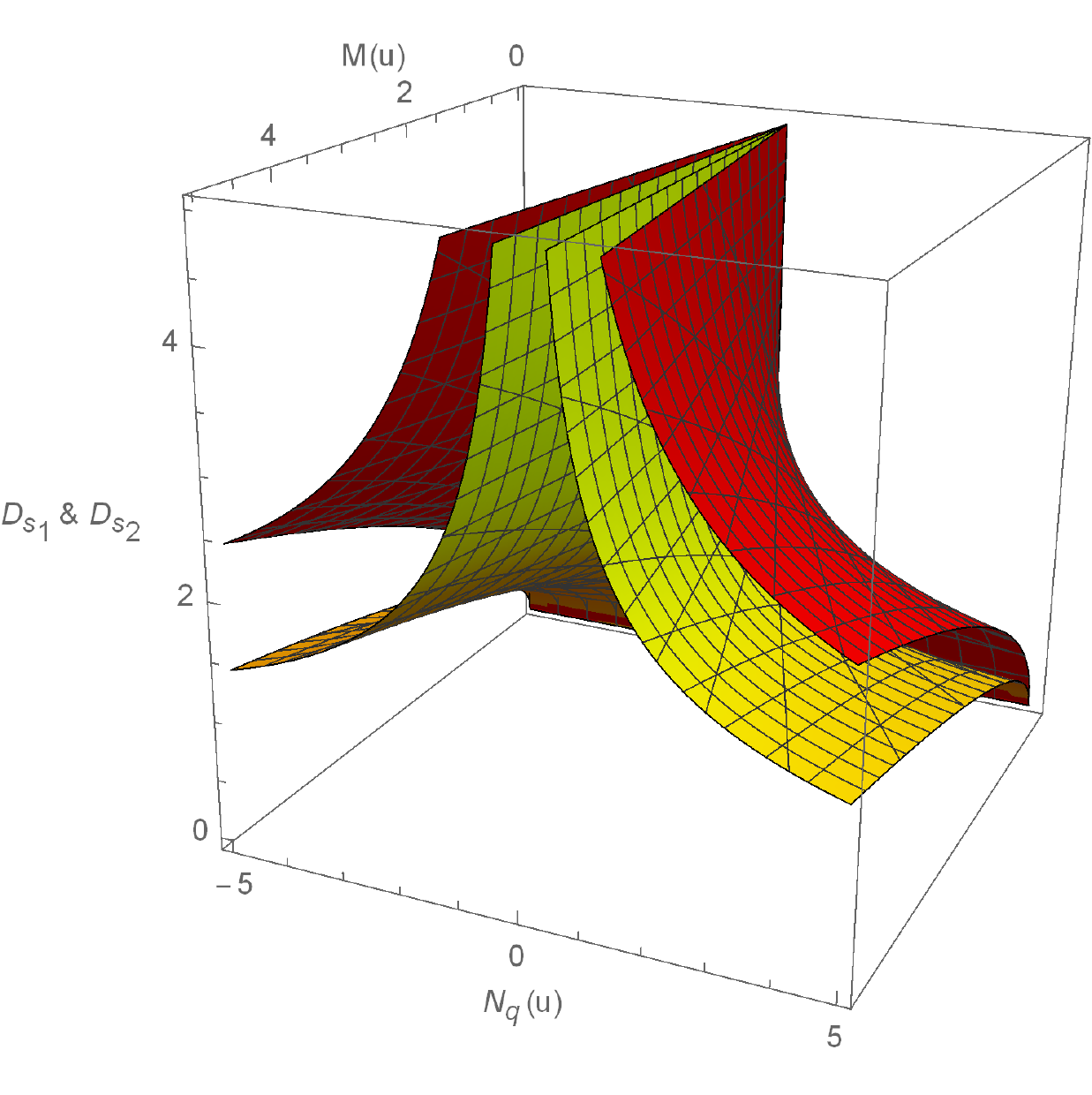}
\end{center}
\caption{\label{rqq} The variation of $D_{s_1}$
(yellow plot) and $D_{s_2}$ (red plot) versus typical
values of the $M(u)$ and $N_q(u)$ parameters for the quintessence background.}
\end{figure}

\begin{figure}
\begin{center}
\includegraphics[scale=0.35]{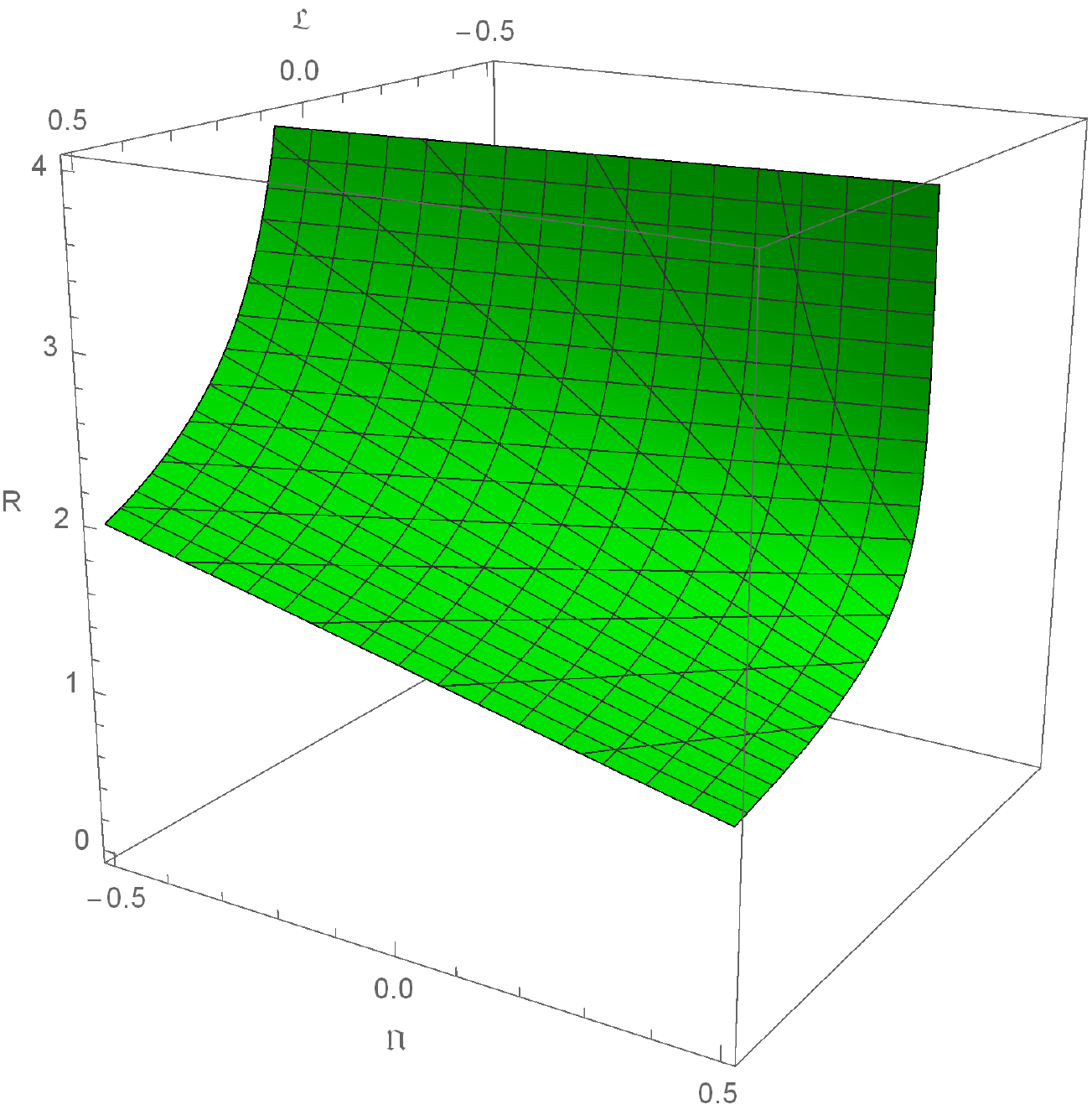}
\caption{\label{gqq} The variation of $R$ versus typical
values of the $\mathfrak{L}$ and $\mathfrak{N}$ parameters in (\ref{eqd})
for the quintessence background. We have set $M=1$ without loss of generality.}
\end{center}
\end{figure}

The equation (\ref{eq}) associated with $a_{i}\simeq
{a_{N}}$ for this case takes the following form
\begin{equation}\label{eqq}
\mathfrak{L} R^{-2}+\frac{1}{2}\mathfrak{N}\simeq MR^{-3}.
\end{equation}
Then, we obtain the following solutions 
\begin{eqnarray}
&&R_1\simeq- \frac{2\mathfrak{L}}{3^{\frac{1}{3}} \left(9M\mathfrak{N}^{2}+\sqrt{3}\sqrt{8\mathfrak{L}^{3}\mathfrak{N}^{3}} +27M^2 \mathfrak{N}^4 \right)^{\frac{1}{3}}}+ \frac{\left(9M\mathfrak{N}^{2}+\sqrt{3}\sqrt{8\mathfrak{L}^{3}\mathfrak{N}^{3}} +27M^2 \mathfrak{N}^4 \right)^{\frac{1}{3}}}{3^{\frac{2}{3}}\mathfrak{N}},\nonumber\\
&&R_2\simeq \frac{(1+i\sqrt3)\mathfrak{L}}{3^{\frac{1}{3}} \left(9M\mathfrak{N}^{2}+\sqrt{3}\sqrt{8\mathfrak{L}^{3}\mathfrak{N}^{3}} +27M^2 \mathfrak{N}^4 \right)^{\frac{1}{3}}}- \frac{(1-i\sqrt3)\left(9M\mathfrak{N}^{2}+\sqrt{3}\sqrt{8\mathfrak{L}^{3}\mathfrak{N}^{3}} +27M^2 \mathfrak{N}^4 \right)^{\frac{1}{3}}}{2\times 3^{\frac{2}{3}}\mathfrak{N}},\nonumber\\
&&R_3\simeq \frac{(1-i\sqrt3)\mathfrak{L}}{3^{\frac{1}{3}} \left(9M\mathfrak{N}^{2}+\sqrt{3}\sqrt{8\mathfrak{L}^{3}\mathfrak{N}^{3}} +27M^2 \mathfrak{N}^4 \right)^{\frac{1}{3}}}- \frac{(1+i\sqrt3)\left(9M\mathfrak{N}^{2}+\sqrt{3}\sqrt{8\mathfrak{L}^{3}\mathfrak{N}^{3}} +27M^2 \mathfrak{N}^4 \right)^{\frac{1}{3}}}{2\times 3^{\frac{2}{3}}\mathfrak{N}}.\nonumber\\
\end{eqnarray}
Then, one finds that the location of this particular distance depends on the parameters $\mathfrak{L}, \mathfrak{N}$ and $M$.
In Figure \ref{gqq}, we have plotted the solutions of  (\ref{eqq}) for some typical ranges of $\mathfrak{L}$ and $\mathfrak{N}$ parameters.
This figure shows that depending the parameter values, there are locations
where the induced force, resulting from the radiation-accretion phenomena
in the quintessence background, is equal to the Newtonian
gravitational force.
%%%%%%%%%%%%%%%%%%%%%%%%%%%%%%%%%%%%%%%%%%%%%%%%%%%%%%%%%%%%%%%%%%%
%
\section{Evaporating-Accreting Vaidya Black Hole Surrounded by the Cosmological Constant }
\subsection{Naked Singularity or Black Hole Formation Analysis}
\begin{figure}
\begin{center}
\includegraphics[scale=0.35]{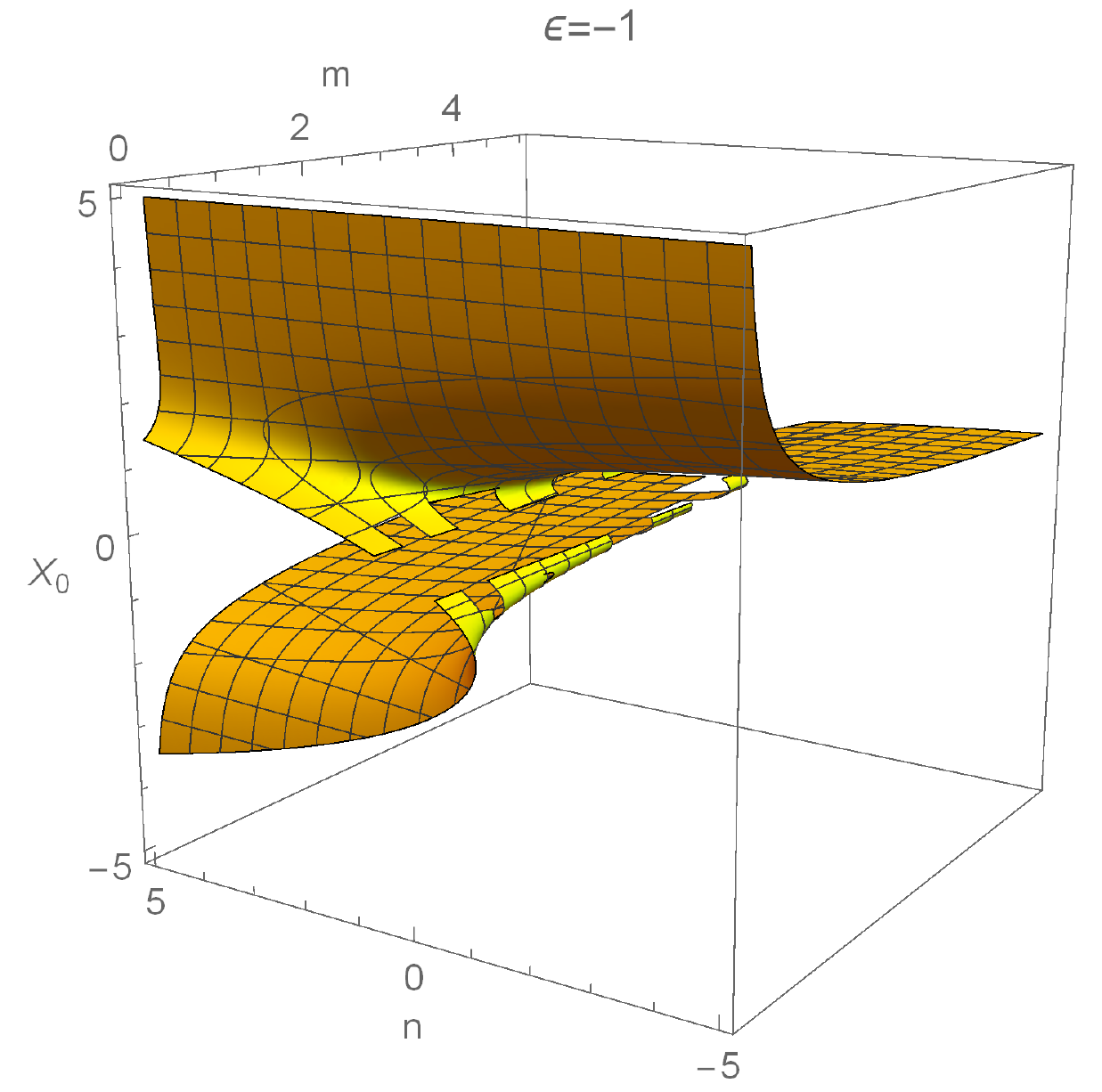}
\includegraphics[scale=0.35]{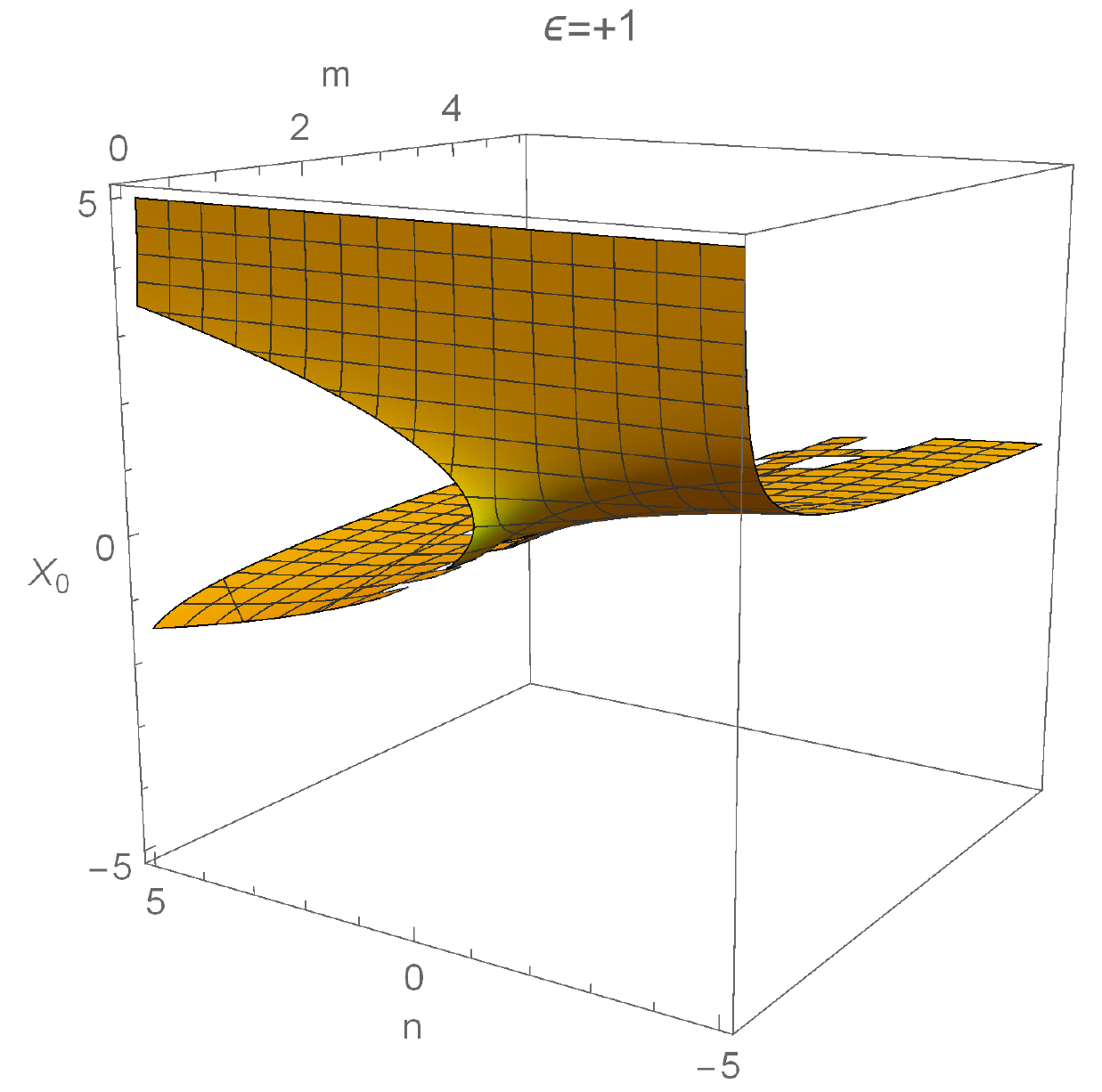}
\caption{ \label{coc} The variation $X_0$ versus typical
values of the $m$ and $n$  parameters in (\ref{ccc}) for the cosmological
constant-like background.}
\end{center}
\end{figure}

For this case, the equation (\ref{28}) takes the following form
\begin{equation}\label{ccc}
nX_{0}^{-1} +2m X_{0}^{2}-X_{0}+2\epsilon=0.
\end{equation}
Then, one can find the following solutions to (\ref{ccc})
\begin{eqnarray}
&&X_{01}=\frac{1}{6m}-\frac{-1-12m}{3\times 2^{\frac{2}{3}}m\Delta}+\frac{\Delta}{6\times 2^{\frac{1}{3}}m}, \nonumber\\
&&X_{02}=\frac{1}{6m}+\frac{(1+i\sqrt3)(-1-12m)}{6\times 2^{\frac{2}{3}}m\Delta}-\frac{(1-i\sqrt3)\Delta}{12\times 2^{\frac{1}{3}}m}, \nonumber\\
&&X_{03}=\frac{1}{6m}+\frac{(1-i\sqrt3)(-1-12m)}{6\times 2^{\frac{2}{3}}m\Delta}-\frac{(1+i\sqrt3)\Delta}{12\times 2^{\frac{1}{3}}m},
\end{eqnarray}
where $\Delta$ is given by 
\begin{eqnarray}
&&\Delta=\Delta_{-}=\left(2+36m-108m^{2}n+\sqrt{4(-1-12m)^3+(2+36m-108m^{2}n)^{2}}\right)^{\frac{1}{3}},
~~\epsilon=-1,\nonumber\\
&&\Delta=\Delta_{+}=\left(2-36m-108m^{2}n+\sqrt{4(-1+12m)^3+(2-36m-108m^{2}n)^{2}}\right)^{\frac{1}{3}},
~~\epsilon=+1.
\end{eqnarray}
Then, one see that some particular conditions on the parameters $m$ and
$n$ are required for having positive
or negative solutions. In Figure \ref{coc}, we have plotted the solutions of  (\ref{ccc}) for some typical ranges of $m$ and $n$ parameters.  Then, regarding this figure, one realizes the possibility of the formation of the both the naked singularities and black
holes in the cosmological constant-like background depending on the value of parameters.

%%%%%%%%%%%%%%%%%%%%%%%%%
%
\subsection{Black Hole-Cosmological Background Field Interactions}
For the cosmological constant-like surrounding field, we set the equation
of state parameter of the cosmological field as $\omega_c=-1$
 \cite{Kiselev, AVikman}. Then, the metric (\ref{mjoon})
takes the following form
\begin{equation}
ds^{2}=-\left(1-\frac{2M(u)}{r}-N_c(u)r^2\right)du^2+2\epsilon dudr+r^2d\Omega^2,
\end{equation}
where $N_c(u)$ is the normalization parameter for the cosmological field surrounding the black hole, with
the dimension of $[N_c(u)] = l^{-2}$. This result indicates the non-trivial contribution of the characteristic feature of the surrounding cosmological constant to the metric of the Vaidya
black hole. The presence of the background cosmological field
changes the causal structure and Penrose diagrams of this black hole solution in comparison to the black hole in an empty background.
This is similar to the case of the static Schwarzschild black
hole in a static de Sitter background such that the Penrose diagram
changes from Schwarzschild to Schwarzschild-(anti) de Sitter.  Then, in our
case, the Penrose diagram changes from Vaidya to Vaidya-de Sitter case with  dynamical cosmological causal boundaries.

Regarding the positive energy condition for this case, represented by the relation
(\ref{WEC}), it is required to have $N_c(u)\geqslant 0$. In this case, $N_{c}(u)$ plays the role of a positive dynamical
cosmological constant. Then, this case may describes the dynamical black holes
in more general cosmological scenarios considering a time varying cosmological term, which have been recently proposed in the literature. The main
purpose of these cosmological models is to provide an explanation for the  recent accelerating phase of the universe
 \cite{lambdavariying1}-\cite{pereira}. These
models are well known as the $\Lambda(t)$, where $t$ is the cosmic time.
 For the case of $N_c=constant=\Lambda$, we recover the
solution of the Vaidya black hole embedded in a de Sitter space  obtained in \cite{mallet}. The evolutionary behaviour  of such an evaporating black hole including the structures,
locations and dynamics of the apparent and event horizons are studied in \cite{Mallet2}.

In this case, the radiation-accretion density is given by
\begin{equation}\label{sigma****}
\sigma(u,r)=\epsilon\left(\frac{2\dot M(u)}{r^{2}}+\dot N_{c}(u)r\right).
\end{equation}
Then, the dynamical behaviour of the background cosmological constant-like field is
governed by
\begin{equation}
\begin{cases}\dot N_{c}(u)\leq -\frac{2}{r^3}\,\dot M(u),~~~ & \epsilon=-1,
\\
\\
\dot N_{c}(u)\geq -\frac{2}{r^3}\,\dot M(u), &\epsilon=+1.
\end{cases}
\end{equation}
\\
Consequently, at any distance $r$ from the black hole, the
surrounding cosmological field must obey the above conditions. Similar to
the previous solution, for the special case of $\dot N_{c}(u)= -\frac{2\dot M(u)}{r^3}$,
there is no pure radiation-accretion density, i.e $\sigma(u,r)=0$.
  This case corresponds to two possible physical situations. The first one is related to the situation where observer can be located at
any distance $r$ such that the cosmological background's and black hole's contributions cancel
out each others  leading to $\sigma(u,r)=0$ for a moment or even a period of time. Then, it is required that for a radiating
black hole, we have an equal absorbing cosmological background or for an accreting black hole we
have an equal accreted cosmological background field.  The second situation is related to
the case that for the given dynamical behaviors of the black hole and its cosmological background, one can find the particular distance
 $r_{*}=\left(-\frac{2\dot M(u)}{\dot N_{c}(u)}\right)^{\frac{1}{3}}$ possessing zero energy density. For $|\dot N_c(u)|\ll|\dot M(u)|$, we have $r_*\rightarrow\infty$. This indicates that for an evolving black hole in an almost static cosmological
background, the positive energy condition is respected everywhere. Here also, the positivity of $r_*$ also guarantees that
$\dot M(u)$ and $\dot N_c(u)$ have opposite signs.
Then, if one realize the black and its surrounding cosmological filed behaviors, i.e  $\dot M(u)$ and $\dot N_c(u)$ values, he can find a distance at which we have no any radiation-accretion energy density contribution.
Based on these possibilities, the various situations  in the Table \ref{table6} can be realized.
\vspace{0.5cm}
\begin{table}[ht]
\begin{center}
\tabcolsep=0.08cm
\begin{tabular}{|c|c|c|c|c|c|c|c|}\hline
$\epsilon$ &  $\dot M$ & $\dot N_c$ & $r_*$ &$\sigma({r<r_*})$ &$\sigma({r=r_*})$&
$\sigma({r>r_*})$  & Physical Process \\\hline
-1  & - & - & - & +  &+ & + &Accretion/Decay of SF by Evaporating/Vanishing BH \\\hline
-1  & + & - & + & -  &0 & + & Accretion of SF by BH  \\\hline
-1  & - & + & + & +  &0 & - & Absorbtion of BH's radiation by SF \\\hline
+1   & - & + & + &- & 0& + & Absorbtion of BH's radiation by SF \\\hline
+1   & + & - & + &+ & 0& - & Accretion of SF by BH  \\\hline
+1  & + & + & - & +  &+ & + & Accretion of BH and SF  \\\hline
\end{tabular}
\end{center}
\caption{ BH and its surrounding cosmological field parameters for $\epsilon=\pm1$.
For the cosmological background, the positive energy condition may be completely
or partially respected regarding to the above situations.}
\label{table6}
\end{table}
\vspace{0.5cm}

Regrading the Table \ref{table6}, the behaviour of radiation-accretion
density $\sigma$ in (\ref{sigma****}) is plotted for some typical values of $\dot M$ and $ \dot
N_c$ in Figure 4.
Using these plots, one can compare the  radiation-accretion densities for the various situations.  
\begin{figure}
\centering
\includegraphics[scale=0.62]{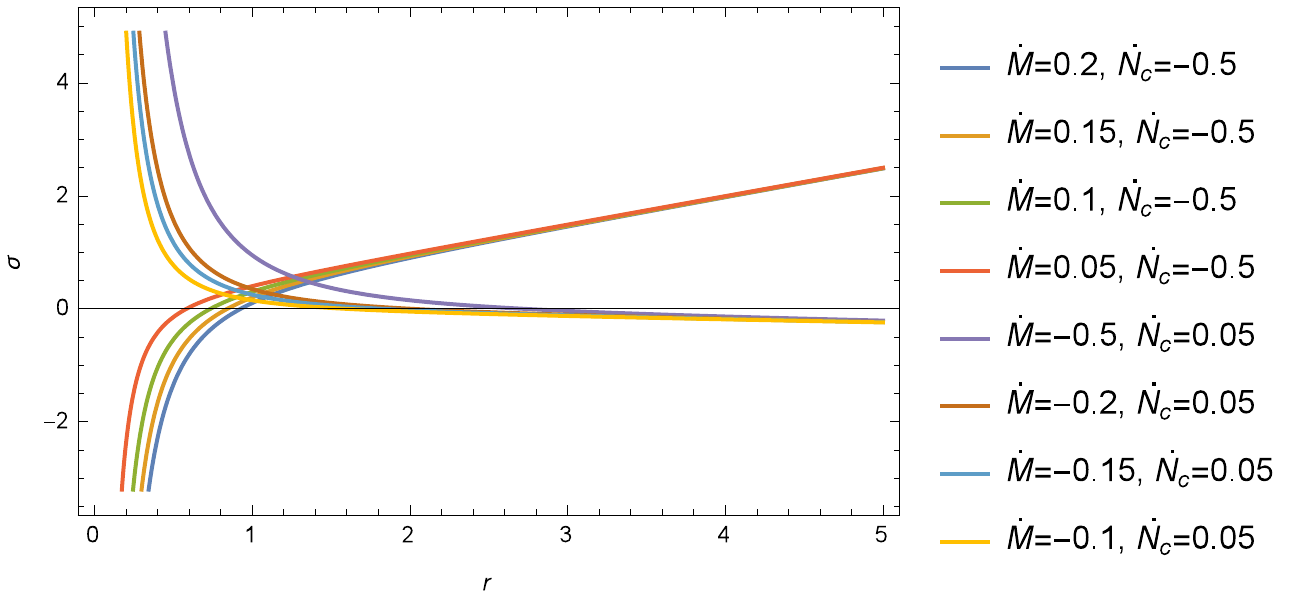}
\includegraphics[scale=0.62]{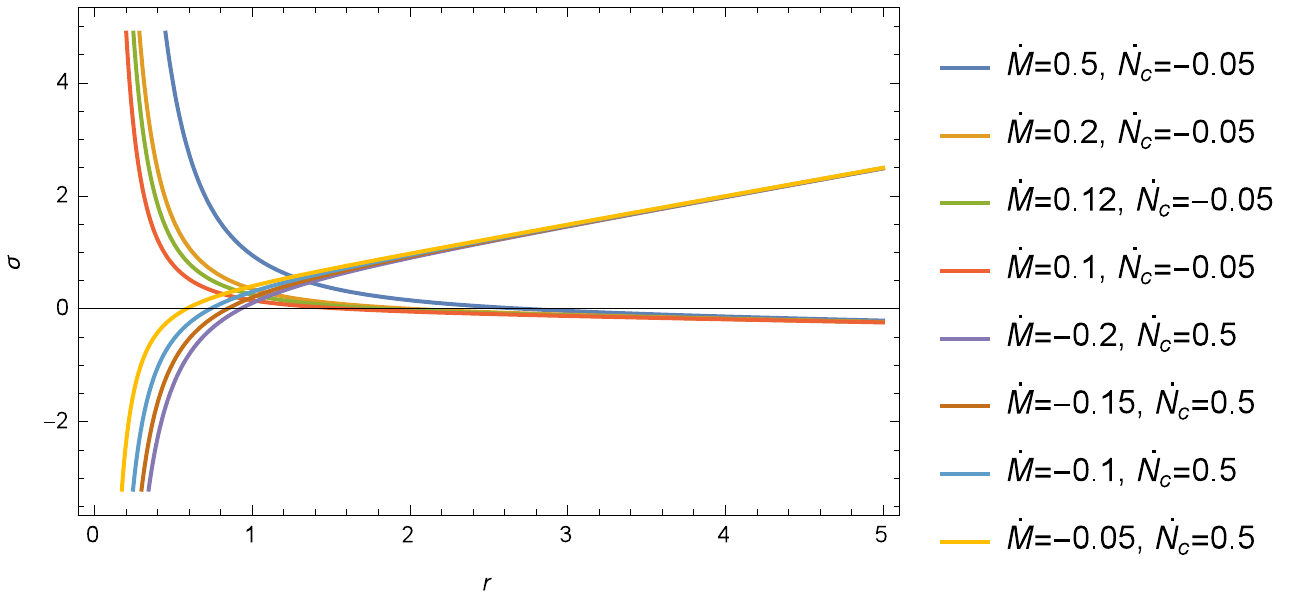}
\caption{\label{coco}
\newline
\textbf{\textit{Left Fig.}} The radiation-accretion  density $\sigma$ versus the distance $r$ for some typical constant values
$\dot M$ and $\dot N_{c}$ values for $\epsilon=-1$ in the cosmological background.
In  the four upper cases, the accretion density  is an increasing function from negative to  positive values.
In the four lower cases,   the radiation
density is a decreasing function  from positive values to negative values. Then, for a dynamical cosmological background, if the  condition  $|\dot N_c(u)|\ll|\dot M(u)|$ is not met, the positive energy condition is violated in some regions of spacetime.
\newline
\textbf{\textit{Right Fig.}} The radiation-accretion  density $\sigma$ versus the distance $r$ for some typical constant values
$\dot M$ and $\dot N_{c}$ values for $\epsilon=+1$ in the cosmological  background.
In  the four upper cases, the accretion density  is a decreasing function from positive to  negative values.
In the four lower cases,   the radiation
density is an increasing function  from negative  to positive values. Then, for a dynamical cosmological background, if the  condition  $|\dot N_c(u)|\ll|\dot M(u)|$ is not met, the positive energy condition is violated in some regions
of spacetime.}
\end{figure}
\subsection{Timelike Geodesics for the Black Hole in the Cosmological Constant-Like Field Background}
For this case, the distances $D_{s_{1}}$
and $D_{s_{2}}$ associated with  $|\frac{a_{s_1}}{a_{N}}|\simeq1$ and $|\frac{a_{s_2}}{a_{L}}|\simeq1$,
respectively, read as 
\begin{equation}
D_{s_{1}}^3=\frac{M(u)}{|-N_q(u)|},~~~~~
D_{s_{2}}\rightarrow\infty.
\end{equation}
The case of $D_{s_2}\rightarrow\infty$ is resulting from the fact that, in contrast to black hole itself,  the cosmological
constant-like field does not couple to angular momentum $L$, see $a_{s_2}$
in (\ref{mhs}). This shows that there is no similar effect to the GR
correction term for the cosmological constant-like field.  In Figure \ref{rqq}, we have plotted the location of the particular distance $D_{s_1}$ for some typical ranges of the black hole mass $M(u)$ and background
cosmological constant-like field $N_c(u)$ parameters.
Then, one finds that there are   possibilities for the 
equality of the Newtonian force  to the corresponding cosmological constant-like background field contributions.
\begin{figure}
\begin{center}
\includegraphics[scale=0.4]{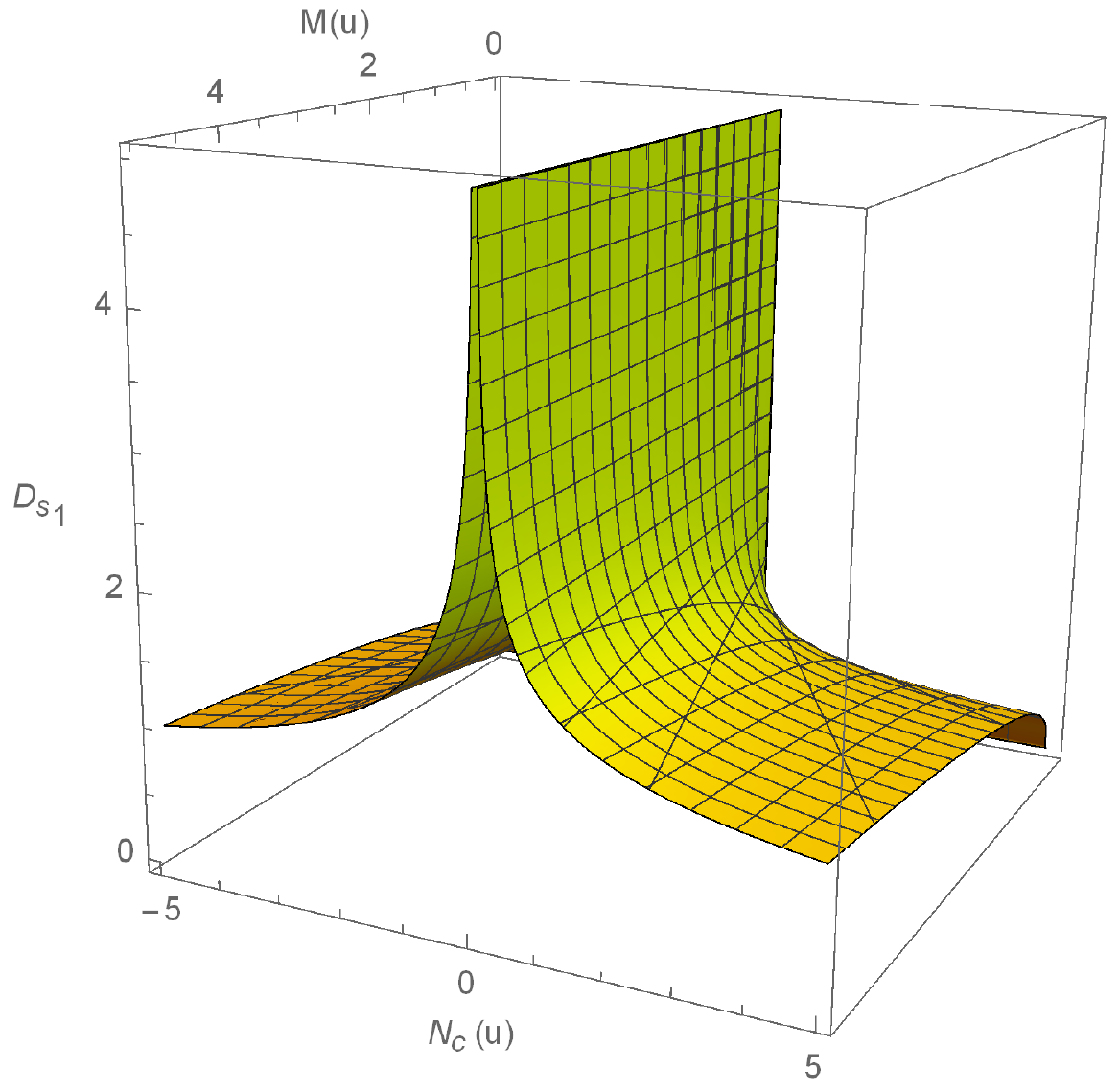}
\end{center}
\caption{\label{rqq} The variation of $D_{s_1}$
 versus typical values of the $M(u)$ and $N_c(u)$ parameters for the cosmological constant-like background.}
\end{figure}

\begin{figure}
\begin{center}
\includegraphics[scale=0.35]{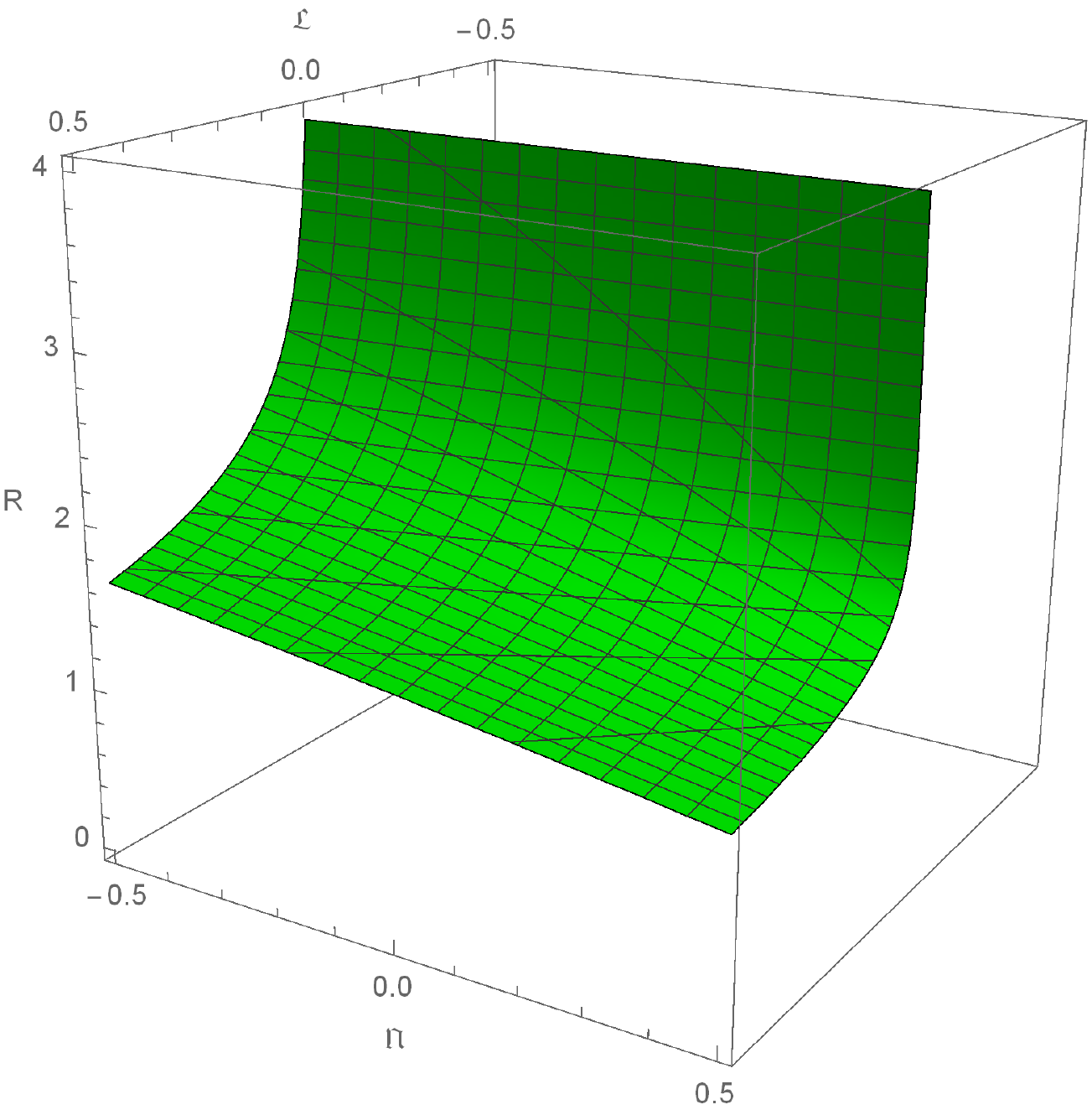}
\caption{\label{gc} The variation of $R$ versus typical
values of the $\mathfrak{L}$ and $\mathfrak{N}$ parameters in (\ref{eqd})
for the cosmological constant-like background. We have set $M=1$ without loss of generality.}
\end{center}
\end{figure}

For this case, the equation (\ref{eq}) associated with $a_i\simeq a_N$ takes the form of
\begin{equation}\label{eqc}
\mathfrak{L} R^{-3}+\frac{1}{2}\mathfrak{N}\simeq MR^{-4}.
\end{equation}
Then, we arrive at the following solutions \begin{eqnarray}
&&R_1\simeq \frac{1}{2}\Delta^{\frac{1}{2}}-\frac{1}{2}\sqrt{-\Delta-\frac{4\mathfrak{L}}{\mathfrak{N}\Delta^{\frac{1}{2}}}},\nonumber\\
&&R_2\simeq \frac{1}{2}\Delta^{\frac{1}{2}}+\frac{1}{2}\sqrt{-\Delta-\frac{4\mathfrak{L}}{\mathfrak{N}\Delta^{\frac{1}{2}}}},\nonumber\\
&&R_3\simeq- \frac{1}{2}\Delta^{\frac{1}{2}}-\frac{1}{2}\sqrt{-\Delta+\frac{4\mathfrak{L}}{\mathfrak{N}\Delta^{\frac{1}{2}}}},\nonumber\\
&&R_4\simeq- \frac{1}{2}\Delta^{\frac{1}{2}}+\frac{1}{2}\sqrt{-\Delta+\frac{4\mathfrak{L}}{\mathfrak{N}\Delta^{\frac{1}{2}}}},\nonumber\\
\end{eqnarray}
where $\Delta$ is
\begin{equation}
\Delta=-\frac{4\times 2^{\frac{2}{3}} M}{3^{\frac{1}{3}} \left(9\mathfrak{L}^{2}\mathfrak{N}+\sqrt{3}\sqrt{27\mathfrak{L}^{4}\mathfrak{N}^{2}} +128M^3 \mathfrak{N}^3 \right)^{\frac{1}{3}}}+\frac{2^{\frac{1}{3}}\left(9\mathfrak{L}^{2}\mathfrak{N}+\sqrt{3}\sqrt{27\mathfrak{L}^{4}\mathfrak{N}^{2}} +128M^3 \mathfrak{N}^3 \right)^{\frac{1}{3}}}{3^{\frac{2}{3}}\mathfrak{N}}.
\end{equation}
Again, we see that how the solutions of this particular distance depends on the parameters $\mathfrak{L}, \mathfrak{N}$ and $M$.
In Figure \ref{gc}, we have plotted the solutions of  (\ref{eqc}) for some typical ranges of $\mathfrak{L}$ and $\mathfrak{N}$ parameters.
This figure shows that depending the parameter values, there are locations
where the induced force, resulting from the radiation-accretion phenomena
in the cosmological background, is equal to the Newtonian
gravitational force.
%%%%%%%%%%%%%%%%%%%%%%%%%%%%%%%%%%%%%%%%%%%%%%%%%%%%%%%%%%%%%%%%%%%%%%%%%%%%%
%
\section{Evaporating-Accreting  Vaidya Black Hole Surrounded by the Phantom Field }
\subsection{Naked Singularity or Black Hole Formation Analysis}
\begin{figure}
\begin{center}
\includegraphics[scale=0.35]{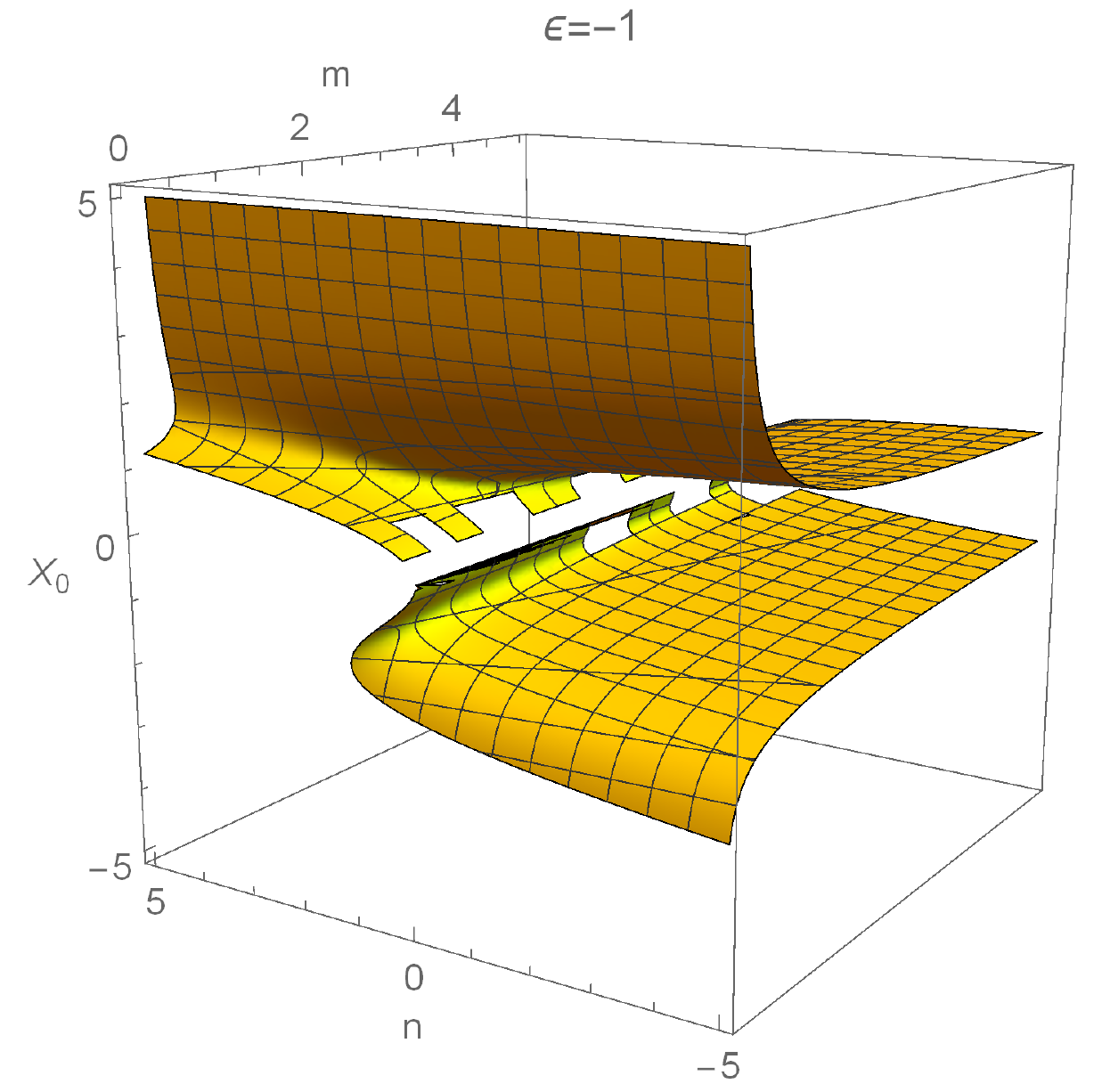}
\includegraphics[scale=0.35]{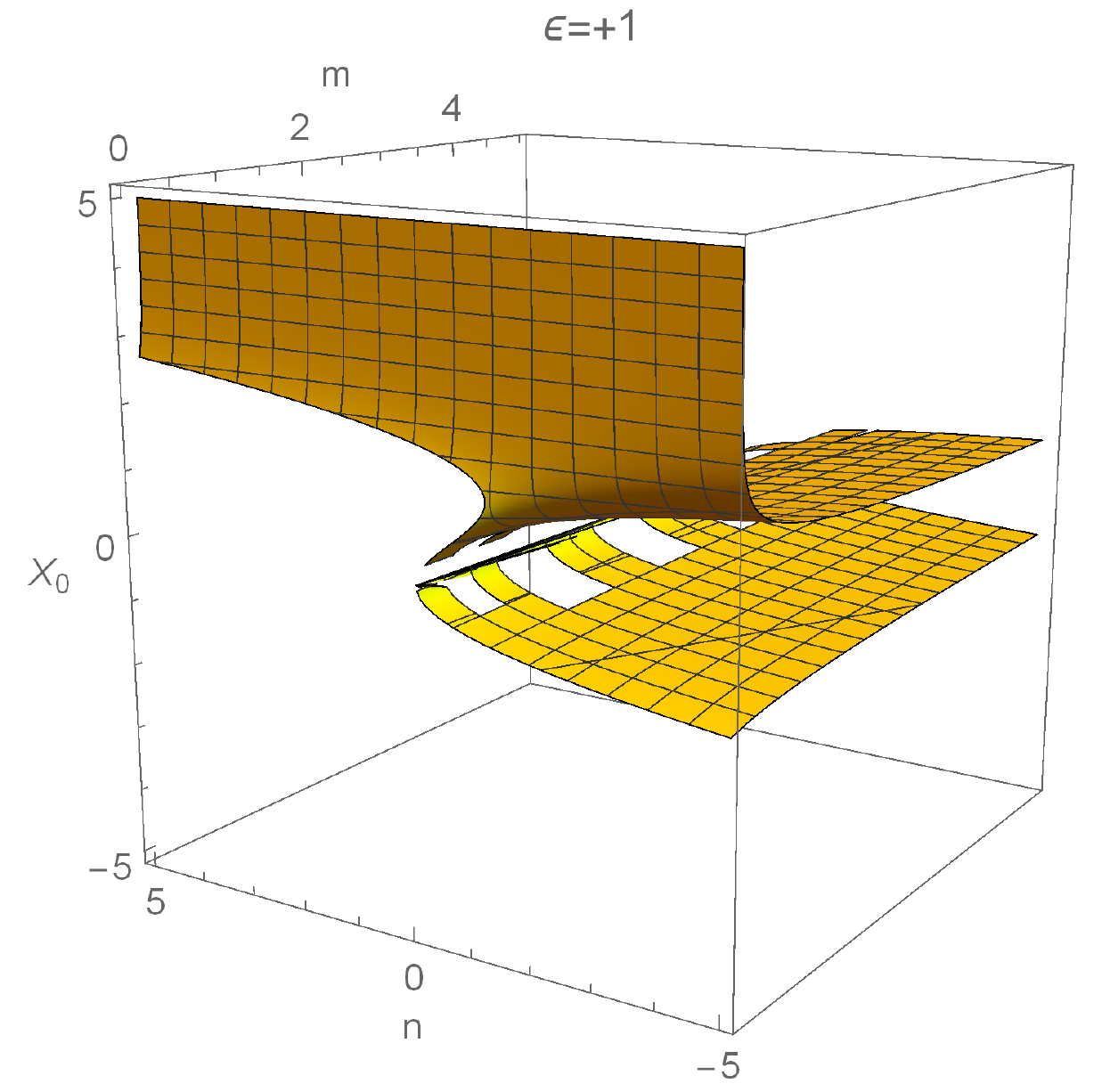}
\caption{\label{pha} The variation $X_0$ versus typical
values of the $m$ and $n$  parameters in (\ref{ppp1}) for the phantom background.}
\end{center}
\end{figure}

For this case, the equation (\ref{28}) takes the following form
\begin{equation}\label{ppp1}
nX_{0}^{-2} +2m X_{0}^{2}-X_{0}+2\epsilon=0.
\end{equation}
Then, one can find the following solutions to this equation 
\begin{eqnarray}
&&X_{01}=\frac{1}{8m}-\frac{1}{2}\Delta^{\frac{1}{2}}-\frac{1}{2}\sqrt{\frac{3}{16m^2}
+\frac{6}{3m}-\Delta-\frac{\frac{1}{8m^3}-\frac{2}{m^2}}{4\Delta^{\frac{1}{2}}}}, \nonumber\\
&&X_{02}=\frac{1}{8m}-\frac{1}{2}\Delta^{\frac{1}{2}}+\frac{1}{2}\sqrt{\frac{3}{16m^2}
+\frac{6}{3m}-\Delta-\frac{\frac{1}{8m^3}-\frac{2}{m^2}}{4\Delta^{\frac{1}{2}}}}, \nonumber\\
&&X_{03}=\frac{1}{8m}+\frac{1}{2}\Delta^{\frac{1}{2}}-\frac{1}{2}\sqrt{\frac{3}{16m^2}
+\frac{6}{3m}-\Delta+\frac{\frac{1}{8m^3}-\frac{2}{m^2}}{4\Delta^{\frac{1}{2}}}}, \nonumber\\
&&X_{04}=\frac{1}{8m}+\frac{1}{2}\Delta^{\frac{1}{2}}+\frac{1}{2}\sqrt{\frac{3}{16m^2}
+\frac{6}{3m}-\Delta+\frac{\frac{1}{8m^3}-\frac{2}{m^2}}{4\Delta^{\frac{1}{2}}}},
\end{eqnarray}
where $\Delta$ is given by 
\begin{eqnarray}
\Delta&=&\Delta_{-}=\frac{1}{16 m^2}+\frac{2}{3 m}+\frac{2\times  2^{\frac{1}{3}} (1+6 m n)}{3 m \left(-16+27n+288 m n+\sqrt{(-16+27 n+288 m n)^2-4 (4+24 m n)^3}\right)^{\frac{1}{3}}}
\nonumber\\
&&+\frac{\left(-16+27n+288 m n+\sqrt{(16+27 n+288 m n)^2-4 (4+24 m n)^3}\right)^{\frac{1}{3}}}{6\times
2^{\frac{1}{3}}m},~~\epsilon=-1,\nonumber\\
\Delta&=&\Delta_{+}=\frac{1}{16 m^2}-\frac{2}{3 m}+\frac{2\times  2^{\frac{1}{3}} (1+6 m n)}{3 m \left(16+27n-288 m n+\sqrt{(16+27n-288 m n)^2-4 (4+24 m n)^3}\right)^{\frac{1}{3}}}
\nonumber\\
&&+\frac{\left(16+27n-288 m n+\sqrt{(16+27n-288 m n)^2-4 (4+24 m n)^3}\right)^{\frac{1}{3}}}{6\times
2^{\frac{1}{3}}m},~~\epsilon=+1.
\end{eqnarray}
Then, similar to the previous cases, some particular conditions on the parameters $m$ and
$n$ are required for having positive
or negative solutions. In Figure \ref{pha}, we have plotted the solutions of  (\ref{ppp1}) for some typical ranges of $m$ and $n$ parameters.  Regarding this figure, we find the possibility of the formation of the both the naked singularities and black
holes in the phantom background depending on the value of parameters.

\subsection{Black Hole-Phantom Background Field Interactions}
For the phantom surrounding field, we set the equation of state parameter of phantom
field as $\omega_p=-\frac{4}{3}$
\cite{ AVikman}. Then, the metric (\ref{mjoon})
takes the following form
\begin{equation}
ds^{2}=-\left(1-\frac{2M(u)}{r}-N_p(u)r^3\right)du^2+2\epsilon dudr+r^2d\Omega^2,
\end{equation}
where $N_p(u)$ is the normalization parameter for the phantom field surrounding the black hole, with
the dimension of $[N_{p}(u)] = l^{-3}$. Similarly, this result is interpreted as the non-trivial contribution of the characteristic feature
of the surrounding phantom field  to the metric of the Vaidya black hole. The presence of the background phantom filed
changes the causal structure and Penrose diagrams of this black hole solution in comparison to the Vaidya black hole in an empty background.

Regarding the weak energy condition for this case, represented by the relation
(\ref{WEC}), it is required to have $N_p(u)\geqslant0$.
In this case, the radiation-accretion density is given by
\begin{equation}\label{sigma5}
\sigma(u,r)=\epsilon\left(\frac{2\dot M(u)}{r^{2}}+\dot N_{p}(u)r^{2}\right).
\end{equation}
Then, the dynamical behaviour of the background field is
governed by
\begin{equation}
\begin{cases}\dot N_{p}(u)\leq -\frac{2}{r^4}\,\dot M(u),~~~ & \epsilon=-1,
\\
\\
\dot N_{p}(u)\geq -\frac{2}{r^4}\dot M(u), &\epsilon=+1.
\end{cases}
\end{equation}
\\
Consequently, at any distance $r$ from the black hole, the
surrounding phantom field must obey the above conditions. Similar to the
previous solutions, for the special case of $\dot N_{p}(u)= -\frac{2\dot M(u)}{r^4}$,
there is no pure radiation-accretion density, i.e $\sigma(u,r)=0$.
  This case corresponds to two possible physical situations. The first one is related to the situation where observer can be located at
any distance $r$ such that the phantom background's and black hole's contributions cancel
out each others  leading to $\sigma(u,r)=0$ for a moment or even a period of time. Then, it is required that for a radiating
black hole, we have an equal absorbing phantom  background or for an accreting black hole we
have an equal accreted phantom background.  The second situation is related to
the case that for the given dynamical behaviors of the black hole and its
phantom background, one can find the particular distance
 $r_{*}=\left(-\frac{2\dot M(u)}{\dot N_{p}(u)}\right)^{\frac{1}{4}}$ possessing zero energy density. Similarly, for $|\dot N_p(u)|\ll|\dot M(u)|$, we have $r_*\rightarrow\infty$. This indicates that for an evolving black hole in an almost static phantom
background, the positive energy condition is satisfied everywhere.
Also, the positivity of $r_*$ also requires that
$\dot M(u)$ and $\dot N_p(u)$ must have opposite signs.
Then, if one realize the black and its surrounding phantom filed behaviors, i.e  $\dot M(u)$ and $\dot N_p(u)$ values, he can find a distance at which we have no any radiation-accretion energy density contribution.
Based on these possibilities, the various situations  in the Table
\ref{table7} can be realized.
\vspace{0.5cm}
\begin{table}[ht]
\begin{center}
\tabcolsep=0.08cm
\begin{tabular}{|c|c|c|c|c|c|c|c|}\hline
$\epsilon$ &  $\dot M$ & $\dot N_p$ & $r_*$ &$\sigma({r<r_*})$ &$\sigma({r=r_*})$&
$\sigma({r>r_*})$  & Physical Process \\\hline
-1  & - & - & Imaginary & +  &+ & + & Accretion/Decay of SF by Evaporating/Vanishing BH\\\hline
-1  & + & - & + & -  &0 & + & Accretion of SF by BH  \\\hline
-1  & - & + & + & +  &0 & - & Absorbtion of BH's radiation by SF \\\hline
+1   & - & + & + &- & 0& + & Absorbtion of BH's radiation by SF \\\hline
+1   & + & - & + &+ & 0& - & Accretion of SF by BH  \\\hline
+1  & + & + & Imaginary & +  &+ & + & Accretion of BH and SF  \\\hline
\end{tabular}
\end{center}
 \caption{BH and its surrounding phantom field parameters for $\epsilon=\pm1$.
For the phantom background, the positive energy condition may be completely
or partially respected regarding to the above situations.}
\label{table7}
\end{table}
\vspace{0.5cm}
\\
Specific scenarios involving the accretion of phantom
energy and resulting in the area decrease of black hole \cite{accretion4, accretion5, accretion6, gao} are related to the first
case in the above table. For example, in \cite{accretion4}, it is shown that the black holes will gradually vanish as the universe approaches a cosmological big rip state with a phantom field.  One should
note that the astrophysically ``observed'' infall of quintessence/phantom fields onto black holes are not detected till now.
 In the present work, we have just introduced a new dynamical solution to the Einstein field equations which can provide a classical model for the ``possible'' black hole accretion and evaporation (presumably very tiny) in different cosmological backgrounds. Such  theoretical studies of the accretion of exotic fields to the black holes are well motivated
by cosmology in which these exotic fields can be  responsible for the current accelerating expansion of the universe. 
Regrading the Table \ref{table7}, the behavior of radiation-accretion
density $\sigma$ in (\ref{sigma5}) is plotted for some typical values of $\dot M$ and $ \dot
N_p$ in Figure 5.
Using these plots, one can compare the  radiation-accretion densities for the various situations.
\begin{figure}
\centering
\includegraphics[scale=0.62]{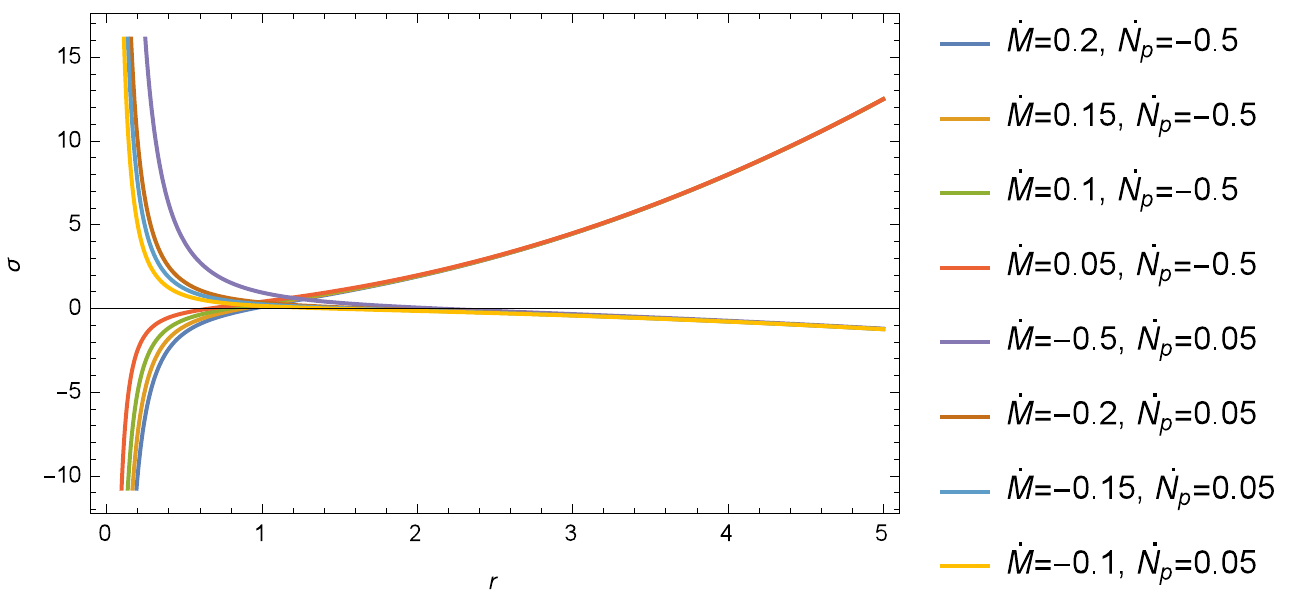}
\includegraphics[scale=0.62]{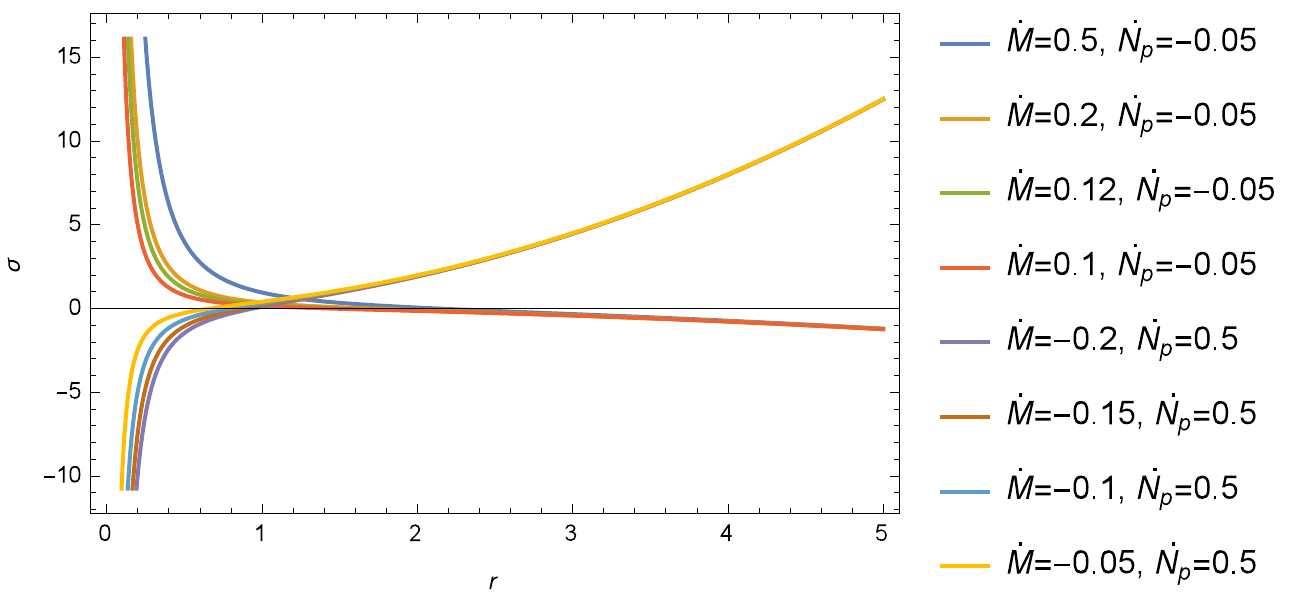}
\caption{\label{phantii}
\newline
\textbf{\textit{Left Fig.}} The radiation-accretion  density $\sigma$ versus the distance $r$ for some typical constant
$\dot M$ and $\dot N_{p}$ values for $\epsilon=-1$ in the phantom background.
In  the four upper cases, the accretion density  is an increasing function from negative to  positive values. In  the four lower cases,   the radiation
density is a decreasing function  from positive  to negative values. Then, for a dynamical phantom background, if the  condition  $|\dot N_p(u)|\ll|\dot M(u)|$ is not met, the positive energy condition is violated in some regions
of spacetime.
\newline
\textit{\textbf{Right Fig.}} The radiation-accretion  density $\sigma$ versus the distance $r$ for some typical constant $\dot M$ and $\dot N_{p}$ values for $\epsilon=+1$ in the phantom  background.
In  the four upper cases, the accretion density  is a decreasing function from positive to  negative values. In  the four lower cases, the radiation
density is an increasing function  from negative  to positive values. Then, for a dynamical phantom background, if the  condition  $|\dot N_p(u)|\ll|\dot M(u)|$ is not met, the positive energy condition is violated in some regions
of spacetime.}
\end{figure}
%%%%%%%%%%%%%%%%%%%%%%%%%%%%%%%%%%%%%%%%%%%%%%%%%%%%%%%%%%%%%%%%%%%%%%
%%%%%%%%%%%%%%%%%%%%%%%%%%%%%%%%%%%%%%%%%%%%%%%%%%%%%%%%%%%%%%%%%%%%%%%%%%%%%%%%%%%%

\subsection{Timelike Geodesics for the Black Hole in the Phantom Field Background}
For this case, the distances $D_{s_{1}}$
and $D_{s_{2}}$ associated with  $|\frac{a_{s_1}}{a_{N}}|\simeq1$ and $|\frac{a_{s_2}}{a_{L}}|\simeq1$,
respectively, are given as 
\begin{equation}
D_{s_{1}}^4=\frac{2M(u)}{3|-N_p(u)|},~~~~~
D_{s_{2}}^4=\frac{6M(u)}{|-N_p(u)|}.
\end{equation}
In   Figure \ref{rqq}, we have plotted the location of these particular
distances for some typical ranges of the black hole mass $M(u)$ and background
phantom field $N_p(u)$ parameters.
Then, one realizes  the   possibilities of the 
equality of the Newtonian force and GR correction terms to the corresponding phantom background
field contributions.
\begin{figure}
\begin{center}
\includegraphics[scale=0.4]{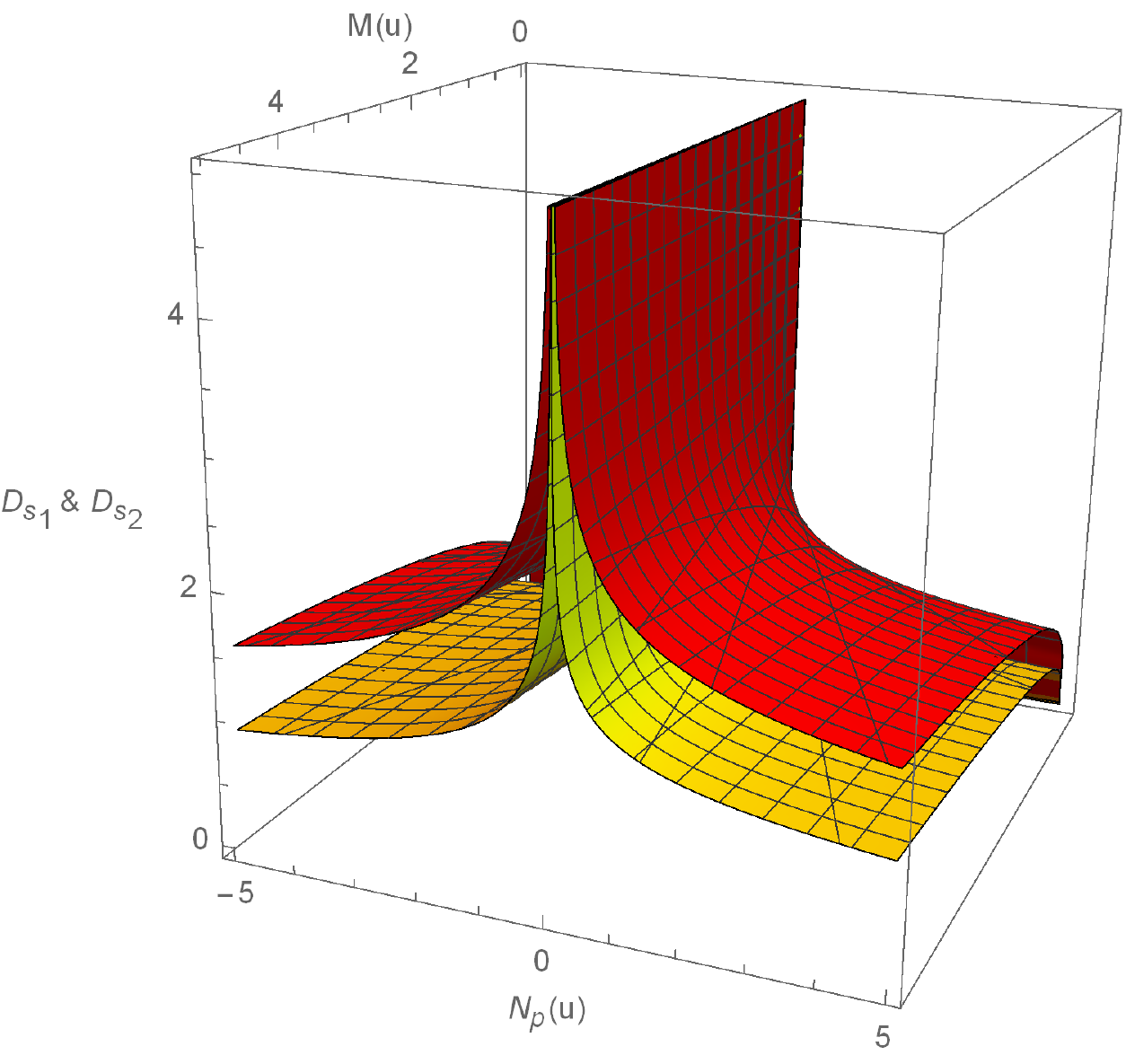}
\end{center}
\caption{\label{rqq} The variation of $D_{s_1}$
(yellow plot) and $D_{s_2}$ (red plot) versus typical
values of the $M(u)$ and $N_q(u)$ parameters for the quintessence background.}
\end{figure}

\begin{figure}
\begin{center}
\includegraphics[scale=0.35]{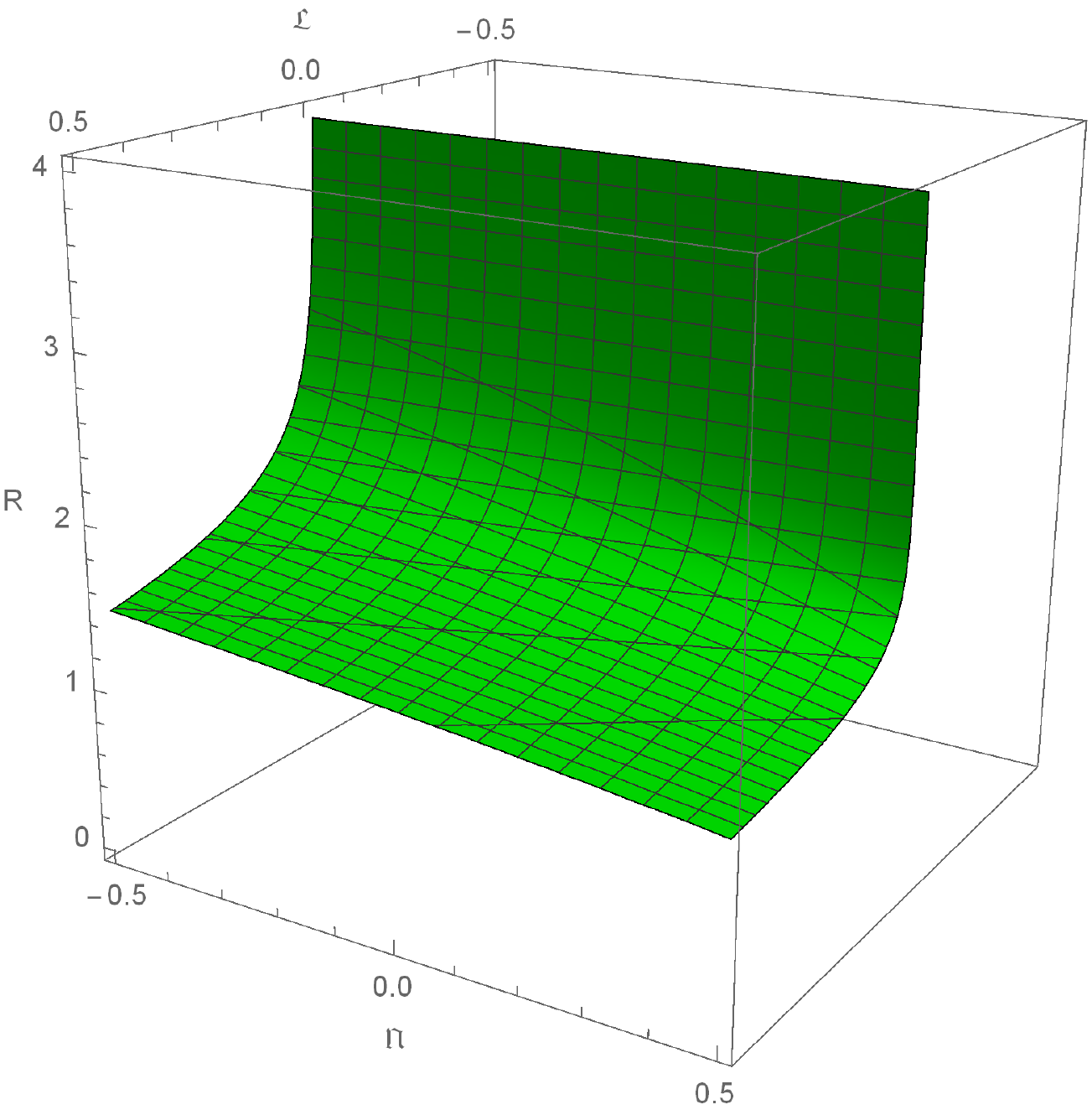}
\caption{\label{gp} The variation of $R$ versus typical
values of the $\mathfrak{L}$ and $\mathfrak{N}$ parameters in (\ref{eqp})
for the phantom background. We have set $M=1$ without loss of generality.}
\end{center}
\end{figure}

Moreover, the equation (\ref{eq}) for
this case takes the following form
\begin{equation}\label{eqp}
\mathfrak{L} R^{-4}+\frac{1}{2}\mathfrak{N}\simeq MR^{-5}.
\end{equation}
Then, we see that this produces a fifth order equation in which finding its analytical solutions is not simple. However, in Figure \ref{gp}, we have shown that there are numerical solutions to  (\ref{eqp}) for some typical ranges of $\mathfrak{L}$ and $\mathfrak{N}$ parameters. This figure indicates that depending the parameter values, there are locations
where the induced force, resulting from the radiation-accretion phenomena
in the phantom background, is equal to the Newtonian
gravitational force.

%%%%%%%%%%%%%%%%%%%%%%%%%%%%%%%%%%%%%%%%%%%%%%%%%%%%%%%%%%%%%%%%%%%%%%%%%%%%%%%%%%
%%%%%%%%%%%%%%%%%%%%%%%%%%%%%%%%%%%%%%%%%%%%%%%%%%%%%%%%%%%%%%%%%%%%%%%%%%%%%%%%%%%%%%%%%%%%%%
\section{Conclusion}
In this work, we have studied the general surrounded Vaidya solution with the cosmological
fields of dust,
radiation, quintessence, cosmological constant-like and phantom,  and investigated
its nature describing the possibility of the formation of naked singularities or black holes. We have obtained the general equation describing the nature of the
solution under a collapse, and have shown that depending on the parameter values, the
formation of both
naked singularity and black hole as the end state of the collapse are possible.  We have given the corresponding analytical solutions as well as
some plots indicating these possibilities. Then,
motivated by the fact that real astrophysical black holes as non-stationary and non-isolated objects are
living in non-empty backgrounds, we have focused on the black hole subclasses of the obtained general solution, namely the ``\textit{surrounded Vaidya black hole}'',  describing a dynamical evaporating-accreting black holes in the mentioned
dynamical cosmological backgrounds.  In the following, we summarize some of our obtained results
for this solution.

\begin{itemize}
\item Some of the subclasses of the obtained general solution for both
the dynamical and stationary limits have been addressed. In particular, we have shown that the original Vaidya solution can be recovered by
turning off the background field, and that the Kiselev static solution
can be obtained  in the stationary limit with an appropriate coordinate transformation. Also, the Schwarzschild solution
can be obtained in the stationary limit with a turned off background.

\item  We have shown that for the background field possessing  $\omega_s>0$,  if $\dot M(u)$ and $\dot N_s(u)$ have a same order of magnitude,   the surrounding background field contribution to the total density $\sigma(u,r)
$ is dominant near the black
hole while at far distances from the black hole it decreases faster than the
contribution  of the black hole mass changing term.
In contrast,  for the background field possessing  $\omega_s<0$,    the surrounding background field contribution is dominant at large distances while the black hole contribution is dominant near the black hole itself.
Then, from  astrophysical point of view, the detected amount of the  radiation-accretion density by the observer not only depends on the distance from the black but
also depends on the nature of background field.

\item  
We have discussed that positive energy condition for the surrounding field is met by the constraint $\omega_s N_{s}(u)\leq0$,
which determines the gravitational nature of the
term associated to surrounding field  in the metric function $f(u,r)$.
The positive energy condition for the radiation-accretion density is met by  the constraints $\dot N_{s}(u)\leq -2\,{r}^{3\omega_s }\,\dot M(u)$ 
and $\dot N_{s}(u)\geq -2\,{r}^{3\omega_s }\,\dot M(u)$ at any distance $r$ for $\epsilon=-1$ and $\epsilon=+1$, respectively. One astrophysical importance
of such  physical constraints is that the observer knows the dynamical range
of the background field at any distance and then prior to any observation,
he knows how to include or remove the background
field contribution, if he is only interested  in black hole's contributions,
or vice vera.

\item We have addressed  the solutions of the black hole in the dust 
($\omega_s=0$),
radiation ($\omega_s=\frac{1}{3}$), quintessence ($\omega_s=-\frac{2}{3}$), cosmological constant-like ($\omega_s=-1$) and phantom ($\omega_s=-\frac{4}{3}$) fields 
in detail. We have found that   the effectively evaporating-accreting black hole in the dust background appears as an effectively
evaporating-accreting black hole with an effective mass $M_{eff}(u) = 2M(u)+N_{d}(u)$.
Then, the presence of
new mass term changes the causal structure  just up to a re-scaling
in the original Vaidya solution.  For the radiation background, the spacetime metric looks like  the Bonnor-Vaidya and radiating dyon solutions with
the dynamical charge $Q(u) = \sqrt{\mathcal{N}_r(u)}$. A similar effect in the causal structure of spacetime
here happens when one adds charge to the static Schwarzschild black hole leading to Reissner-Nordström black
hole.
For  the black hole in the dust and radiation backgrounds, the spacetime
metrics are asymptotically flat while for the the quintessence, cosmological-like
and phantom backgrounds, spacetime metrics are asymptotically non-flat quintessence,
de Sitter-like and phantom, respectively. Consequently, the causal structure of these
three latter spacetimes are quite different from the original Vaidya spacetime
where the background is turned off.
\item Regarding the obtained radiation-accretion density $\sigma(u,r)$, we have found that there are particular distances $r_*$  where
$\sigma(u,r)$ vanishes, i.e $\sigma(u,r_*)=0$.   These distances are given
by  $r_{*}(u)=\left(-\frac{\dot N_{s}(u)}{2\dot M(u)}\right)^{\frac{1}{3\omega_{s}}}$ and $r_*=\infty$. Then, one realizes that in
the first case, for $(i)$ $-\frac{2}{3}<\omega_s<0$ with $|\dot N_s(u)|\ll 2|\dot M(u)|$ and for $(ii)$ $\omega_s\geq0$ with $2|\dot M(u)|\ll |\dot N_{s}(u)|$, we
have $r_*\rightarrow \infty$.  This means that for $(i),$ the black hole evolves very faster than its background while for $(ii)$,
the black hole evolves very slow relative to its background.  
 Also, for $(iii)$  $\omega_s\leq-\frac{2}{3}$ with
 $|\dot N_s(u)|\ll 2|\dot M(u)|$, representing a rapidly evolving black hole relative to its background,  we have $r_*\rightarrow \infty$.
Then, by satisfaction
of these dynamical conditions to hold $r_*\rightarrow\infty$, the positive energy density is respected everywhere in the spacetime. The case $(ii)$ includes
the  black hole surrounded by the rapidly evolving dust and radiation fields while
the  cases $(i)$ and $(iii)$ imply an evolving black hole in an almost static   cosmological backgrounds
(quintessence, cosmological constant-like or phantom fields) responsible for the accelerating expansion of the universe. In practice,
an astrophysical observer detects a radiation-accretion density resulting from the interaction
of the black hole with its surrounding field even at far distances, in which
for $(ii)$ the main contribution in
the detected radiation-accretion density belongs to  surrounding
field while for $(i)$ and $(iii)$,  it is the black hole which has the main contribution in the radiation-accretion density.

\item In the case that there is a  real, positive and finite value for $r_*$,
the positive energy condition is violated in some regions
of spacetime such that real features of those regions are hidden by the weak energy condition.
Then, it is physically reasonable
to do any astrophysical experiments in the regions respecting the energy condition. For the cases in which $r_*$ is not positive and real, the  interpretation
is as follows: the radiation-accretion density $\sigma(u,r)$ never and nowhere vanishes.
 
\item We have classified the possible situations respecting or violating
the energy condition for all the solutions of black hole in dust, radiation,
quintessence, cosmological constant-like and phantom backgrounds in the Tables
1-7. It is shown that there are cases for all the backgrounds in which the
positive
energy condition is respected in whole spacetime under the determined behaviors of black hole and
its surrounding field. Also, we have given some plots for radiation-accretion density versus some typical values of black hole and its
surrounding fields in Figs 1-5. Using these plots, one realizes
for the dust and radiation backgrounds, although the radiation-accretion
density is a decreasing function but is always positive, and consequently the positive energy condition is satisfied everywhere in spacetime. This is while for
the quintessence, cosmological constant-like and phantom backgrounds,  if the  condition  $|\dot N_{q,c,p}(u)|\ll|\dot M(u)|$ is not met, the positive energy condition is violated in some regions
of spacetime. Comparing the plots with common values of the parameters, we observe that the radiation-accretion
density $\sigma(r)$ for the radiation background is larger than  the dust background at any distance $r$, i.e $\sigma_{r}(r)>\sigma_{d}(r)$. Similarly,
for
the quintessence, cosmological constant-like and phantom backgrounds, we have
 $\sigma_q(r)<\sigma_c(r)<\sigma_p(r)$ for the radiation-accretion density.
\item  We have analyzed the timelike geodesics associated with the obtained surrounded black holes
and have found that two kinds of new correction terms arise relative to the case of Schwarzschild black hole. The first kind of corrections are 
due to the presence of the background fields
which surround the Vaidya black hole. This corrections include two terms in which its first term is similar to the term of Newtonian gravitational
potential, while its second term is similar to the relativistic correction of GR. For the various background fields, we have discussed that there are
possibilities for  the equality of Newtonian and GR correction terms to the
corresponding background fields contributions. We have given some plots denoting
these possibilities for each case. The second kind of corrections is also a non-Newtonian correction
resulting from the dynamics of black hole and its surrounding field. We  have shown that depending on the dynamical features
of black hole and its background, there are also possibilities  that dynamical
correction
terms  can be equal to the Newtonian case. Some plots representing
these situations are given for each case. Then,
 one realizes that for the more realistic non-empty and non-static backgrounds, the geodesic equation of
any object depends strictly not only on the mass of the central object of the system and the angular momentum of the orbiting body, but also on the ($i$) background field
type
and ($ii$) black hole and its background field dynamics. \end{itemize}
We have reported elsewhere on the causal structures and thermodynamical properties of our obtained solutions here \cite{our}.  We also aim to generalize this work
to the case of charged solutions. 
%%%%%%%%%%%%%%%%%%%%%%%%%%%%%%%%%%%%%%%%%%%%%%%%%%%%%%%%%%%%%%%%%%%%%%%%%%%

\end{document}